\documentclass[12pt]{article}

\usepackage{lscape}
\usepackage{dcolumn}
\newcolumntype{d}[1]{D{.}{.}{#1}}

\usepackage{epsfig}
\usepackage{amsmath}
\usepackage{hhline}
\usepackage{amssymb}
\usepackage{times}
\usepackage{cite}

\newlength{\dinwidth}
\newlength{\dinmargin}
\begin{document}  

\newcommand {\gapprox}
   {\raisebox{-0.7ex}{$\stackrel {\textstyle>}{\sim}$}}
\newcommand {\lapprox}
   {\raisebox{-0.7ex}{$\stackrel {\textstyle<}{\sim}$}}
\newcommand{\mx}{M_{_{\rm X}}}                                          
\newcommand{\my}{M_{_{\rm Y}}}                                          
\newcommand{\dd}{\mathrm{d}}
\newcommand{\pom}{{I\!\!P}}
\newcommand{\reg}{{I\!\!R}}
\newcommand{\alphapom}{\alpha_{_{\rm I\!P}}}
\newcommand{\alphareg}{\alpha_{_{\rm I\!R}}}
\newcommand{\xpom}{x_{_{I\!\!P}}}
\newcommand{\etamax}{\eta_{\rm max}}
\newcommand{\sigrd}{\sigma_r^{D(3)}}
\newcommand{\sigrdarg}{\sigma_r^{D(3)}(\xpom,\beta, Q^2)}

\def\Journal#1#2#3#4{{#1} {\bf #2} (#3) #4}
\def\NCA{\em Nuovo Cimento}
\def\NIM{\em Nucl. Instrum. Methods}
\def\NIMA{{\em Nucl. Instrum. Methods} {\bf A}}
\def\NPB{{\em Nucl. Phys.}   {\bf B}}
\def\NPA{{\em Nucl. Phys.}   {\bf A}}
\def\PLB{{\em Phys. Lett.}   {\bf B}}
\def\PRL{\em Phys. Rev. Lett.}
\def\PRD{{\em Phys. Rev.}    {\bf D}}
\def\ZPC{{\em Z. Phys.}      {\bf C}}
\def\EJC{{\em Eur. Phys. J.} {\bf C}}
\def\CPC{\em Comp. Phys. Commun.}

\newcommand{\etal}{{\em et al.}}                    


\begin{titlepage}

\noindent
\begin{flushleft}
DESY 06-049\hfill ISSN 0418-9833\\
May 2006
\end{flushleft}

\vspace*{2cm}

\begin{center}
\begin{Large}

{\bf Measurement and QCD Analysis of the 
Diffractive Deep-Inelastic Scattering Cross Section at HERA \\
}

\vspace{2cm}

H1 Collaboration

\end{Large}
\end{center}

\vspace{2cm}

\begin{abstract}
\noindent
A detailed analysis is presented of the 
diffractive deep-inelastic scattering process $ep\rightarrow eXY$, where
  $Y$ is a proton or a low mass proton excitation carrying a fraction
  $1 - \xpom > 0.95$ of the incident proton longitudinal momentum and
the squared four-momentum transfer at the proton vertex satisfies 
$|t|<1 \ {\rm GeV^2}$. Using data taken by the H1 experiment, 
the cross section is measured 
for photon virtualities in the range 
$3.5 \leq Q^2 \leq 1600 \rm\ GeV^2$,
triple differentially in $\xpom$, $Q^2$ and
$\beta = x / \xpom$, where $x$ is the Bjorken scaling variable.
At low $\xpom$,
the data are consistent with a factorisable
$\xpom$ dependence, which can be described by the exchange of an
effective pomeron trajectory with intercept 
$\alphapom(0)= 1.118 \pm 0.008 \ {\rm (exp.)} \
^{+0.029}_{-0.010} \ {\rm (model)}$.
Diffractive parton distribution functions and their 
uncertainties are determined
from a next-to-leading order DGLAP QCD analysis of the $Q^2$ and 
$\beta$ dependences of the cross section. The
resulting gluon distribution carries
an integrated fraction of around $70 \%$ of the exchanged momentum
in the $Q^2$ range studied.
Total and differential cross sections are also measured 
for the diffractive charged current process
$e^+ p \rightarrow \bar{\nu}_e XY$
and are found to be 
well described by predictions based on the diffractive parton
distributions. 
The ratio of the diffractive to the inclusive 
neutral current $ep$ cross sections is studied.
Over most of the kinematic range, this ratio shows no significant
dependence on $Q^2$ at fixed $\xpom$ and $x$ or on $x$ at fixed 
$Q^2$ and $\beta$.
\end{abstract}

\vspace{0.5cm}

\begin{center}
Submitted to \EJC
\end{center}

\end{titlepage}



\begin{flushleft}

A.~Aktas$^{9}$,                
V.~Andreev$^{25}$,             
T.~Anthonis$^{3}$,             
B.~Antunovic$^{26}$,           
S.~Aplin$^{9}$,                
A.~Asmone$^{33}$,              
A.~Astvatsatourov$^{3}$,       
A.~Babaev$^{24, \dagger}$,     
S.~Backovic$^{30}$,            
A.~Baghdasaryan$^{37}$
% ...

\bigskip{\it
% ...
$ ^{3}$ Inter-University Institute for High Energies ULB-VUB, Brussels;
          Universiteit Antwerpen, Antwerpen; Belgium$^{ c}$ \\
% ...
$ ^{9}$ DESY, Hamburg, Germany \\
% ...
$ ^{24}$ Institute for Theoretical and Experimental Physics,
          Moscow, Russia$^{ k}$ \\
 $ ^{25}$ Lebedev Physical Institute, Moscow, Russia$^{ e}$ \\
 $ ^{26}$ Max-Planck-Institut f\"ur Physik, M\"unchen, Germany \\
% ...
$ ^{30}$ Faculty of Science, University of Montenegro,
          Podgorica, Serbia and Montenegro$^{ e}$ \\
% ...
$ ^{33}$ Dipartimento di Fisica Universit\`a di Roma Tre
          and INFN Roma~3, Roma, Italy \\
% ...
$ ^{37}$ Yerevan Physics Institute, Yerevan, Armenia \\
% ...
$ ^{\dagger}$ Deceased \\

\bigskip
% ...
$ ^c$ Supported by FNRS-FWO-Vlaanderen, IISN-IIKW and IWT
      and  by Interuniversity
Attraction Poles Programme,
      Belgian Science Policy \\
% ...
$ ^e$ Supported by the Deutsche Forschungsgemeinschaft \\
% ...
$ ^k$ Partially Supported by Russian Foundation
      for Basic Research,  grants  03-02-17291
      and  04-02-16445 \\
}

\end{flushleft}
\newpage


\section{Introduction}

Quantum Chromodynamics (QCD) is well established as the gauge
field theory of the strong interaction. However,
it is only able to provide reliable predictions for scattering processes
if they involve
short distance partonic interactions, where 
perturbative methods may be applied.
In contrast, hadronic scattering 
cross sections 
are dominated by soft
interactions, to which perturbation theory is not applicable. 
In a large fraction of these soft interactions, 
often termed `diffractive',
one or both of the interacting hadrons remains intact.
Such processes are commonly discussed in terms of exchanges  
with net vacuum quantum numbers, 
though the exact nature of these exchanges is not well known.

\begin{figure}[h] \unitlength 1mm
 \begin{center}
 \begin{picture}(100,65)
  \put(-28,-3){\epsfig{file=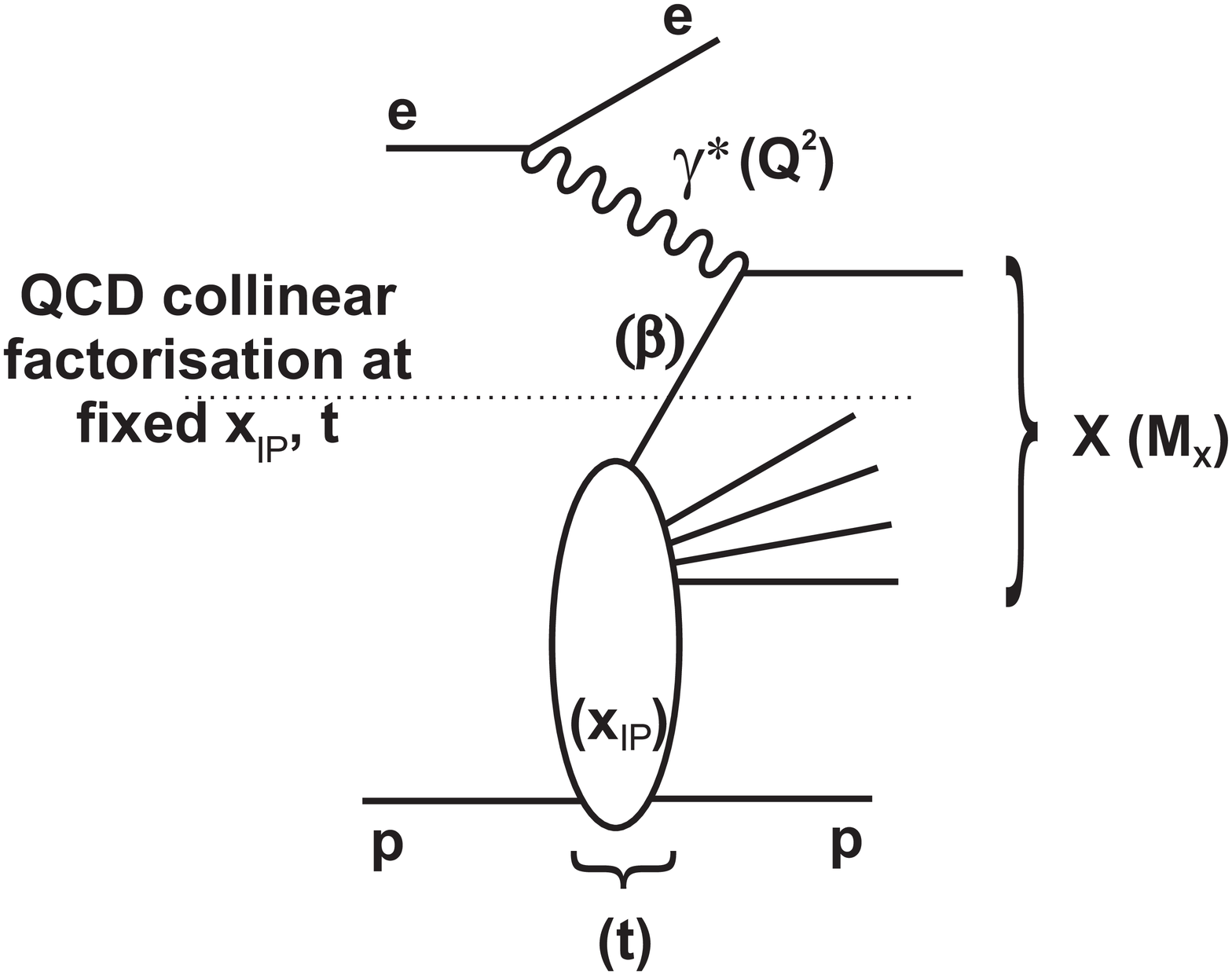,width=0.49\textwidth}}
  \put(61,3){\epsfig{file=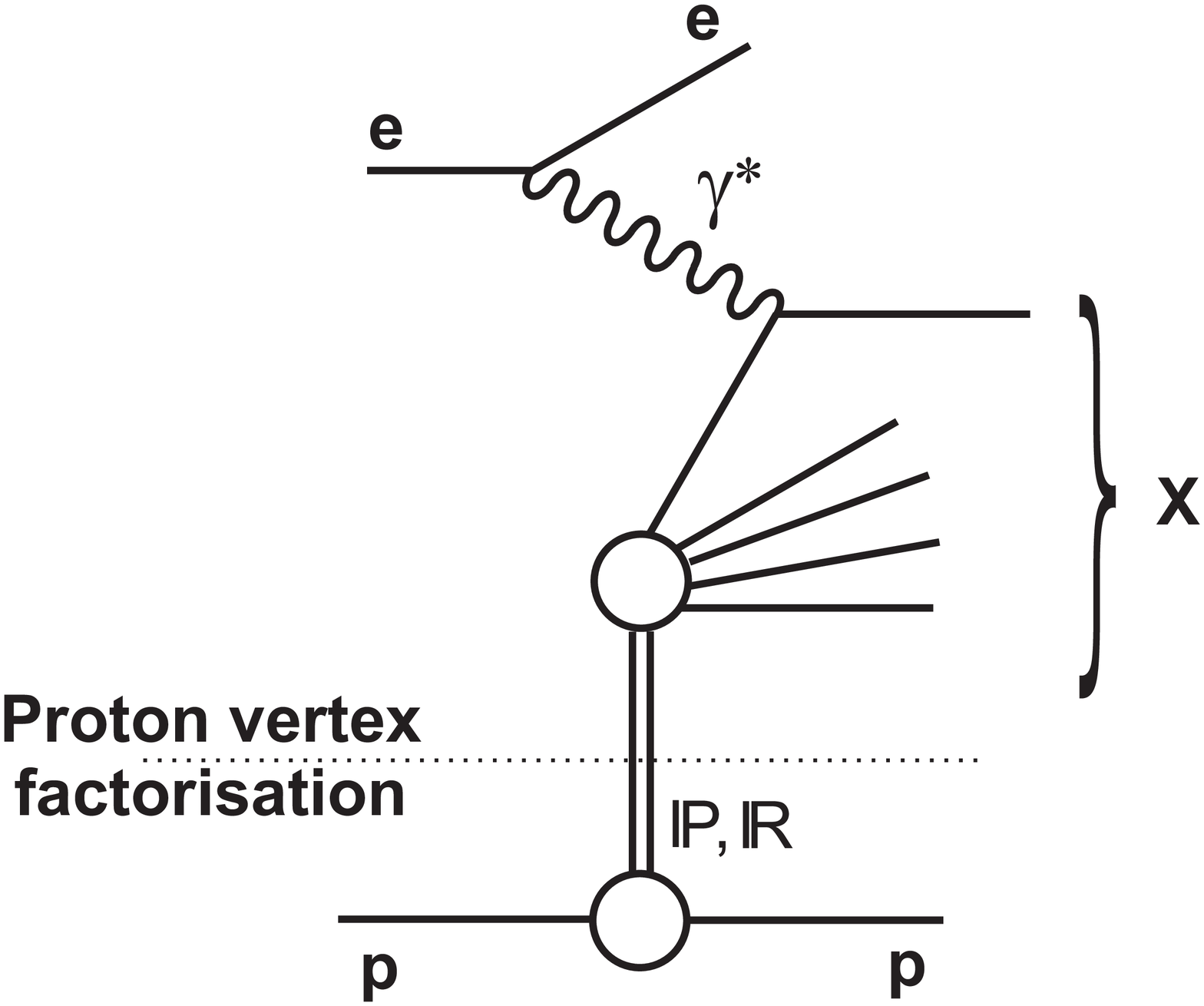,width=0.425\textwidth}}
  \put(-30,60){\Large{\bf{(a)}}}
  \put(63,60){\Large{\bf{(b)}}}
 \end{picture}
 \end{center}
 \caption{Schematic illustration of the neutral current diffractive
DIS process $ep \rightarrow eXp$, proceeding via virtual photon exchange.
The dotted lines in (a) and (b) show the points at which
the diagram can be divided under the assumptions of
QCD hard scattering collinear factorisation and proton
vertex factorisation, respectively.  
The kinematic variables defined in section~\ref{sec:kine} are also
indicated in (a).}
\label{process}
\end{figure}

The observation of high transverse momentum jet production
in diffractive $p \bar{p}$ scattering \cite{UA8} introduced the
possibility of understanding the diffractive exchange in terms of
partons.
The presence of  
processes of the type $ep \rightarrow e X p$ (figure~\ref{process})
in deep-inelastic scattering (DIS) at low Bjorken-$x$ 
at the HERA collider \cite{obsdiff}
offers a uniquely well controlled environment in which to study the
QCD properties and structure 
of diffraction. Several measurements of the 
semi-inclusive cross section 
for this `diffractive DIS' process have been made by the
H1 \cite{f293,h1f2d93,h1f2d94,H1FPS} and 
ZEUS \cite{ZEUS:93,ZEUS:94,zeus02,zeuslps,ZEUS:97}
collaborations. 

The detailed explanation of hard diffraction has
become a major challenge in the development of our understanding of
the strong interaction at high energies and 
low $x$ values \cite{hebecker:review}.
A wide variety of models has been put forward to interpret the
dynamics of diffractive DIS as well as its relationships
to inclusive DIS and to diffractive
hadron-hadron 
scattering \cite{nikolaev:zakharov,semicl,sci,soper,bekw,kgb,absorb}. 
A general theoretical framework is provided by
the proof \cite{collins} of 
a hard scattering QCD collinear factorisation
theorem \cite{berera:soper1,berera:soper2,kunszt:stirling}
for semi-inclusive DIS cross sections such as that for 
$ep \rightarrow eXp$.  As illustrated in figure~\ref{process}a,
this theorem implies
that the concept of `diffractive 
parton distribution functions' (DPDFs)~\cite{berera:soper2,facold}
may be introduced, representing conditional proton parton probability
distributions under the constraint of a leading final state proton with a
particular four-momentum.  
Empirically, a further factorisation has been found to apply to good
approximation, whereby
the variables which describe the proton vertex factorise
from those describing the hard interaction \cite{h1f2d94,H1FPS}, 
as illustrated in figure~\ref{process}b. According to this
`proton vertex' factorisation,
the shape of the DPDFs is independent of the
four-momentum of the final state proton. 
The dependence of the DPDF normalisation on the proton
four-vector can be parameterised conveniently using Regge asymptotics,
which amounts to a description of
diffraction in terms of the exchange of a factorisable 
`pomeron' ($\pom$) \cite{pomeron}
with universal parton densities \cite{ingschl}. 

Several authors have analysed diffractive DIS data
to extract 
DPDFs 
\cite{h1f2d94,zeus:gpjets,kgb:kw,stirling,actw,soper,zeuslps,royonbartelsfits,watt}.
In \cite{h1f2d94,H1FPS}, H1 measurements are studied, 
with the conclusion that the data are
compatible with proton vertex factorisation at low
fractional proton energy losses, $\xpom$.
At larger $\xpom$, 
a separately factorisable sub-leading exchange ($\reg$),
with a different $\xpom$ dependence and partonic composition, is present.
The DPDFs extracted in \cite{h1f2d94} are
dominated by the gluon distribution.
Further tests of the
factorisation properties of diffractive DIS have been made by
comparing predictions using these DPDFs 
with hadronic final state observables such as diffractive jet
\cite{h1jets} and heavy quark \cite{h1dstar} cross sections. These tests
have shown a remarkable internal consistency within
the HERA DIS data. 
In contrast,
the DPDFs extracted in DIS are not expected to be directly applicable to
hadron-hadron 
scattering \cite{fac:break,berera:soper1,berera:soper2,collins}. 
Indeed diffractive
factorisation breaks down spectacularly when the DPDFs 
from \cite{h1f2d94} are applied
to diffractive $p \bar{p}$ interactions at the Tevatron \cite{tevatron}. 
However, with the introduction of 
an additional `rapidity gap survival probability' 
factor to account for secondary interactions between the
beam remnants \cite{gap:survival}, the HERA DPDFs remain an essential
ingredient in the phenomenology of 
diffraction at the Tevatron and
the LHC \cite{lhcdiff}.

In this paper,
a new measurement\footnote{The results presented here are in
agreement with the previous H1 measurement using similar 
techniques \cite{h1f2d94}
throughout most of the measured kinematic range. 
The new measurement and DPDFs supersede the old due to the 
improvements in precision, kinematic coverage,
theoretical modelling and detector understanding.} 
of the diffractive neutral current DIS
cross section is presented. This is
based upon H1 data 
for which there is an absence of hadronic activity in a large 
rapidity region extending close to the outgoing proton beam 
direction. It is thus complementary to measurements such 
as \cite{H1FPS}, in which the
leading proton is detected and measured.
The proton vertex factorisation property is tested and 
the dependence of the diffractive cross section on $\xpom$
is expressed in terms of an effective pomeron intercept $\alphapom(0)$.  
The dependence on $x$ and $Q^2$ is interpreted
through a QCD analysis using the DGLAP \cite{dglap} evolution equations at
next-to-leading order (NLO) \cite{dglapnlo},
from which new DPDFs are determined.
The kinematic range of validity of the DPDFs
is tested systematically and,
for the first time, an assessment is made of the experimental
and theoretical uncertainties.
A first measurement is also made of the
diffractive charged current cross section, 
which is compared
with a prediction based on the DPDFs extracted from the neutral current data.
The ratio of diffractive to inclusive cross sections and its kinematic
dependences are also investigated.
Section~\ref{sec:kine} introduces the formalism adopted for the paper.
Section~\ref{sec:expproc} describes the cross section measurement. The 
results are presented in sections~\ref{sec:results}--\ref{sec:incdif} 
and are followed by a 
summary in section \ref{sec:summary}.


\section{Diffractive DIS Kinematic Variables and Observables}
\label{sec:kine}

The data studied here are subsets of inclusive H1 neutral and
charged current DIS samples, arising from the
processes $e^+ p \rightarrow e^+ X^\prime$ and 
$e^+ p \rightarrow \bar{\nu}_e X^\prime$, where the positron 
(with four-momentum $k$) couples to an electroweak gauge boson ($q$), 
which interacts with the proton
($P$). The usual DIS kinematic variables are defined as
\begin{equation}
Q^2=-q^2 \ , \qquad x=\frac{-q^2}{2 P\cdot q} \ , \qquad
y=\frac{P\cdot q}{P \cdot k} \ ,
\end{equation}
where $Q^2$ is the boson virtuality, $x$ is the
longitudinal momentum fraction of the proton carried by the
struck quark 
and $y$ measures the inelasticity of the process.  The squared invariant
masses of the positron-proton and gauge boson-proton systems 
are $s =  (k+P)^2$ and $W^2 = (q+P)^2$, respectively.

The hadronic final state of any 
DIS event may be broken down into two systems $X$ and $Y$, separated
by the largest gap in the rapidity distribution of the hadrons 
relative to an axis defined by the exchanged boson and the 
proton in their centre of mass frame \cite{h1f2d94}. If the masses 
$\mx$ and $\my$ of these two systems are small compared with 
the mass $W$ of the full hadronic final state, the two systems are
expected to be
separated by a large rapidity gap and a colourless exchange of 
well defined four-momentum may be considered to have taken place between them. 
As illustrated in figure~\ref{process}a, 
the longitudinal momentum fractions,
$\xpom$ of the colourless exchange with respect to the incoming
proton, and $\beta$ of the struck quark with respect to the colourless
exchange, are then defined by
\begin{eqnarray}
\xpom = \frac{q \cdot (P - p_Y)}{q \cdot P} \ ,
\qquad 
\beta = \frac{Q^2}{2 q \cdot (P - p_Y)} .
\end{eqnarray}
Here, $p_Y$ is the four-momentum of the $Y$ system
and $\beta \xpom = x$. 
The squared four-momentum transferred at 
the proton vertex is
\begin{eqnarray}
t=(P-p_Y)^2 \ .
\end{eqnarray}
The rapidity gap selection (section~\ref{difsel}) implies that the
data analysed in this paper are dominated by the case where
$Y$ is a lone proton and $|t|$ is relatively small. 
However, since the system $Y$ is not
detected directly, a small admixture of proton excitations and other
systems such as leading neutrons is also present (see section~\ref{pdiss}).

The neutral current data
are presented in the form of a 
`diffractive reduced cross section' $\sigma_r^{D(3)}$, 
integrated over the ranges of $t$ and $M_Y$ specified in 
section~\ref{pdiss} and
related to the
differential cross section measured experimentally by
\begin{equation}
\frac{{\rm d}^3\sigma^{ep \rightarrow e X Y}}{\dd \xpom \ \dd x \ \dd Q^2} = \frac{2\pi
  \alpha^2}{x Q^4} \cdot Y_+ \cdot \sigma_{r}^{D(3)}(\xpom,x,Q^2) \ ,
\label{sigmar}
\end{equation}
where $Y_+ = 1 + (1-y)^2$.  
Similarly to inclusive DIS \cite{h1nc9900},
the reduced $e^+ p$ cross section depends on the 
diffractive structure functions
$F_2^{D(3)}$ and $F_L^{D(3)}$ 
in the one-photon exchange approximation
according to
\begin{equation}
\sigma_r^{D(3)} = 
F_2^{D(3)} - \frac{y^2}{Y_+} F_L^{D(3)}.
\label{sfdef}
\end{equation}
For $y$ not too close to unity, 
$\sigma_r^{D(3)} = F_2^{D(3)}$ holds to very good approximation. 
In previous measurements of inclusive diffractive DIS at HERA, 
the data were presented in terms of
$F_2^{D(3)}$ instead of $\sigma_r^{D(3)}$.

Due to the smaller available data sample,
the charged current measurements must be integrated over some or all of the
kinematic variables. They are presented as a 
total cross section and single differentially in either $\xpom$, $\beta$
or $Q^2$.


\section{Experimental Procedure}
\label{sec:expproc}

\subsection{H1 Apparatus}
\label{sec:h1det}

The H1 coordinate system 
is defined such that the origin is at the nominal $ep$ interaction point and
the polar angle
$\theta = 0$ corresponds to the direction of the outgoing proton beam.
The region $\theta < 90^\circ$, which has
positive pseudorapidity $\eta = - \ln \tan
\theta /2$, is referred to as the `forward' hemisphere.  

The interaction region is surrounded by the central tracking 
system, which consists of a silicon vertex detector, drift chambers
and multi-wire proportional chambers,
all located within a solenoidal magnetic
field of $1.15 \rm\ T$.
The trajectories of charged
particles are measured in the range $-1.7<\eta<1.7$ with a transverse
momentum resolution of $\sigma(p_T)/p_T \simeq 0.005 \,
p_T \, /\mathrm{GeV} \oplus 0.015$. 
The Forward Tracking Detector and the 
Backward Drift Chamber (BDC) provide
track segments of charged particles at smaller and larger $\theta$
than the central tracker, respectively.

A highly segmented Liquid Argon (LAr) sampling calorimeter, covering the range
$-1.5<\eta<3.4$, surrounds the tracking chambers and consists of
electromagnetic and hadronic sections.  The energy resolution is 
$\sigma(E)/E \simeq 11\%/\sqrt{E/\mathrm{GeV}}$ for electrons and
$\sigma(E)/E \simeq 50\%/\sqrt{E/\mathrm{GeV}}$ for hadrons, as
obtained from test beam measurements \cite{lartestbeam}.  
The backward direction ($-4.0<\eta<-1.4$) is covered by a lead /
scintillating fibre calorimeter (SpaCal), 
which also has both
electromagnetic and hadronic sections. The energy resolution for
electrons is $\sigma(E)/E \simeq 7\%/\sqrt{E/\mathrm{GeV}}$.  

In addition to the `central detectors' described above, a set of 
`forward detectors' are also used in the present analysis. 
The copper / silicon Plug calorimeter, the Forward Muon Detector (FMD)
and the Proton Remnant Tagger (PRT) 
are sensitive to hadronic activity at large pseudorapidity, near to the
outgoing proton beam.  The 
Plug enables energy measurements to be made in the
pseudorapidity range $3.5<\eta<5.5$.  The FMD is 
a series of drift chambers 
covering the range $1.9<\eta<3.7$. Primary
particles
produced at larger $\eta$ are often detected indirectly in the FMD 
if they undergo a secondary
scattering with the beam-pipe or other material. 
The PRT, a set of scintillators
surrounding the beam pipe at $z=26 \ \mathrm{m}$, detects charged
particles in the region $6.0 \ \lapprox \ \eta \ \lapprox \ 7.5$.  

The $ep$
luminosity is determined by measuring the
rate of the QED Bremsstrahlung process
$ep\rightarrow ep\gamma$ using a photon calorimeter close to the 
backward beam
pipe at $z=-103 \ \mathrm{m}$.
A much more detailed description of the H1 apparatus can be found 
in \cite{h1det,spacal}.


\subsection{Data Samples}

Different event samples are used 
for different $Q^2$ ranges of the measurement,
as summarised in table \ref{tab:datasets}.  
For the interval $3<Q^2<13.5 \rm\ GeV^2$,
a `minimum bias' sample (`1997 MB') corresponding to an 
integrated luminosity of $2.0 \ {\rm pb^{-1}}$ is used,
which was recorded during a dedicated data 
taking period in 1997
with unbiased triggers. For intermediate photon virtualities
($13.5<Q^2<105 \ {\rm GeV^2}$), data taken throughout 1997
are used (`1997 all'), corresponding to an
integrated luminosity of $10.6 \ {\rm pb^{-1}}$. 
The kinematic range $Q^2>133 \rm\ GeV^2$
is covered by a sample (`1999-2000') corresponding to 
$61.6 \ {\rm pb^{-1}}$, taken in the years 1999 and 2000. 

For all three samples, HERA collided positrons with protons, the positron
beam energy being $E_e = 27.5 \ {\rm GeV}$ in each case. The proton beam
energy was $E_p = 820 \ {\rm GeV}$ in 1997 and 
$E_p = 920 \ {\rm GeV}$ in 1999 and 2000, leading to $ep$
centre of mass energies of $\sqrt{s} = 301 \ {\rm GeV}$ and 
$\sqrt{s} = 319 \ {\rm GeV}$, respectively. 
The `1997 MB' and `1997 all' samples are used to study neutral current
interactions with the scattered 
electron\footnote{The 
scattered positron is referred to as an electron throughout this paper.}  
detected in the SpaCal calorimeter.
More details of the analysis of diffractive data 
with SpaCal electrons may be found in \cite{carrie:thesis}. 
The `1999-2000' sample is used for the
study of both neutral and charged current interactions. In the neutral
current case, the scattered electron is detected in the LAr
calorimeter. These measurements
are further described in \cite{paul:thesis}.

\renewcommand{\arraystretch}{1.35} 
\begin{table}[h]
\centering
\begin{tabular}{|r@{$Q^2$}l|l|c|r@{$.$}l|}
\hline 
\multicolumn{2}{|c|}{$Q^2$ range} & \multicolumn{1}{c|}{Data Set} & Proton Energy $E_p$  & \multicolumn{2}{c|}{Luminosity}  \\
\hline \hline
$3 <$ \ & \ $< 13.5 \rm\ GeV^2$ & 1997 MB  & $820 \rm\ GeV$ & $2$ & $0 \ {\rm pb^{-1}}$ \\
$13.5 <$ \ & \ $< 105 \rm\ GeV^2$ & 1997 all & $820 \rm\ GeV$ & $10$ & $6 \ {\rm pb^{-1}}$ \\
& \ $> 133 \ {\rm GeV^2}$ & 1999-2000 & $920 \rm\ GeV$ & $61$ & $6 \ {\rm pb^{-1}}$ \\
\hline
\end{tabular} 
\caption{Summary of the data samples used in the analysis.}
\label{tab:datasets}
\end{table}
\renewcommand{\arraystretch}{1.} 


\subsection{Selection and Reconstruction of DIS Events}

The trigger conditions, detector alignment and calibration, and 
inclusive DIS selection criteria 
are very similar to those used in 
the analogous fully inclusive H1 analyses at low \cite{h1f29697}
and high \cite{h1nc9900} $Q^2$. The selection 
criteria are summarised below.

Neutral current
DIS events are triggered by the detection of an energetic
electromagnetic calorimeter cluster attributed to the scattered electron. 
Inefficient regions of the calorimeters, for example due
to cracks between modules or poorly performing trigger cells, are not
included in the analysis. The trigger efficiency is then close to $100\%$ for
the ranges in electron energy considered here, namely 
$E_e^\prime > 6.5 \ {\rm GeV}$ for electrons detected in the SpaCal and
$E_e^\prime \ \gapprox \ 11 \ {\rm GeV}$ for LAr electrons.
To suppress photoproduction background, in which the scattered electron
escapes undetected in the backward direction and a hadron fakes the
electron signature, cuts are applied
on the lateral extent and isolation
of the cluster forming the electron candidate and its
containment within the electromagnetic part of the calorimeter.

An event vertex, reconstructed by the central or forward tracker, 
is required within $30 \ (35) \rm\ cm$ 
of the nominal
interaction point for the SpaCal (LAr) electron
samples. To suppress background where a photon fakes the scattered electron, 
a charged particle 
track segment must be associated to the electron candidate. This track is 
reconstructed in the BDC for SpaCal electron candidates and in the
central tracking system for LAr electron candidates.
The polar angle $\theta_e'$ of the scattered electron 
is calculated from the interaction vertex and the BDC track
(the LAr cluster) for SpaCal (LAr) electrons. 
In order to ensure good acceptance for the electron in the calorimeters
and associated trackers, the scattered electron polar angle range
considered is $156^\circ < \theta_e' < 176.5^\circ$ for the 
SpaCal electron samples and $\theta_e' < 153^\circ$ for the LAr
electron sample.

Hadronic final state objects are reconstructed from the
LAr and SpaCal calorimeters and the central tracking system using
an energy flow algorithm 
which combines charged particle tracks with calorimeter deposits
without double counting \cite{fscomb}.  Isolated low energy
calorimeter deposits are classified as noise and are rejected from the
analysis. To further suppress backgrounds, a 
minimum of two remaining hadronic final state objects
is demanded.
Consistency is required between the variable $y$, as reconstructed
according to
\begin{eqnarray*}
y_e & = & 1-E_e'/E_e \cdot \sin^2 (\theta_e^\prime /2) 
        \hspace*{2.625cm} {\rm (Electron \ method)} \ , \\
y_h & = & (E-p_z)_h / 2 E_e 
        \hspace*{3.95cm} {\rm (Hadron \ method)} \ , \\
y_d & = & \tan(\gamma/2)/\left[ \tan(\theta_e^\prime/2)+\tan(\gamma/2) \right]
        \ \ \ \ \ \ {\rm (Double \ angle \ method)} \ ,
\end{eqnarray*}
using, respectively, the scattered electron only, the 
hadronic final state\footnote{Here and in the following, the
four-vector of the reconstructed hadronic final state is denoted $(E, p_x, 
p_y, p_z)_h$. In diffractive events, $ep \rightarrow eXY$,
where the leading baryonic system $Y$
is not observed, this corresponds to the four-vector of the $X$ system.}
only \cite{hadmeth},
and the electron and hadronic final state polar angles 
$\theta_e^\prime$ and $\gamma$ \cite{dameth}.
The criteria $|y_e - y_h| < 0.25$ and 
$|y_e - y_d| < 0.25$ remove badly reconstructed events, further suppress
photoproduction background 
and reduce QED radiative corrections due to
photon emission from the initial state positron.

The variables $y$, $Q^2$ and $x$ are reconstructed by
combining information from the scattered electron and the
hadronic final state 
using the method introduced in \cite{h1f2d94}:
\begin{equation}
  y = y_e^2 + y_d \ (1 - y_d) \ ; \qquad
  Q^2 = \frac{4 E_e^2 
  \ ( 1 - y)}{\tan^2 (\theta^\prime_e / 2)} \ ; \qquad x = \frac{Q^2}{s y} \ .
\end{equation}
This reconstruction method interpolates between 
the electron method at large $y$ where it has
the best performance and the double angle method at
low $y$. For diffractive events, where the $X$ component of the hadronic 
final state is well
contained in the central detectors, the polar 
angle $\gamma$
is well measured and the double angle method has an improved
resolution compared with that for non-diffractive events. This method
yields a resolution of $5-15\%$ in $y$,
improving with increasing $y$.
The resolution in $Q^2$ is around $3\%$.

The principal selection criterion for the charged current sample
is a large missing transverse
momentum, corresponding to the unobserved final state neutrino. This
is identified at the trigger level 
mainly using the Liquid Argon calorimeter. 
The trigger efficiency exceeds 60\% 
throughout the kinematic range studied 
here. For the final selection, the missing transverse 
momentum must exceed $12 \ {\rm GeV}$. A reconstructed event vertex is required
as for the neutral current case. To suppress backgrounds, further selection 
criteria are applied on the event topology, as described in \cite{h1nc9900}.
The inclusive kinematic variables 
are reconstructed using the final state hadrons \cite{hadmeth}.

The selection $y > 0.04$ is applied to all data samples to ensure 
reasonable 
containment of the hadronic final state in the central detectors. For the 
neutral current LAr electron data, the sample is
restricted to $y_e <0.63$ for $Q^2 < 890 \ {\rm GeV^2}$ and to
$y_e < 0.9$ for $Q^2 > 890 \ {\rm GeV^2}$, which
suppresses photoproduction background.


\subsection{Selection and Reconstruction of Diffractive Events}
\label{difsel}

Diffractive events are selected on the basis of a large rapidity gap,
separating the leading baryonic system $Y$ from the 
system $X$. The rapidity gap is inferred from the absence
of activity in detectors sensitive to forward energy flow. 
The pseudorapidity of the most forward energy deposit
above a noise threshold of $400 \ {\rm MeV}$ 
in the LAr calorimeter must satisfy 
$\eta_{\rm max} < 3.3$. This requirement ensures that the
forward extent of the
$X$ system lies within the acceptance range of the main detector 
components and thus that $\mx$
can be reconstructed reliably. There must
also be no activity above 
noise thresholds in the Plug, FMD and PRT detectors. 
Studies of the correlations between the activity levels in the 
different forward detector
components show that the LAr, Plug, FMD and PRT requirements 
have rejection efficiencies for events with no large rapidity gap of
around 95\%, 80\%, 80\% and 30\%, respectively. These efficiencies are
well described by the simulations of inclusive DIS used in the analysis 
(section~\ref{sec:mc}) and the combined 
efficiency for the rejection of events with hadronic activity in the range
$3.3 < \eta \ \lapprox \ 7.5$ is close to 100\%.
Corrections of around $5\%$, evaluated using randomly triggered events,
are applied to account for the component of the
diffractive signal which is rejected due to electronic noise, synchrotron
radiation or other effects which fake activity in the forward detectors.

The diffractive event kinematics are reconstructed using the 
mass of the system $X$, obtained from
\begin{eqnarray}
  \mx = 
\sqrt{(E^2 - p_x^2 - p_y^2 - p_z^2)_h \cdot
\frac{y}{y_h}} \ \ .
\label{mx:rec}
\end{eqnarray}
Neglecting the transverse momentum of the hadrons,
this method of reconstructing $\mx$ reduces 
at large $y$ to a measurement of
$\sqrt{2 (E + p_z)_h \ (E_e - E_e^\prime \sin^2 \frac{\theta_e^\prime}{2})}$ , thus
improving the resolution
where losses of hadrons in the backward ($-z$) direction become significant. 
The resolution in 
$\mx$ varies between $20\%$ and $30\%$ in the measured kinematic range.
The diffractive variables $\beta$ and $\xpom$ are obtained from
\begin{eqnarray}
   \beta = \frac{Q^2}{Q^2 + \mx^2 } \ ; \qquad
   \xpom = \frac{x}{\beta} \ .
\end{eqnarray}

The sensitivity of the measurement to variations in the details of the 
selection and reconstruction has been tested in detail. For example, there
is no significant change in the results when one or more of the forward
detectors is not used in the measurement.
The final charged current sample is scanned visually and all events are
consistent with production via
$ep$ scattering in the interaction region.


\subsection{Measurement Ranges in {\boldmath {$\my$}} and {\boldmath {$t$}}}
\label{pdiss}

The large rapidity gap selection yields a sample which is dominated by the
single dissociation process $ep \rightarrow eXp$, with the proton transverse
momentum $p_t(p)$, and hence $|t| \simeq p_t^2(p)$, relatively small. 
However, there is an
admixture of proton dissociative events, $ep \rightarrow
eXY$, where the proton dissociation system has a small mass $\my$. 
The ranges of sensitivity of the measurement in $\my$ and $t$ are determined
by the acceptances of the forward detectors which are used to 
identify the large rapidity gap (section~\ref{difsel}).

In order to keep the uncertainties arising from proton dissociation
small and to ease comparisons with 
previous data \cite{h1f2d94,h1jets,h1dstar}, the
measurement is integrated over the region
\begin{eqnarray}
\my < 1.6 \ {\rm GeV}
\ , \qquad
|t| < 1 \ {\rm GeV^2} \ .
\end{eqnarray}
The correction
factors applied to account for the net migrations about these 
limits are determined by evaluation of the
forward detector response to elastic proton and proton dissociative 
processes\footnote{Only proton dissociation
to low $\my$ states is considered in this procedure. Proton dissociation
with $\my > 5 \ {\rm GeV}$ is simulated using the inclusive Monte
Carlo models as described in section~\ref{sec:mc}.},
using the DIFFVM \cite{diffvm} simulation.
The ratio of the full generated
proton dissociation cross section to the generated
elastic cross section is taken to be
1:1. Proton dissociation is simulated using an approximate
${\rm d} \sigma / {\rm d} \my^2 \propto 1 / \my^2$ dependence, with
explicit simulations of the most important
resonances at low $\my$ \cite{pdg}. The $t$
dependence for proton dissociation follows 
${\rm d} \sigma / {\rm d} t \propto e^{B_{PD} t}$ with
a slope parameter 
$B_{PD} = 1 \ {\rm GeV^{-2}}$.
The uncertainties
are evaluated by varying
the details of this simulation as described in section~\ref{sec:syst}.
The resulting correction factors are $-8.2 \pm 5.8 \%$ (`1997 MB'), 
$-8.6 \pm 5.8 \%$ (`1997 all') and $-12.0 \pm 7.4 \%$ (`1999-2000').

Comparison of the current data with
a similar measurement in which the leading proton is directly 
measured \cite{H1FPS} yields a ratio of cross sections for 
$\my < 1.6 \ {\rm GeV}$ and $\my = m_p$ of 
$1.23 \pm 0.03 \ {\rm (stat.)} \ \pm 0.16 \ {\rm (syst.)}$
which is consistent with the DIFFVM prediction of $1.15^{+0.15}_{-0.08}$.
Neither the comparison in \cite{H1FPS}, 
nor further studies sensitive to the larger
$\my$ region \cite{carrie:thesis}, show any evidence for a dependence 
of the ratio of proton dissociation to elastic cross sections on 
any of the measured kinematic variables in the region under study.


\subsection{Simulations and Corrections to the Data}
\label{sec:mc}

Corrections for detector inefficiencies, acceptances and migrations
between $\xpom$, $\beta$ and $Q^2$ 
measurement intervals are evaluated using a Monte Carlo
simulation which combines several models. 
The RAPGAP 
\cite{rapgap} event generator simulates the processes
$e^+ p \rightarrow e^+ Xp$ and $e^+ p \rightarrow \bar{\nu}_e Xp$ 
with $\xpom < 0.15$, assuming proton vertex factorisation.
Both pomeron and sub-leading exchanges are included,
based on
the DPDFs from a
leading order QCD fit to previous H1 data 
(`fit 2' in \cite{h1f2d94}).
The parton densities are evolved using $Q^2$ as 
a scale and are convoluted with leading order QCD matrix elements.
Higher order QCD radiation is modelled using initial and final state
parton showers in the leading $\log(Q^2)$
approximation \cite{meps}.
Hadronisation is simulated 
using the Lund string model \cite{lund} as implemented in
JETSET \cite{jetset}.  QED radiative effects, including virtual loop
corrections, are taken into account via
an interface to the HERACLES program \cite{heracles}.
Small weighting factors are applied to the neutral current simulation
to ensure that the $t$ dependence 
matches that measured in \cite{H1FPS} 
and to optimise the description of the
current data. 
The DIFFVM model \cite{diffvm} is used to
simulate the exclusive production of the $\rho$, $\omega$,
$\phi$ and $J/\psi$ vector mesons, which contribute 
significantly in the SpaCal electron samples at small
$\mx$ (high $\beta$).

Due to the small inefficiency in the rejection of events with forward
hadronic activity 
using the forward detectors,
small contributions in the selected data arise from
the regions $\xpom > 0.15$ and $\my > 5 \ {\rm GeV}$. These backgrounds are
subtracted using a simulation based on the DJANGO \cite{django} Monte
Carlo model of inclusive DIS for the SpaCal electron sample
and the non-diffractive RAPGAP simulation \cite{rapgap} for the LAr electron
and charged current samples. 
The cross sections in the inclusive simulations 
are obtained from QCD fits to recent H1 
DIS data \cite{h1f29697,h1nc9497}.

Residual photoproduction background, which is 
sizeable only at the highest
$y$ values, is subtracted on a statistical basis. 
Its contribution to the data 
is evaluated using the PHOJET~\cite{phojet} model for
the SpaCal electron data and the PYTHIA \cite{jetset}
Monte Carlo model for the charged current sample. 
For the neutral current LAr data, the background 
is evaluated from the fraction of reconstructed events 
for which the detected lepton candidate has the opposite charge to the beam
lepton, under the assumption that the background is charge 
symmetric \cite{h1nc9900}.
The small backgrounds near $\beta = 1$ from QED-Compton scattering
($ep\rightarrow ep\gamma$) and from di-lepton production via 
photon-photon fusion ($ep \rightarrow ep e^+e^-$) are subtracted using
the COMPTON \cite{compton} and LPAIR \cite{lpair} Monte Carlo 
simulations, respectively.
In the charged current measurement, a
further small background from the production of real electroweak gauge bosons
is simulated using the EPVEC \cite{epvec} Monte Carlo model.
The normalisations and kinematic dependences of each of the background
simulations have been checked using dedicated alternative selections 
designed to enhance the corresponding background.


\subsection{Systematic Uncertainties}
\label{sec:syst}

A detailed systematic error analysis has been performed, in which the
sensitivity of the measurements to variations in the efficiencies and
energy scales of the detector components and to the
details of the correction procedure is tested. 
For the neutral current measurement, the systematic
error sources leading to uncertainties which are correlated between 
data points are listed below.

\begin{itemize}

\item The uncertainty on the SpaCal 
electromagnetic energy scale varies from 2.4\%
at $E_e = 6.5 \ {\rm GeV}$ to 0.5\% at 
$E_e = 27.5 \ {\rm GeV}$ \cite{h1f29697}. For electrons detected in
the LAr calorimeter, the energy scale is known to a precision varying  
between 1\% and 1.5\%, depending on $\theta_e^\prime$ \cite{h1nc9900}. 
The uncertainties
in the relative alignment of the different detector components are 
reflected in possible biases in the electron polar angle measurement
at the level of $0.5 \ {\rm mrad}$
for the SpaCal data \cite{h1f29697} and between 
$1 \ {\rm mrad}$ and $2 \ {\rm mrad}$, depending on $\theta_e^\prime$, 
for the LAr data \cite{h1nc9900}. 
  
\item The hadronic energy scale of the LAr
calorimeter is known to $2\%$ for all samples studied. 
The uncertainty on the hadronic energy scale of the
SpaCal is 7\% and that on the contribution
to the hadronic energy measurement from charged particle tracks is
3\%.

\item Imperfect treatment of calorimeter noise can result in a bias in the
reconstruction of $\mx$. The corresponding uncertainty is evaluated 
by varying the amount of calorimeter energy classified as noise by 
$10 \%$. This level of precision is determined by 
comparing the calorimeter noise subtracted in the data with the
Monte Carlo model, which includes a simulation of noise based on 
randomly triggered events.
  
\item The efficiency with which the FMD registers activity when there
is hadronic energy flow in its acceptance region is varied by 
5\% in the simulation. For the PRT, this efficiency is varied by 20\%. 
The Plug energy scale is varied by 30\%. These levels of
uncertainty are obtained
by comparison of the present data with the Monte Carlo simulation
for samples in which forward detector activity is required to be present
rather than absent.
  
\item The model dependence of the acceptance and migration corrections
and background subtractions
is estimated by varying the details of the Monte Carlo simulation
within the limits permitted by the present data. 
In the RAPGAP simulation of diffraction for the SpaCal electron data,
the $\xpom$ distribution is reweighted by 
$(1/\xpom)^{\pm 0.05}$, the $\beta$ distribution by
$\beta^{\pm 0.05}$ and $(1-\beta)^{\pm 0.05}$, the $t$ distribution by
$e^{\pm t}$ and the $Q^2$ distribution by $(\log Q^2)^{\pm 0.2}$. 
The same systematic shifts are applied for the LAr electron data, except
that the powers by which the $\xpom$, $\beta$ and $1-\beta$ 
distributions are reweighted are increased from $0.05$ to $0.1$, 
reflecting the
weaker constraints on those distributions from the data at high $Q^2$.
The normalisation of the sub-leading meson exchange
in RAPGAP is varied by $\pm 25\%$ and that of the
vector meson production simulation (DIFFVM) is varied by $\pm 50\%$. 
The uncertainty in the
background from high $\xpom$ or $\my$, as simulated by 
the DJANGO and inclusive RAPGAP Monte Carlo models, is taken to be 100\%.
Appropriate
variations are also made in the normalisations of the 
photoproduction, QED-Compton scattering 
and lepton pair production background simulations.

\end{itemize}

Several further uncertainties, listed below, 
affect all data points in an identical
manner and are thus considered as normalisation uncertainties. 

\begin{itemize}

\item The uncertainty on the factor accounting for
smearing about the $\my$ and $t$ boundaries of the measurement
(section~\ref{pdiss}) is 5.8\% for the data taken in 1997 and
7.4\% for the 1999-2000 data. The dominant contribution to this 
uncertainty arises from variations in the 
assumed ratio of proton dissociation
to elastic proton cross sections in the range 0.5 to 2.0, 
which is determined from
studies with alternative forward detector requirements. 
Smaller contributions arise from reweighting the 
$\my$ and $t$ distributions of the proton dissociation
simulation and propagating the uncertainties on the FMD, PRT and
Plug detectors.

\item The uncertainty arising from the luminosity measurement 
is 1.5\% for all samples.

\item The correction factor for diffractive events
  rejected due to noise fluctuations in the forward detectors
is varied by 25\%, which corresponds to
the r.m.s. variation over the different fills of HERA. This leads to  
normalisation uncertainties at the $1\%$ level, varying slightly between
the different data sets.

\end{itemize}

A final class of systematic errors leads to uncertainties which are not
taken to be correlated between data points. 

\begin{itemize}

\item The calculated acceptance of the $\etamax$ cut depends on the modelling
of the hadronic final state topology. The associated uncertainty is 
estimated from the effect of using an alternative model for higher
order QCD processes (the colour dipole 
approach \cite{cdm}
as implemented in ARIADNE \cite{ariadne} in place of parton showers). 
This results in an uncertainty which
depends to good approximation on $\xpom$ only and varies between 1.2\% 
at $\xpom = 0.0003$ and 11\% at $\xpom = 0.03$. 

\item The uncertainty in the trigger efficiency is
1.0\% in the SpaCal data \cite{h1f29697} and 0.3\% for  LAr 
data \cite{h1nc9900}.

\item Uncertainties of 0.5\% arise in all neutral current data sets
due to the uncertainty in the efficiency of the
track-link requirements for the 
electron candidate\cite{h1f29697,h1nc9900}.
  
\end{itemize}

The total systematic uncertainty on each data point is formed by adding
the individual contributions in quadrature.
Away from the boundaries of the kinematic region studied in the
neutral current measurement, the 
systematic error varying from point to point is around 
$5\%$, with 
no single source of uncertainty dominating. The 
systematic error increases to typically $15\%$ at 
the largest $\xpom = 0.03$, where the contribution from the
modelling of the acceptance of the $\etamax$ requirement becomes
important. At small $\mx$ values, the calorimeter noise uncertainty 
becomes the largest.
These point-to-point 
systematic uncertainties are to be compared with normalisation
uncertainties of
6.2\% and 7.6\% for the SpaCal and LAr electron data, respectively,
and statistical errors of between 5\% and 20\%.

For the charged current measurement, the statistical uncertainties are
dominant. 
The systematic error treatment is similar to that used in the 
neutral current case, except
that the model variations are increased in light of the lack
of previous data to constrain the kinematic dependences. The largest 
systematic uncertainties arise from these model uncertainties, in particular
from that on the sub-leading meson exchange contribution, and from the 
modelling of the acceptance of the $\etamax$ requirement. 


\section{The Diffractive Neutral Current Cross Section}
\label{sec:results}

In order to obtain the reduced neutral current cross section defined in 
equation~\ref{sigmar}, 
the data are corrected to fixed values of $Q^2$, $\beta$ and $\xpom$,
the influence of the
finite bin sizes being evaluated using a 
parameterisation of the QCD fit described in section~\ref{sec:qcdfit}. 
The measurements are quoted at the Born level after the corrections for
QED radiative effects described in 
section~\ref{sec:mc}. 
For all data points shown, the acceptance, bin 
purity and bin stability\footnote{Purity (stability) is evaluated using the 
Monte Carlo simulation and
is defined as the fraction of events reconstructed (generated) in a 
measurement bin which are also generated (reconstructed) 
in that bin.}
exceed 30 \%. The results are
given in numerical form in tables~\ref{spacal:data} (SpaCal electron
data) and~\ref{lar:data} (LAr electron data) and are shown graphically
as described in sections~\ref{sec:betaq2} and~\ref{sec:xpom}.
They can also be found at \cite{fit:table}.

\subsection{Dependences on {\boldmath $\beta$} and 
{\boldmath $Q^2$}}
\label{sec:betaq2}

The QCD properties and structure of the diffractive interaction are most
easily interpreted from the dependences on $x$ (or equivalently $\beta$) and 
$Q^2$, with $\xpom$ fixed. According to \cite{collins}, 
DPDFs can then be defined for each fixed $\xpom$ value, independently
of the validity of proton vertex factorisation. 
A binning scheme is therefore adopted with fixed $\xpom$, $x$ and $Q^2$
values. In order to minimise the
statistical uncertainties and the systematic uncertainties associated
with the reconstruction of $\mx$, 
relatively large $\xpom$ intervals are chosen. The binning in the
much better resolved variables, $x$ and $Q^2$,
is chosen to match previous inclusive measurements \cite{h1f29697}.

\begin{landscape} 
\begin{figure}[p]
  \begin{picture}(210,100)
    \put(0,20){\epsfig{file=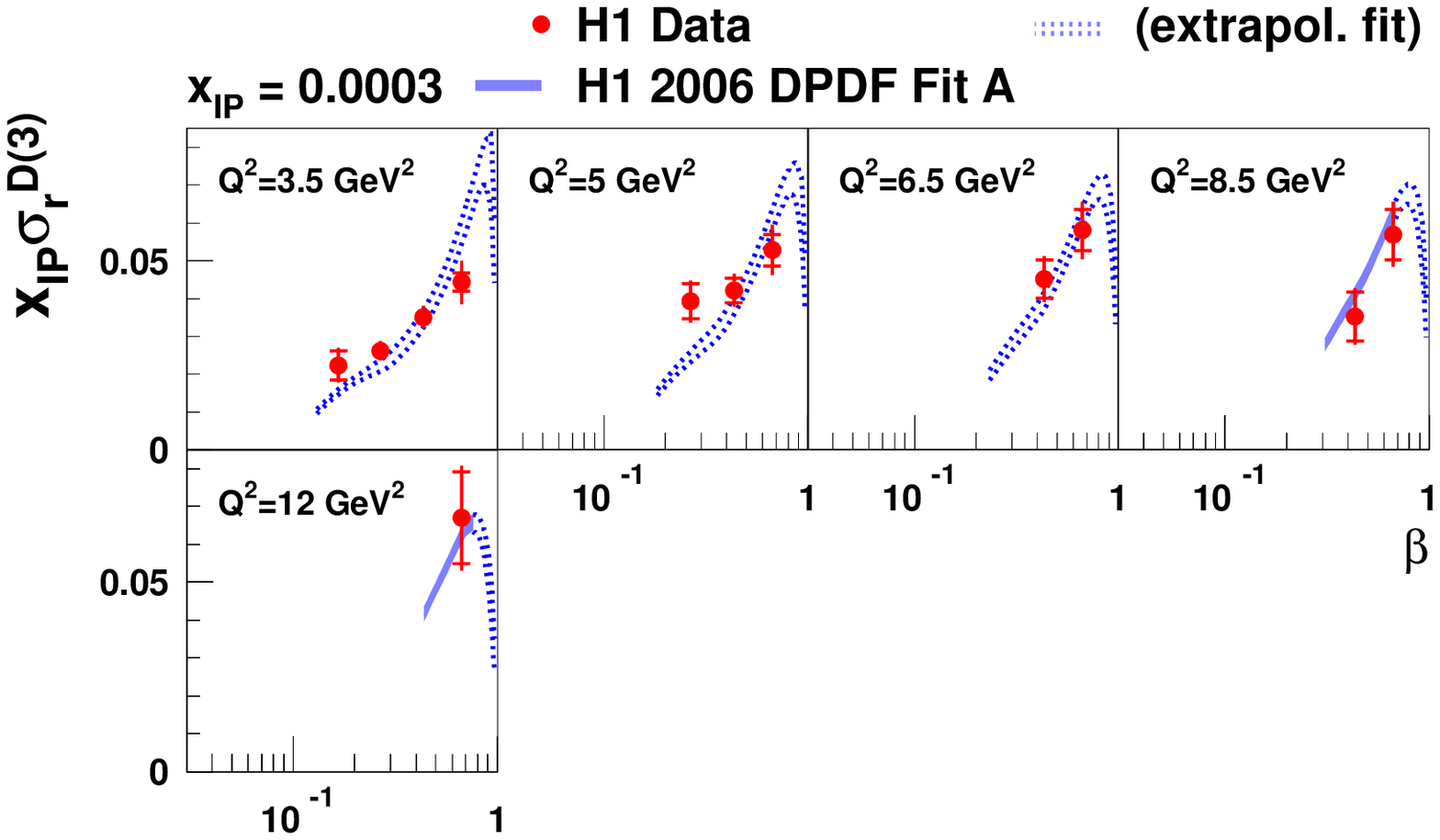,width=0.465\linewidth}}
    \put(120,20){\epsfig{file=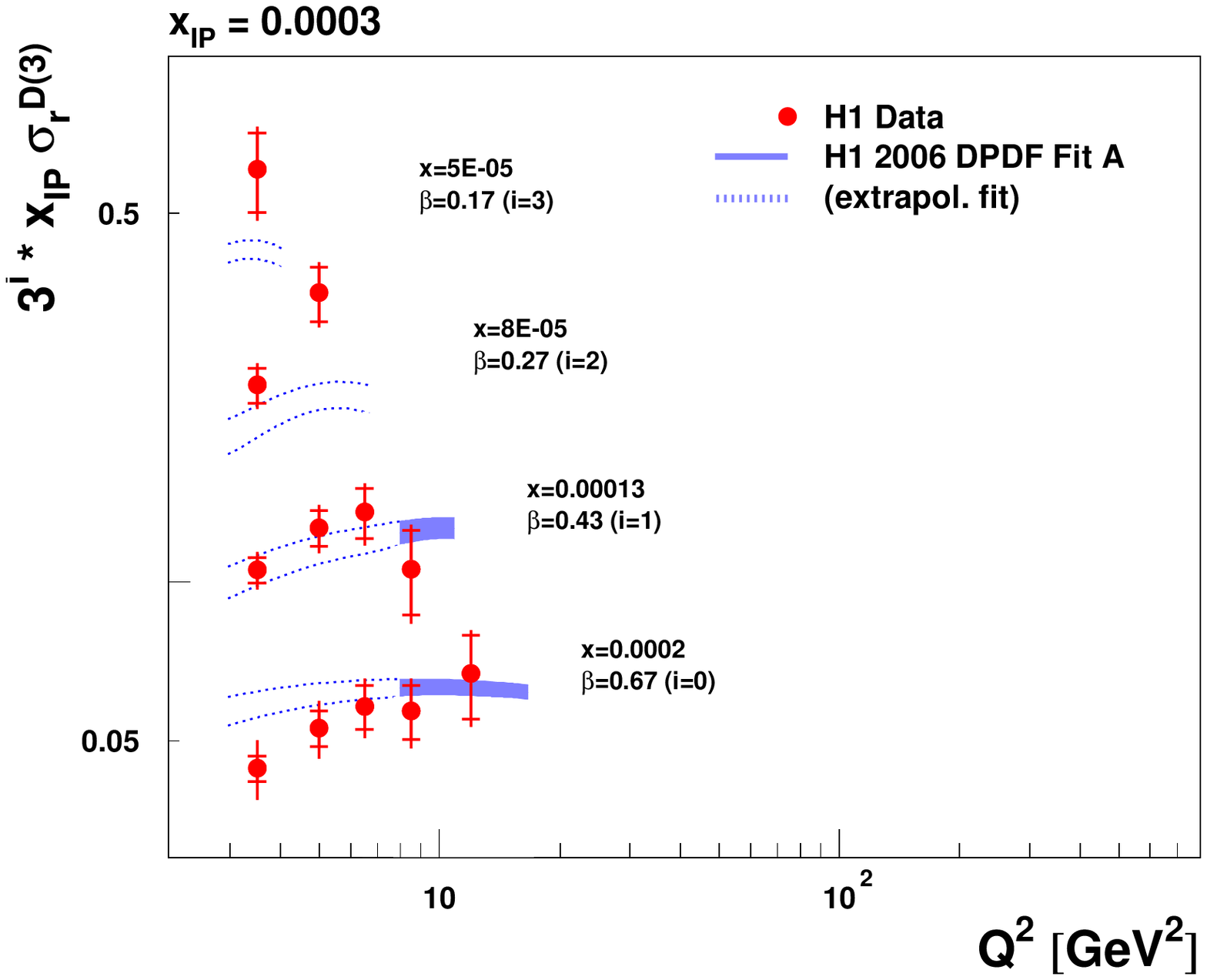,width=0.425\linewidth}}
     \put(10,90){\bf{\Large{(a)}}}
     \put(210,100){\bf{\Large{(b)}}}
  \end{picture}
\caption{The $\beta$ and $Q^2$ dependences of the
diffractive reduced cross section, multiplied by $\xpom$,
at $\xpom = 0.0003$. 
In (b) the data are multiplied by a further factor of $3^i$ for
visibility, with $i$ as indicated.
The inner and outer error bars on the data points
represent the statistical and total uncertainties, respectively. 
Normalisation uncertainties are not shown. The data
are compared with the reduced cross section at $E_p = 820 \ {\rm GeV}$
derived from the results of `H1 2006 DPDF Fit A',
which is shown as a 
shaded error band (experimental uncertainties only) in 
kinematic regions which
are included in the fit and as a pair of dashed lines in regions which
are excluded from the fit.}
\label{q2dep1}
\end{figure}
\end{landscape}

\begin{landscape} 
\begin{figure}[p]
  \begin{picture}(210,100)
     \put(0,20){\epsfig{file=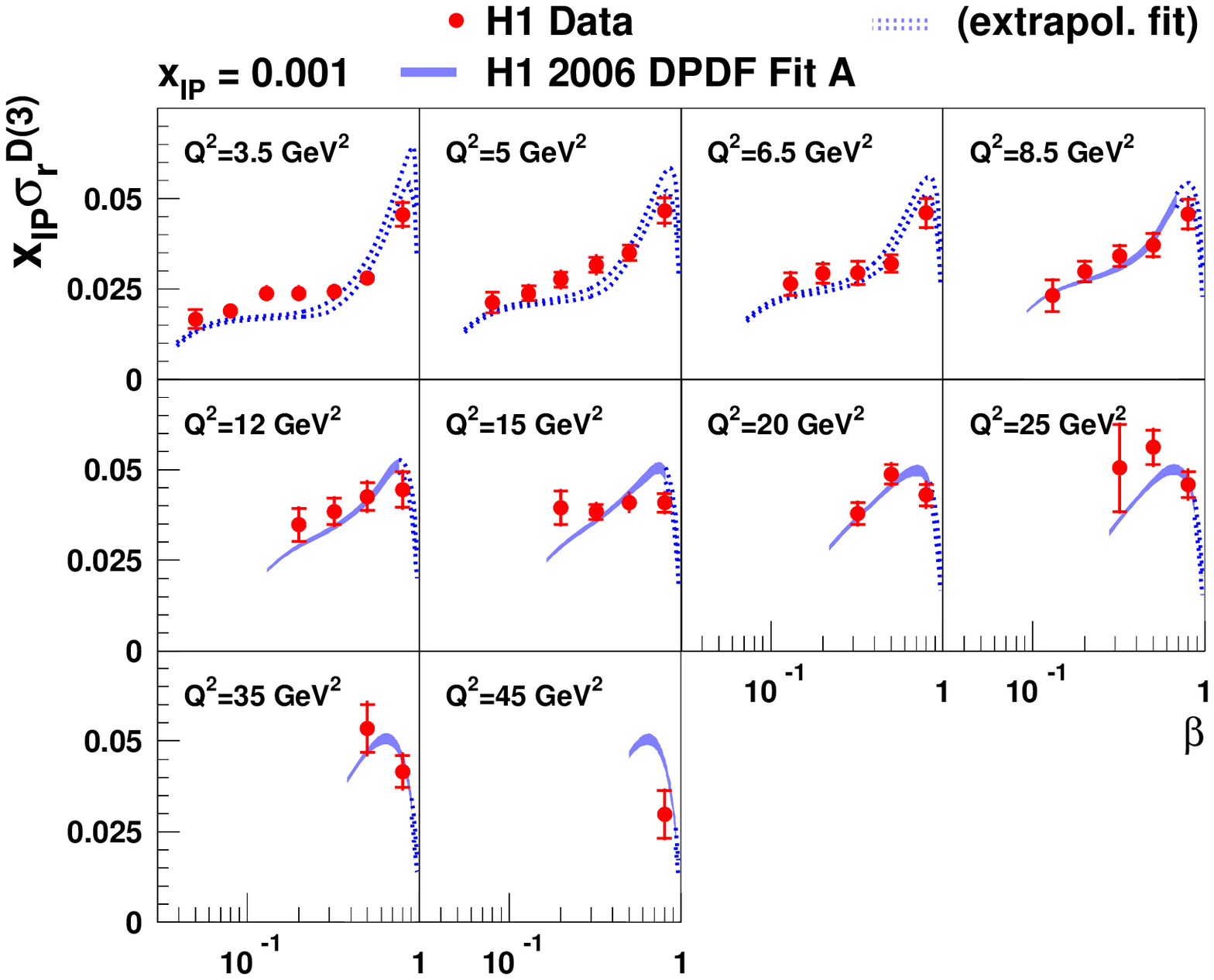,width=0.465\linewidth}}
     \put(120,20){\epsfig{file=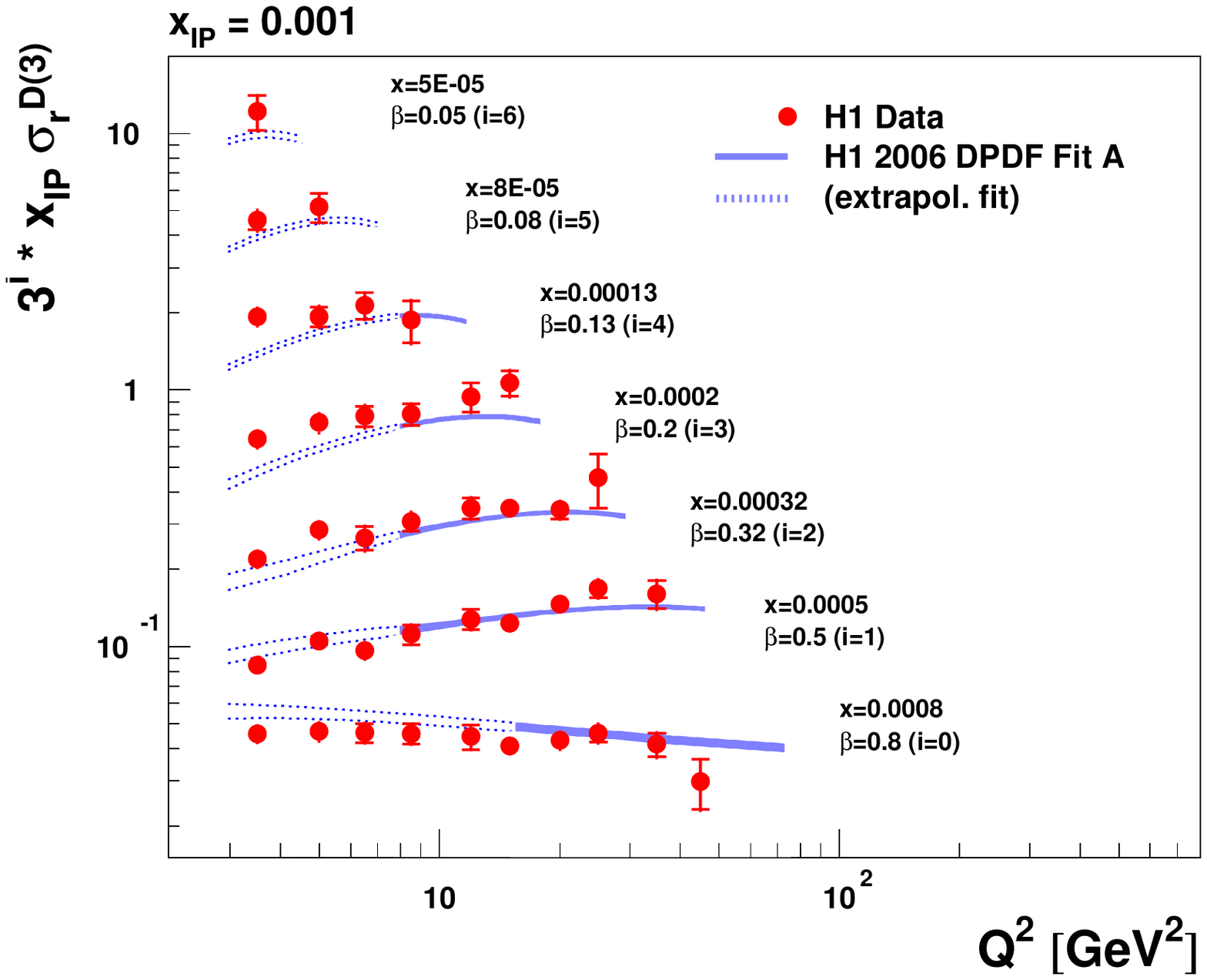,width=0.425\linewidth}}
     \put(10,110){\bf{\Large{(a)}}}
     \put(210,100){\bf{\Large{(b)}}}
  \end{picture}
\caption{The $\beta$ and $Q^2$ dependences of the
diffractive reduced cross section, multiplied by $\xpom$,
at $\xpom = 0.001$. See the caption of figure~\ref{q2dep1} for further
details.}
\label{q2dep2}
\end{figure}
\end{landscape}

\begin{landscape} 
\begin{figure}[p]
  \begin{picture}(210,145)
     \put(0,30){\epsfig{file=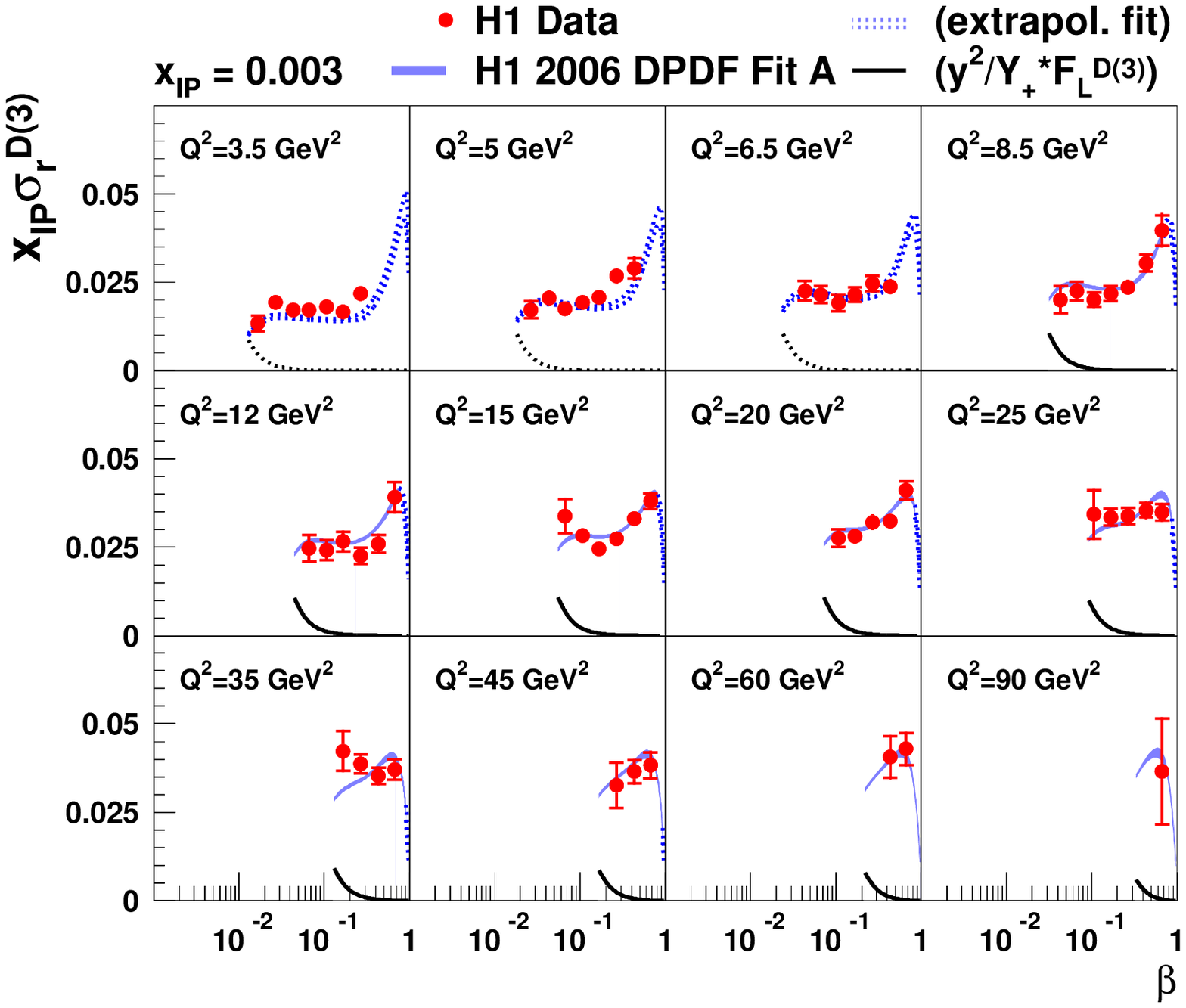,width=0.465\linewidth}}
     \put(120,0){\epsfig{file=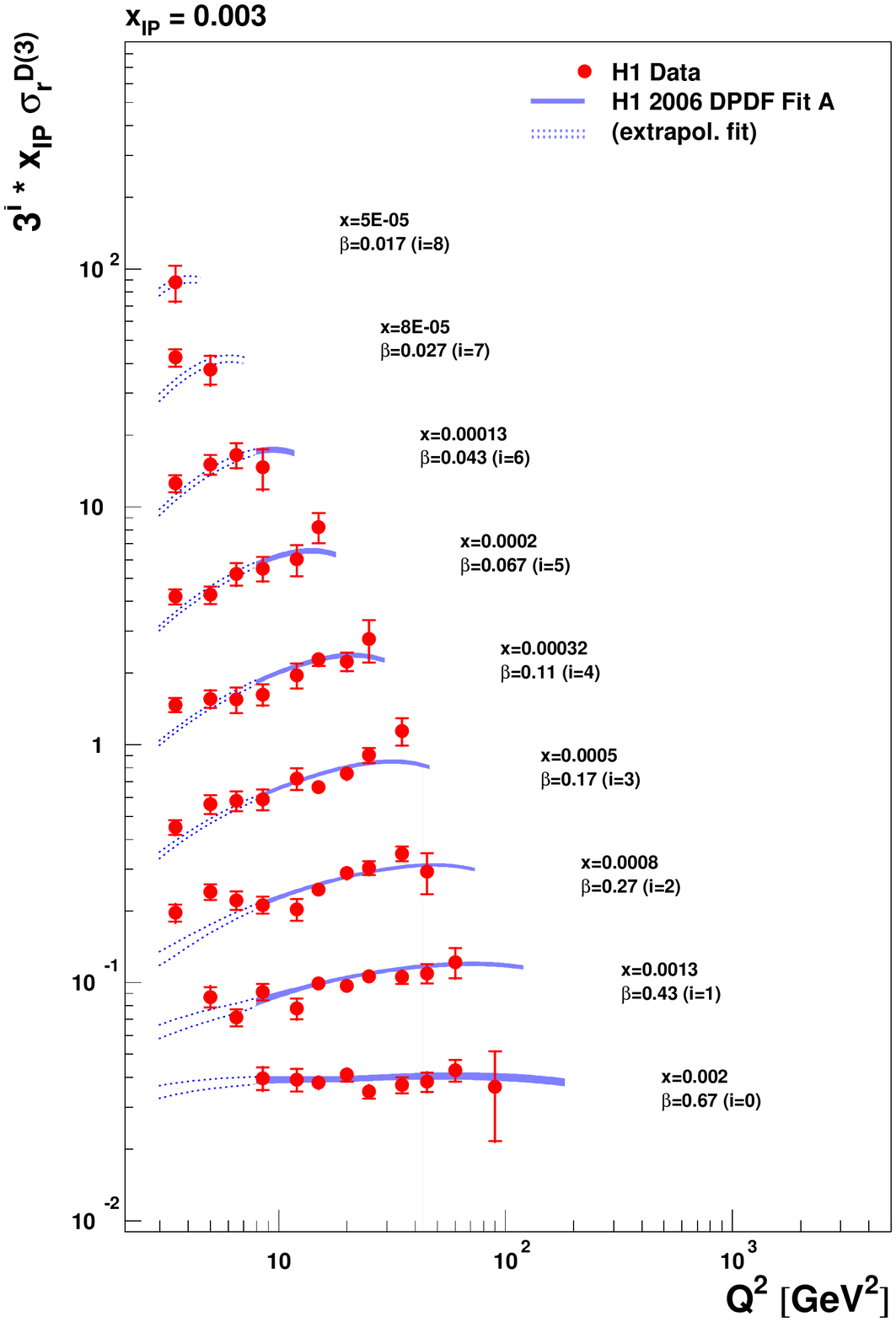,width=0.425\linewidth}}
     \put(10,130){\bf{\Large{(a)}}}
     \put(210,148){\bf{\Large{(b)}}}
  \end{picture}
\caption{The $\beta$ and $Q^2$ dependences of the
diffractive reduced cross section, multiplied by $\xpom$,
at $\xpom = 0.003$. In (a), the quantity 
$y^2 / Y_+ \cdot F_L^{D(3)}$ is also
shown, as extracted from the `H1 2006 DPDF Fit A'. 
Adding this quantity to the reduced 
cross section yields $F_2^{D(3)}$.
See the caption of figure~\ref{q2dep1} for further
details.}
\label{q2dep3}
\end{figure}
\end{landscape}

\begin{landscape} 
\begin{figure}[p]
  \begin{picture}(210,145)
     \put(0,20){\epsfig{file=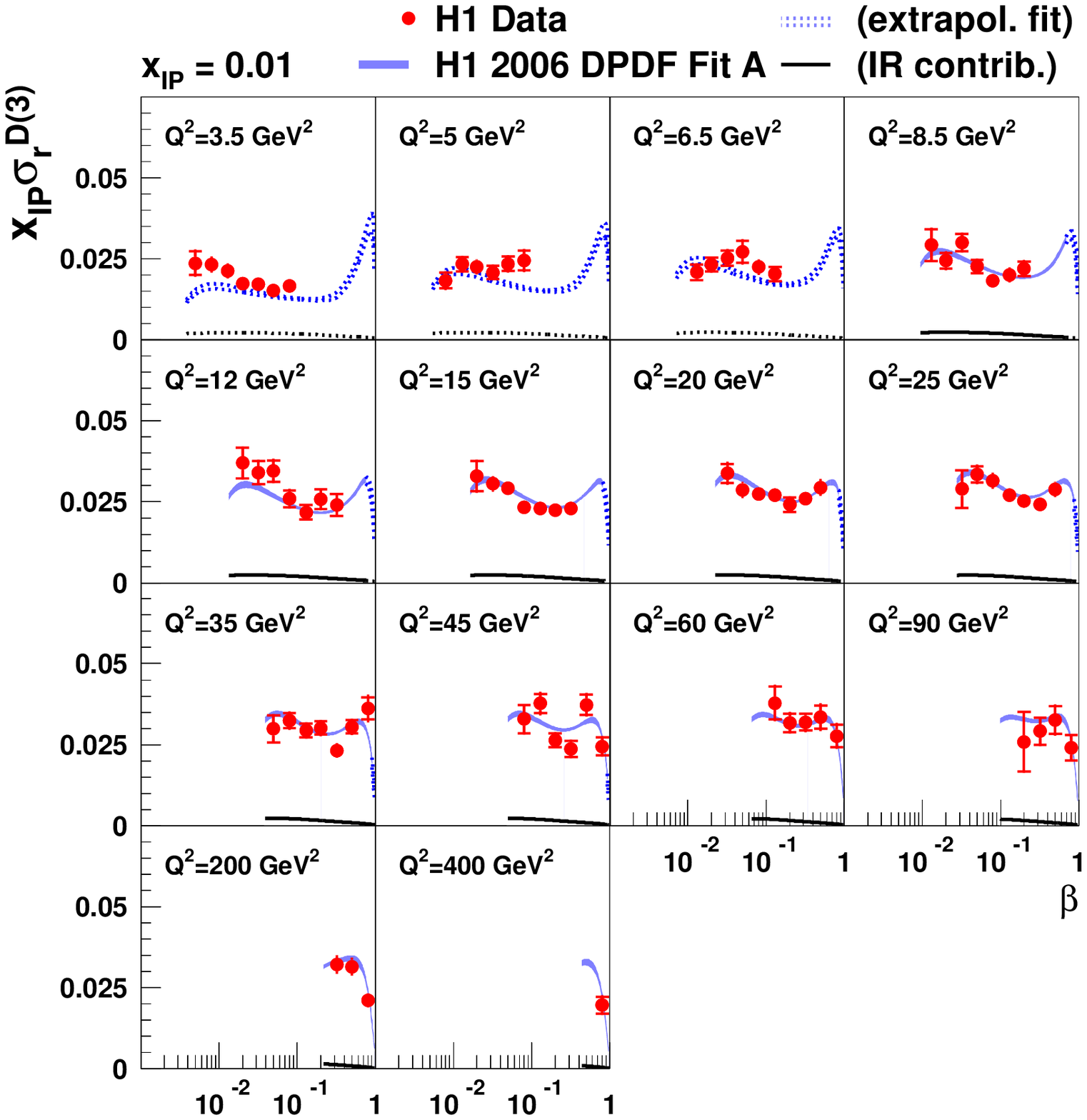,width=0.465\linewidth}}
     \put(120,0){\epsfig{file=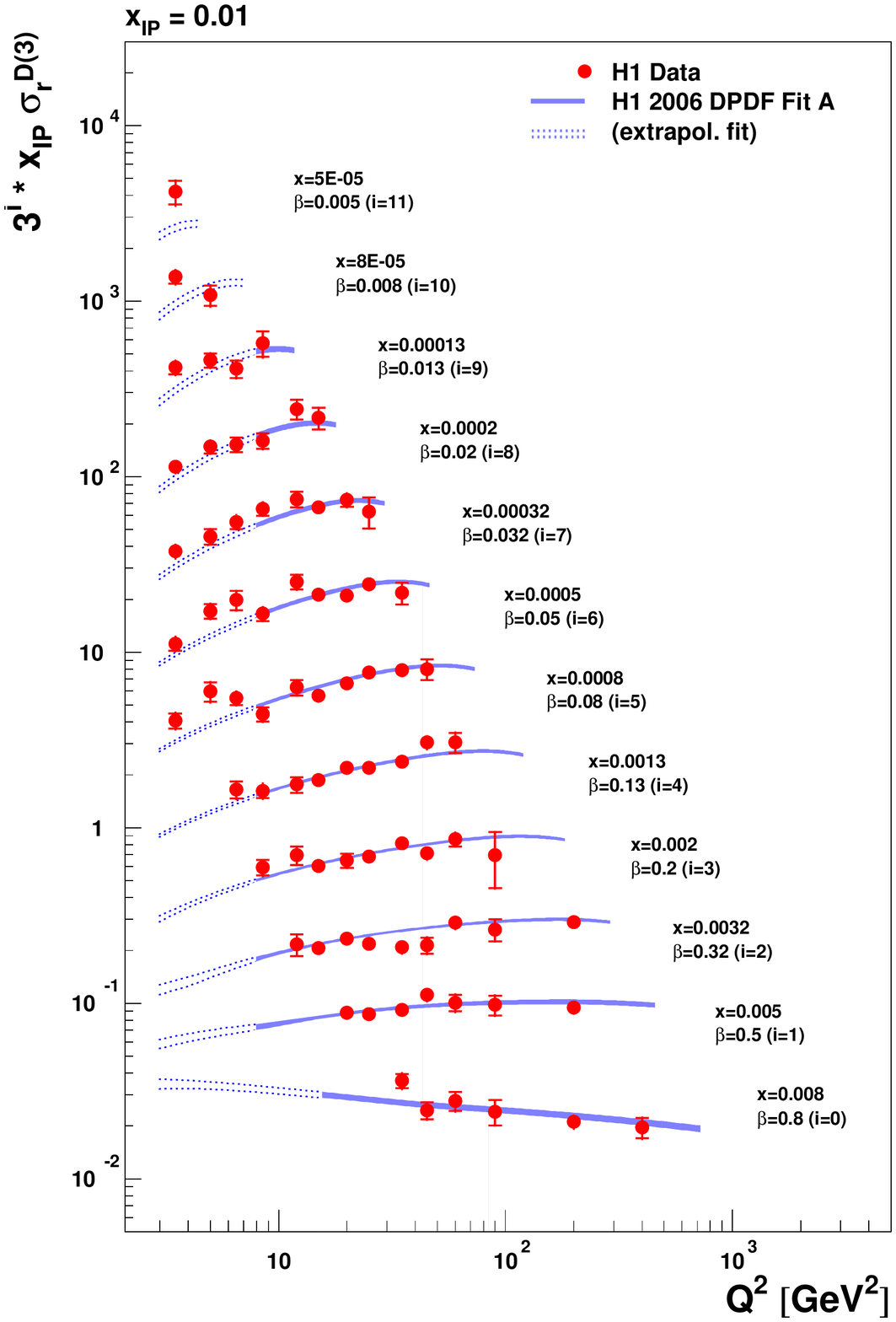,width=0.425\linewidth}}
     \put(10,140){\bf{\Large{(a)}}}
     \put(210,148){\bf{\Large{(b)}}}
  \end{picture}
\caption{The $\beta$ and $Q^2$ dependences of the
diffractive reduced cross section, multiplied by $\xpom$,
at $\xpom = 0.01$. 
In (a), the contribution of the sub-leading exchange
alone according to the `H1 2006 DPDF Fit A' is also shown. The data with
$Q^2 \leq 90 \ {\rm GeV^2}$ 
($Q^2 \geq 200 \ {\rm GeV^2}$) were obtained with 
$E_p = 820 \ {\rm GeV}$ ($E_p = 920 \ {\rm GeV}$). The fit results
are shown for $E_p = 820 \ {\rm GeV}$.
See the caption of figure~\ref{q2dep1} for further
details.}
\label{q2dep4}
\end{figure}
\end{landscape}

\begin{landscape} 
\begin{figure}[p]
  \begin{picture}(210,145)
     \put(0,20){\epsfig{file=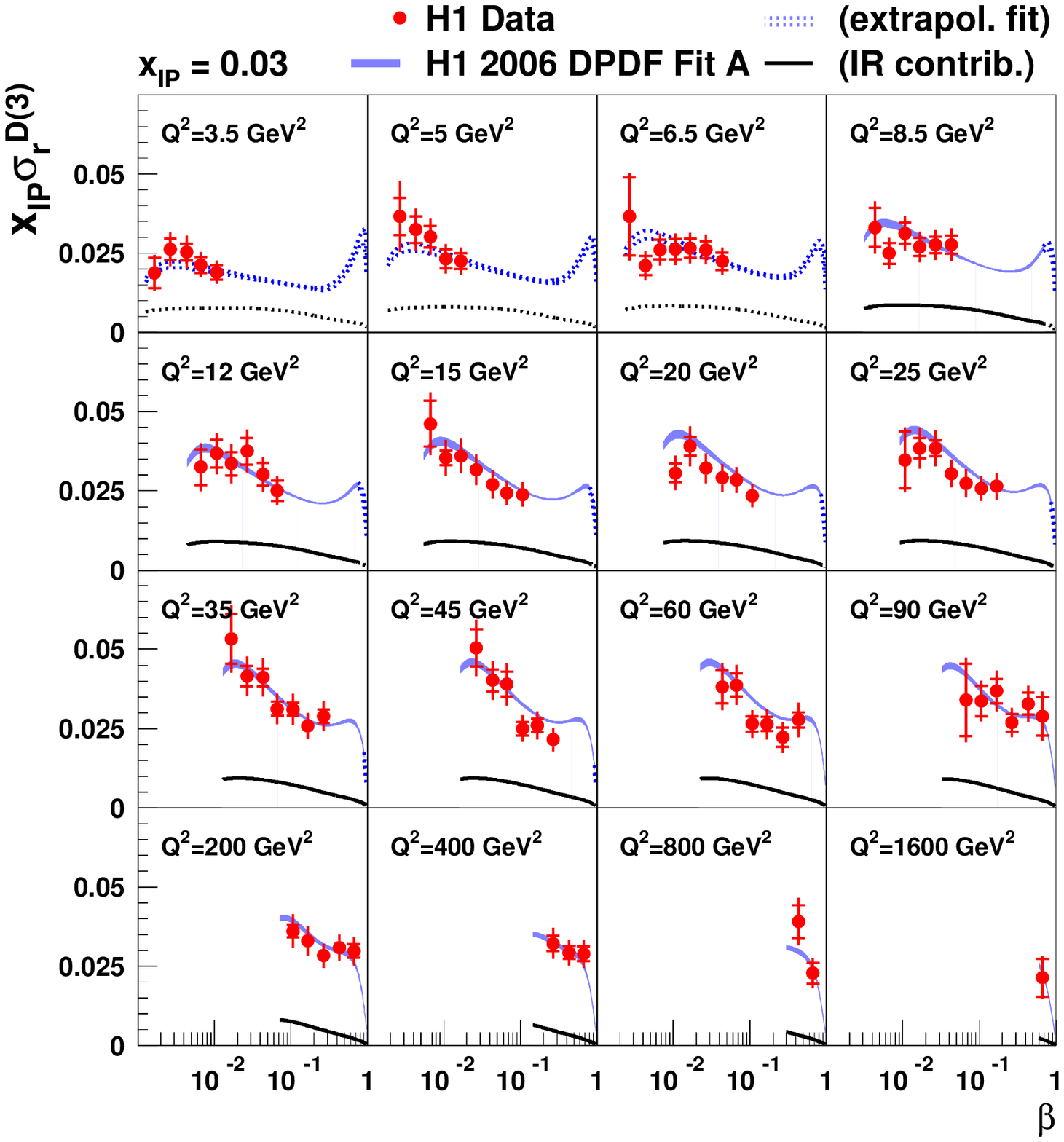,width=0.465\linewidth}}
     \put(120,0){\epsfig{file=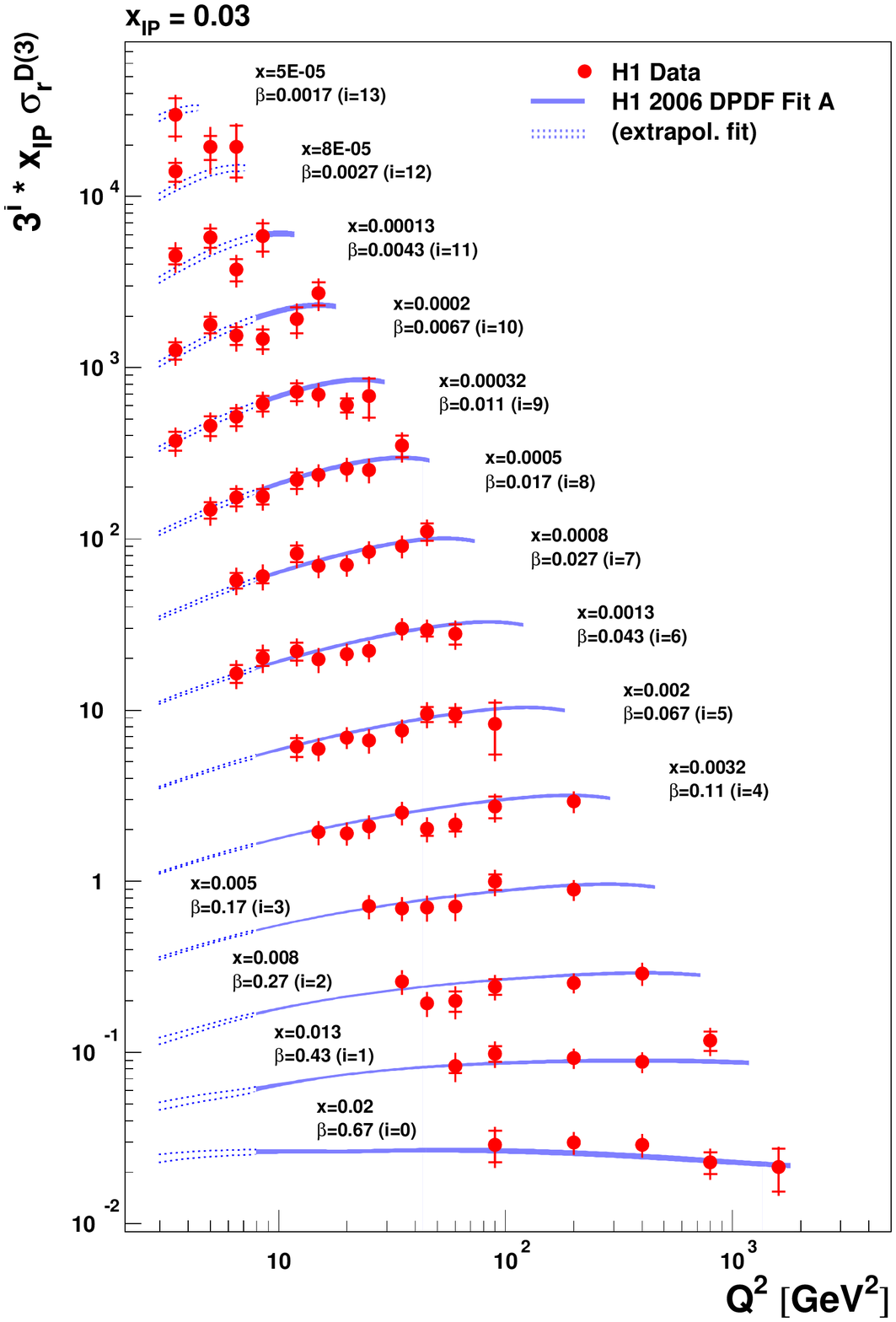,width=0.425\linewidth}}
     \put(10,140){\bf{\Large{(a)}}}
     \put(210,148){\bf{\Large{(b)}}}
  \end{picture}
\caption{The $\beta$ and $Q^2$ dependences of the
diffractive reduced cross section, multiplied by $\xpom$,
at $\xpom = 0.03$. 
See the captions of figures~\ref{q2dep1} and~\ref{q2dep4} for further
details.}
\label{q2dep5}
\end{figure}
\end{landscape}

The $\beta$ and $Q^2$ dependences of the reduced cross section,
multiplied by $\xpom$, are shown in 
figures~\ref{q2dep1}-\ref{q2dep5} at fixed values of $\xpom = 0.0003$,
$0.001$, $0.003$, $0.01$ and $0.03$, respectively. 
The results from different $\xpom$ values complement one another 
in their $\beta$ and $Q^2$ ranges, though there is 
also considerable 
overlap between the coverage at different $\xpom$ values. 
For each $\xpom$ value considered, the data exhibit similar $\beta$
and $Q^2$ dependences.

As can be seen in figures~\ref{q2dep1}a-\ref{q2dep5}a,
the cross section remains large up to the highest accessed values 
of $\beta$ (i.e. where $x \rightarrow \xpom$) at fixed $\xpom$ and $Q^2$.
This behaviour is in marked contrast to that of 
hadron structure functions,
but is qualitatively similar to that of the photon \cite{nissius}.
The $Q^2$ dependence of the data is shown for fixed $\xpom$ and $\beta$
in figures~\ref{q2dep1}b-\ref{q2dep5}b.
The reduced cross section increases with $Q^2$
throughout most of the kinematic range, up to large $\beta \simeq 0.5$. 
These positive scaling violations confirm
earlier observations \cite{h1f2d94} and contrast with
the case of inclusive scattering from hadrons, for which
the cross section at fixed $x$ falls with increasing $Q^2$ for
$x \ \gapprox \ 0.1$ \cite{h1f29697}. 

The data in figures~\ref{q2dep1}-\ref{q2dep5}
are compared\footnote{The curves shown correspond to $E_p = 820 \ {\rm GeV}$.
The predictions 
for $E_p = 920 \ {\rm GeV}$ differ slightly at the lowest $\beta$
values, due to the influence of $F_L^{D(3)}$.}
with the results of the `H1 2006 DPDF Fit A'
described in section~\ref{sec:qcdfit}. This fit assumes
proton vertex factorisation and includes a sub-leading 
exchange contribution, which is important 
at low $\beta$ and large $\xpom$ as shown in 
figures~\ref{q2dep4}a and~\ref{q2dep5}a.
It is clear from the good overall
description that the data are broadly consistent with such
a model.
The $\beta$ dependence of $\sigma_r^{D(3)}$ then
directly reflects the quark structure of the diffractive exchange 
with each quark flavour weighted by its squared electric charge, 
whilst the measured
$Q^2$ dependence is sensitive to the diffractive gluon density. 

\begin{figure}[p] \unitlength=0.9mm
 \begin{picture}(160,215)
    \put(30,115){\epsfig{file=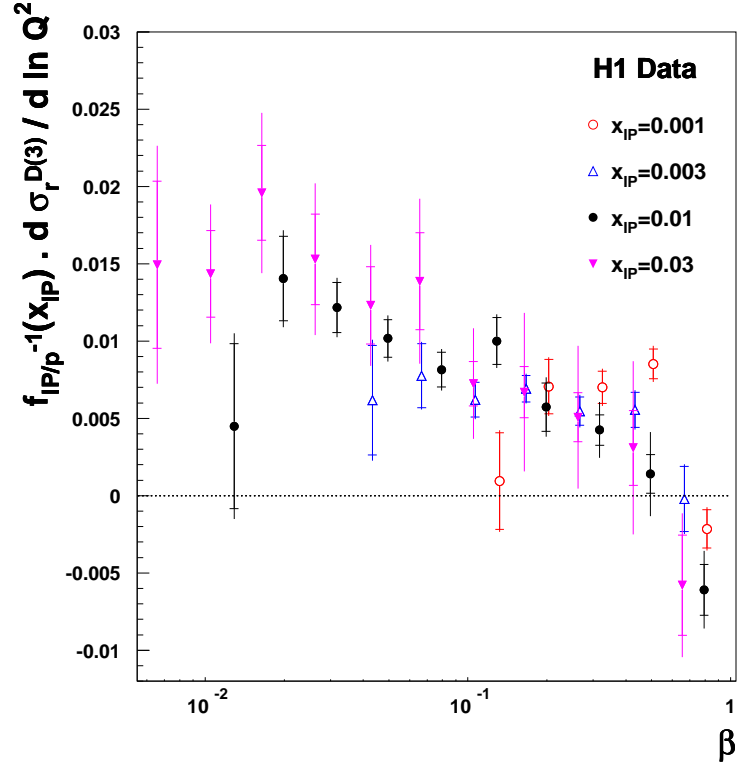,width=0.675\linewidth}}
    \put(30,0){\epsfig{file=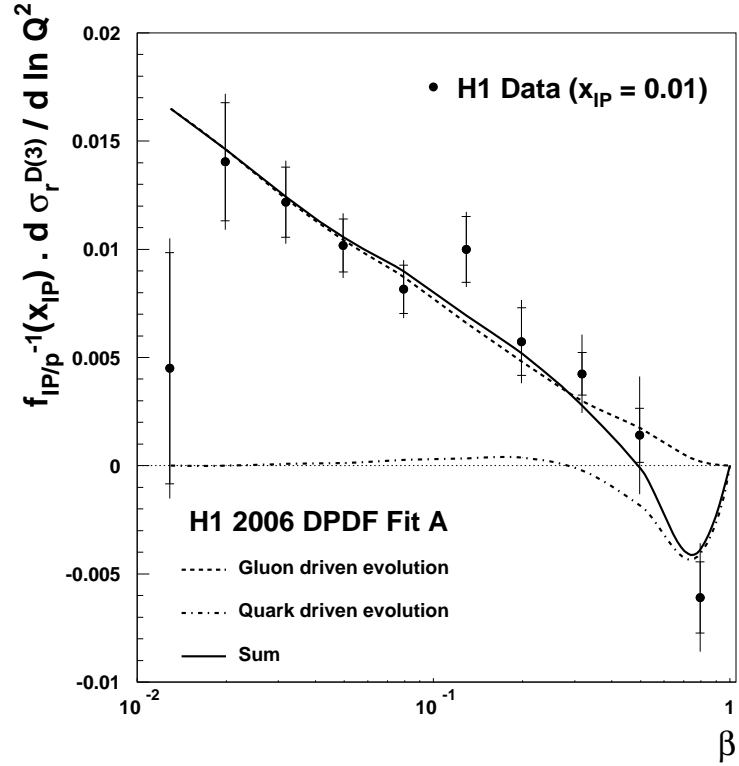,width=0.675\linewidth}}
    \put(155,160){\bf{\Large{(a)}}}
    \put(155,53){\bf{\Large{(b)}}}
  \end{picture}
\caption{(a) Measurements of the logarithmic $Q^2$ derivative of the
reduced diffractive cross section at different fixed values of $\xpom$ and
$\beta$, obtained by fitting equation~\ref{eqn:logderiv} to the
data. 
The derivatives are divided by the diffractive flux factor as defined
in equation~\ref{eq:fluxfac}. (b) The logarithmic $Q^2$ derivative 
at $\xpom = 0.01$, divided by the diffractive flux factor and 
compared with the prediction of the 
`H1 2006 DPDF Fit A'. The prediction
is also decomposed into contributions to the evolution from the splittings
$g \rightarrow q \bar{q}$ (`Gluon driven evolution') and
$q \rightarrow q g$ (`Quark driven evolution'). The inner and outer error bars 
represent the statistical and total uncertainties, respectively.
Normalisation uncertainties are not shown.}
\label{fig:logderiv}
\end{figure}

The $Q^2$ dependence is 
quantified by fitting the data at fixed $\xpom$ and $\beta$ to the form
\begin{eqnarray}
  \sigma_r^{D(3)}(\xpom, Q^2, \beta) = a_{\rm D} (\beta, \xpom) + 
b_{\rm D} (\beta, \xpom) \ln Q^2 \ ,
\label{eqn:logderiv}
\end{eqnarray}
such that 
$b_{\rm D} (\beta, \xpom) = \left[ \partial \sigma_r^{D(3)} / \partial \ln Q^2
\right]_{\beta, \xpom}$ 
is the first logarithmic $Q^2$ 
derivative of the reduced cross section. 
This observable has been used previously 
to discriminate between different
models of diffractive DIS \cite{brazil}.
Equation~\ref{eqn:logderiv} is fitted to
data with $0.001 \leq \xpom \leq 0.03$ from each $\beta$ value 
if there are a minimum of three
available data points\footnote{If only data with $Q^2 \geq 8.5 \ {\rm GeV^2}$ 
are included, as in the fit described in section~\ref{sec:qcdfit}, the 
changes to the logarithmic $Q^2$ derivatives are small and the conclusions
are unaffected.}.
The resulting $\ln Q^2$ derivatives are shown in
figure~\ref{fig:logderiv}a, after dividing $b_{\rm D}(\beta,\xpom)$ by the 
factor $f_{\pom / p} (\xpom)$, defined in equation~\ref{eq:fluxfac}, 
which  
is used to parameterise the $\xpom$ 
dependence so that the results from different $\xpom$ values can be 
compared in normalisation as well as in shape.
Although the logarithmic derivatives at 
different $\xpom$ values probe different $Q^2$
regions, they are remarkably similar when viewed as a function of $\beta$. 
This confirms the applicability of the proton vertex
factorisation framework to the description of the current data.
The lack of any significant change in behaviour at large $\xpom$
indicates that 
the derivatives are not significantly affected by the presence
of sub-leading exchange contributions.

According to the DGLAP evolution equations, the $\ln Q^2$ derivative
of $F_2^{D(3)}$
contains contributions due to the splittings $g \rightarrow q \bar{q}$
and $q \rightarrow q g$, convoluted with the diffractive
gluon and quark densities, respectively.
In figure~\ref{fig:logderiv}b, the derivatives at $\xpom = 0.01$
are shown together
with the decomposition into these two 
contributions according to the 
`H1 2006 DPDF Fit A' described in section~\ref{sec:qcdfit}. 
The curves correspond to $Q^2$ values which
vary in order to match the average $\ln Q^2$ 
of the data which are fitted at each $\beta$ value.
The theoretical calculation is in good agreement with the data.
The $\ln Q^2$ derivative
is determined almost entirely
by the diffractive gluon density up to $\beta \simeq 0.3$.
The large positive $\ln Q^2$ derivatives in this region
can thus be attributed to a large gluonic component in
the DPDFs. 
For $\beta \ \gapprox \ 0.3$, the contribution to the $Q^2$ evolution 
from quark splittings becomes increasingly important and the derivatives
become less sensitive to the gluon density.
The fall in the derivative as $\beta \rightarrow 1$ then arises 
dominantly from 
gluon radiation, $q \rightarrow qg$, shifting quarks from higher to lower
$\beta$ with increasing $Q^2$. 

Considering diffractive DIS 
in terms of the elastic scattering from the proton of 
colour dipoles formed by partonic
fluctuations of the exchanged virtual photon \cite{wusthoff} provides
a complementary framework in which to 
describe diffractive DIS.
Modelling the exchange by two gluons in a 
net colour singlet configuration \cite{low:nussinov},
the data at low and moderate $\beta$ are then 
described in terms of $q \bar{q}$ and $q \bar{q} g$
fluctuations of transversely polarised photons,
whilst the high $\beta$ region contains a $Q^2$-suppressed non-leading twist 
contribution from $q \bar{q}$ fluctuations of longitudinally polarised
photons \cite{bekw,kgb,hebecker:teubner}. 
A significant contribution from this term could explain
the tendency of the data to rise with increasing $\beta$
for $\beta > 0.1$ at low $Q^2$ (figure~\ref{q2dep1}a-~\ref{q2dep3}a).
However, the high $\beta$ data
in figures~\ref{q2dep1}b-\ref{q2dep5}b are adequately described by a single
logarithmic
dependence on $Q^2$ and do not require a sum of leading and $Q^2$-suppressed
terms. 


\subsection{Dependence on {\boldmath $\xpom$} and Comparisons with other Data}
\label{sec:xpom}

For comparison with 
previous measurements of diffractive DIS and for a more detailed study of the
$\xpom$ dependence at fixed $\beta$ and $Q^2$, the reduced cross section
is also measured using
a binning scheme with fixed $\beta$, $Q^2$ and $x = \beta \cdot \xpom$,
as shown in figure~\ref{stampa}. The data are
multiplied by $\xpom$ for visibility and are compared with the results of
the QCD fit described in section~\ref{sec:qcdfit}, which is also
in good agreement with the data obtained using this binning scheme. 
The $\xpom$ dependence is roughly flat for all 
$\beta$ and $Q^2$ values, implying that 
the reduced cross section approximately
follows a $\sigma_r^{D(3)} \propto 1 / \xpom$ dependence.
However, when viewed in detail, there are clear deviations from this
behaviour.
The variations in 
the $\xpom$ dependence as $\beta$ changes are
as expected from the interplay between the leading pomeron and a 
sub-leading trajectory exchange, as discussed in section~\ref{dpdf:results}.

\begin{figure}[p] \unitlength=0.9mm
 \begin{picture}(160,215)
    \put(18,115){\epsfig{file=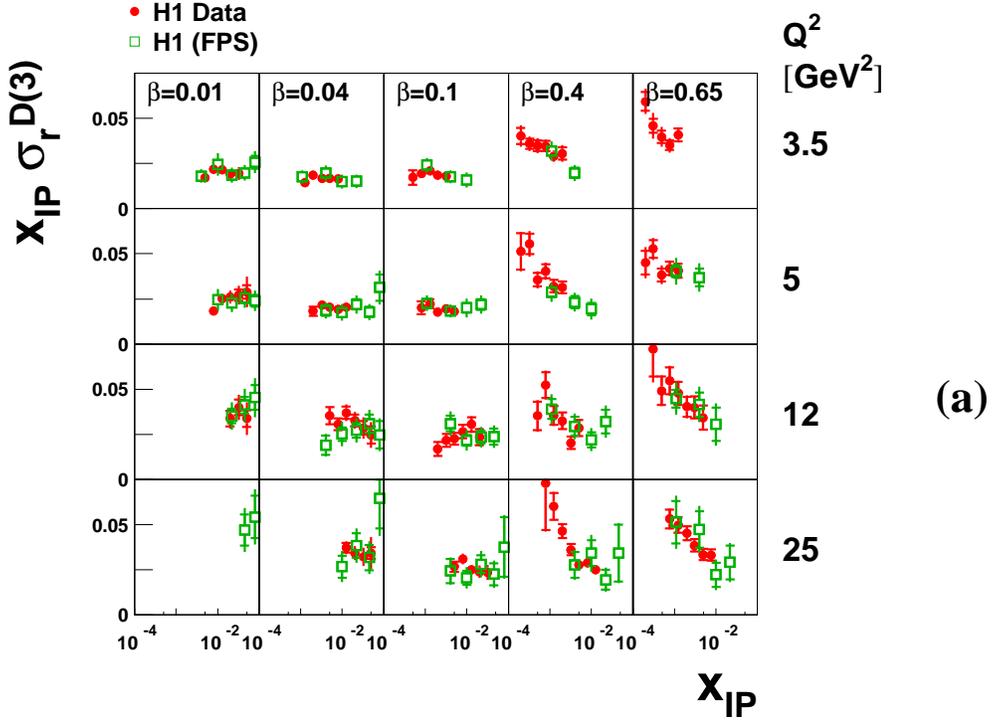,width=0.725\linewidth}}
    \put(30,0){\epsfig{file=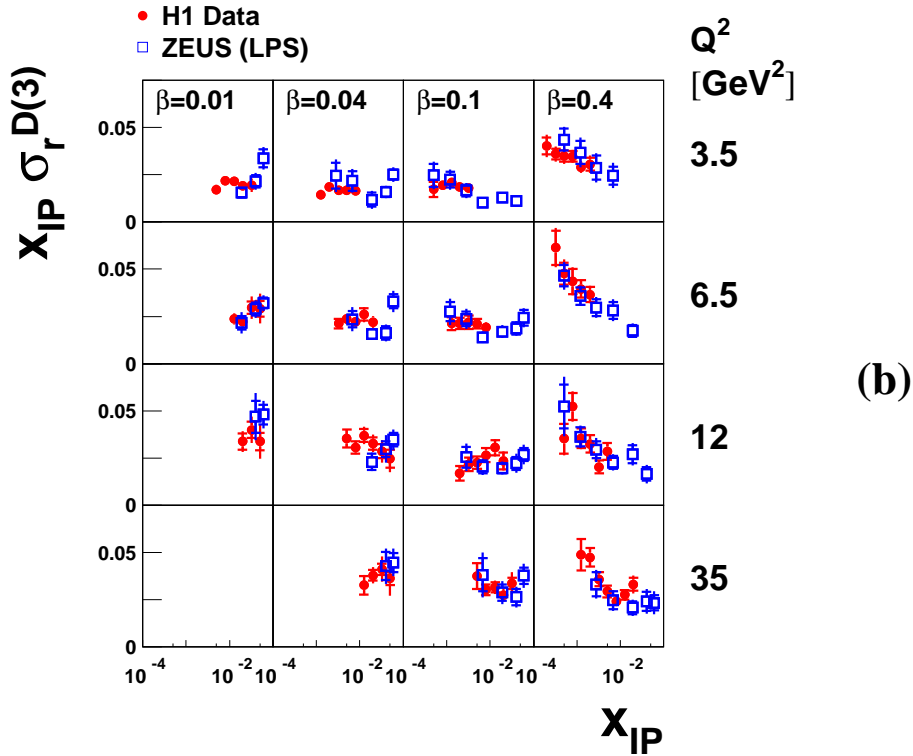,width=0.65\linewidth}}
    \put(155,160){\bf{\Large{(a)}}}
    \put(155,53){\bf{\Large{(b)}}}
  \end{picture}
\caption{Comparisons between subsets of the present data
and results obtained by the direct measurement
of the final state proton using (a) the H1 Forward Proton
Spectrometer (FPS) \cite{H1FPS} and
(b) the ZEUS Leading Proton Spectrometer (LPS) \cite{zeuslps}.
The FPS and LPS data are shifted to the $Q^2$ and $\beta$ values shown 
using small translation factors and are multiplied by a further
universal factor of 1.23 such that they
correspond to $\my < 1.6 \ {\rm GeV}$. 
The inner error bars represent the statistical uncertainties and
the outer error bars represent the combined statistical and systematic
uncertainties. Normalisation uncertainties are not shown.}
\label{stampb}
\end{figure}

\begin{figure}[p]
\centering \epsfig{file=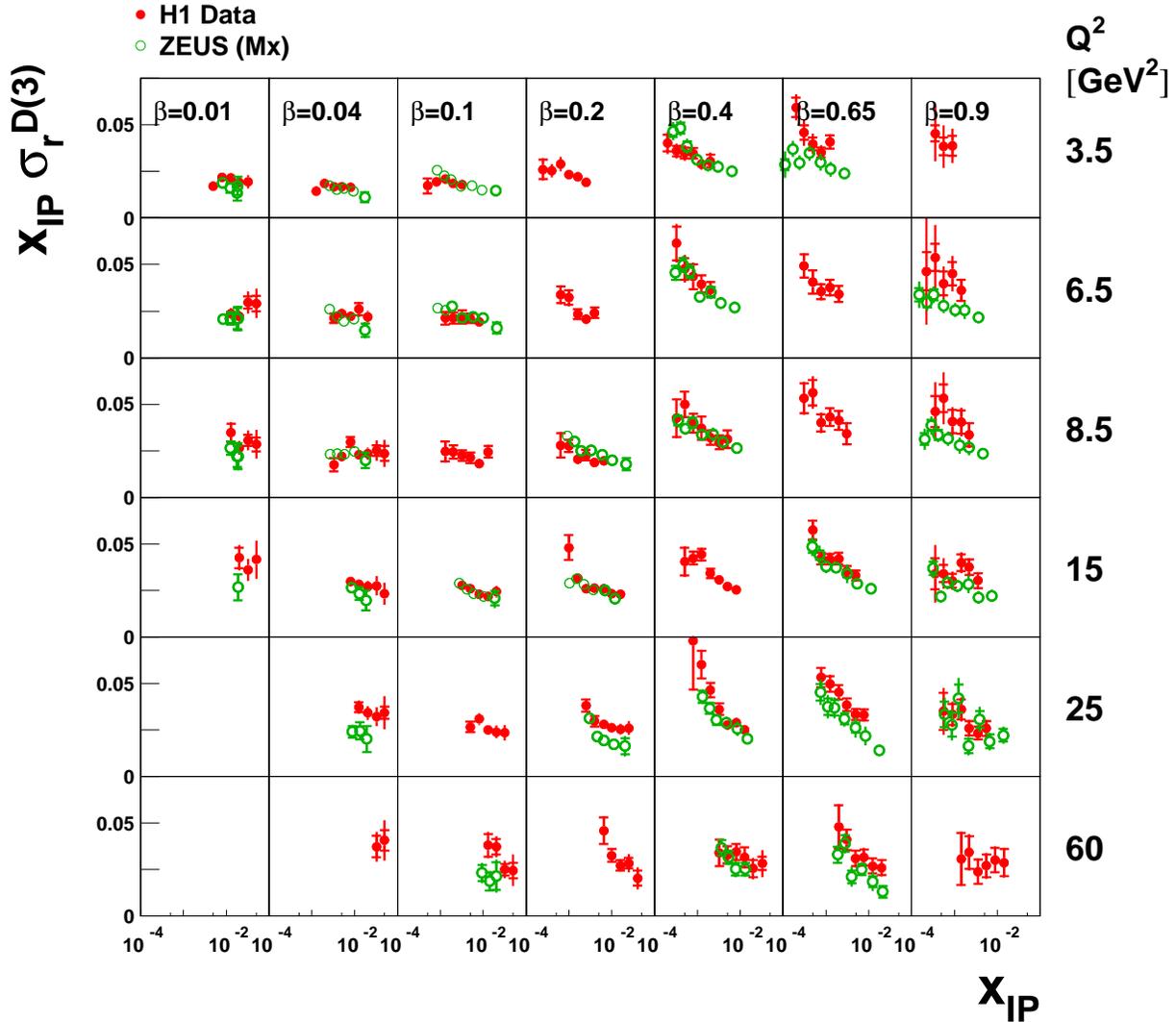,width=\linewidth}
\caption{Comparison between a subset of the present data and the results of
the ZEUS collaboration, obtained by decomposition of the inclusive
$\ln \mx^2$ distribution \cite{ZEUS:97}.
The ZEUS data are shifted to the $Q^2$ and $\beta$ values shown 
using small translation factors and have been multiplied by a further
universal factor of 0.86 so that they
correspond to $\my < 1.6 \ {\rm GeV}$. 
The inner error bars represent the statistical uncertainties and
the outer error bars represent the combined statistical and systematic
uncertainties. Normalisation uncertainties are not shown.}
\label{stampc}
\end{figure}

In figure~\ref{stampb}, the measured reduced cross section is compared with 
results obtained by direct measurement of the final state proton  
using the H1 Forward Proton Spectrometer 
(`H1 (FPS)') \cite{H1FPS} 
and the ZEUS Leading Proton Spectrometer (`ZEUS (LPS)') \cite{zeuslps}.
In figure~\ref{stampc}, a comparison is made with ZEUS data obtained by
decomposition of the inclusive
$\ln \mx^2$ distribution into diffractive and non-diffractive components
(`ZEUS ($\mx$)') \cite{ZEUS:97}. 
Shifts, evaluated using the DPDF fit described
in section~\ref{sec:qcdfit}, are applied to these 
data
in order to transport them to the $\beta$ and $Q^2$ values
of the present measurement. Since
no uncertainties are ascribed to this procedure, only data points for
which the shifts are small and 
relatively insensitive to the choice of DPDFs are 
shown. 
The different
contributions from proton dissociation in the different data sets 
are accounted
for by the application of global factors of $1.23$ to the 
`H1 (FPS)' and `ZEUS (LPS)' data and of 
$0.86$ to the `ZEUS ($\mx$)' data. The former factor
corresponds to the measured ratio 
of cross sections for $\my < 1.6 \ {\rm GeV}$
and $\my = m_p$, for which the uncertainty is 0.16 
(see \cite{H1FPS} and section~\ref{pdiss}). 
The latter factor contains an
additional contribution of $0.7$, corresponding to the
ratio of cross sections for $\my = m_p$
and $\my < 2.3 \ {\rm GeV}$ according to \cite{ZEUS:97}. 

There is broad agreement between all of the data sets on the general
behaviour of the diffractive cross section. 
The `H1 (FPS)' and `ZEUS (LPS)' data are compatible with the present
measurement throughout the kinematic range available for comparison.
A more detailed comparison with the `H1 (FPS)' data is presented 
in \cite{H1FPS}.
The `ZEUS ($\mx$)' data are in good agreement 
with the present measurement in some regions, for example
at low $\beta$ and low $Q^2$. However, there are 
disagreements in the low $Q^2$, high $\beta$ region and in the
high $Q^2$, low $\beta$ region, which correspond to low and high values
of $\mx = \sqrt{Q^2 \ ( 1/ \beta - 1)}$, respectively.
The high $\mx$ discrepancy gives rise to a weaker 
$Q^2$ dependence of the 
`ZEUS ($\mx$)' cross section at low $\beta$ than is the case 
for H1 \cite{HERA:LHC}.


\section{QCD Analysis and Diffractive Parton Distributions}
\label{sec:qcdfit}

The high precision and large kinematic range of the diffractive 
cross section data presented in this paper allow 
detailed tests of the factorisation properties of diffractive DIS and
the extraction of DPDFs, which may be used to predict cross sections for
other diffractive processes at HERA and elsewhere. 

\subsection{Theoretical Framework}
\label{facprop}


QCD hard scattering collinear factorisation, when applied to
diffractive DIS \cite{collins}, implies that the cross section for the process 
$ep \rightarrow eXY$ can be written in terms of convolutions of 
partonic cross sections $\hat{\sigma}^{e i} (x, Q^2)$ with 
DPDFs $f_i^D$ as
\begin{equation}
{\rm d} \sigma^{ep \rightarrow eXY} (x, Q^2, \xpom, t) = \sum_i \ 
f_i^D(x, Q^2, \xpom, t) \ \otimes \
{\rm d} \hat{\sigma}^{ei}(x,Q^2) \ .
\label{equ:diffpdf}
\end{equation}
The partonic cross sections
are the same as those for inclusive DIS.
The DPDFs
represent probability distributions for partons $i$ in the proton
under the constraint that the proton is scattered to a 
particular system $Y$ with a specified four-momentum. 
They are not known
from first principles, but 
can be determined from fits to the data using
the DGLAP \cite{dglap} evolution
equations.
The factorisation formula in equation~\ref{equ:diffpdf} is valid for 
sufficiently large $Q^2$ and fixed
$\xpom$, $t$ and system $Y$. It also applies 
to any cross section which
is integrated over a fixed range in $\my$ and $t$ and may thus be
applied to the present data with $\my < 1.6 \ {\rm GeV}$ and 
$|t| < 1 \ {\rm GeV^2}$.

Due to kinematic constraints,
it is not possible to access the full range of $x$ and $Q^2$
using data from only one value of $\xpom$.
A parameterisation of the $\xpom$ 
dependence of the DPDFs, for which there is no clear procedure in QCD,
is therefore necessary.
The proton vertex factorisation framework is adopted 
here, such that the DPDFs
are factorised into a term depending only on
$\xpom$ and $t$ and a term depending only on $x$ (or $\beta$)
and $Q^2$:
\begin{equation}
f_i^D(x,Q^2,\xpom,t) = f_{\pom/p}(\xpom,t) \cdot
f_i (\beta=x/\xpom,Q^2) \ .
\label{reggefac}
\end{equation}
This is equivalent to treating
the diffractive exchange as a `pomeron'
with a partonic structure given by the parton
distributions $f_i (\beta,Q^2)$, the variable $\beta$ 
corresponding to the fraction of the pomeron longitudinal momentum
carried by the struck quark. The 
`pomeron flux factor' $f_{\pom/p}(\xpom,t)$ represents the
probability that a pomeron with particular values of $\xpom$ and $t$
couples to the proton. 
In the low $\xpom$ region where sub-leading exchange contributions
are negligible, the data presented in section~\ref{sec:results}
are consistent with
factorisation of the $\xpom$ dependence, as required for the proton vertex
factorisation expressed in equation~\ref{reggefac}.
There is also no evidence from the investigations in \cite{zeuslps,H1FPS} 
for any change in the $t$ or $\my$ dependences 
as either $\beta$ or $Q^2$ vary in the
range relevant to the present analysis.

\subsection{Fit Procedure}
\label{fitproc}

To determine the DPDFs,
fits are made to the reduced
cross section data as presented in 
figures~\ref{q2dep1}-\ref{q2dep5},
for which $\beta \leq 0.8$.
In order to avoid regions which are most likely to be influenced
by higher twist contributions or other problems with the chosen 
theoretical framework,
only data with
$\mx > 2 \ {\rm GeV}$ are included in the fit
and the region $Q^2 < 8.5 \ {\rm GeV^2}$ is excluded, as explained 
in section~\ref{fitsysts}.
The total number of fitted data points is 190.

Input parameters 
describing the DPDFs at a starting scale $Q_0^2$ for QCD evolution
are adjusted to obtain the best description of the data
after NLO DGLAP \cite{dglapnlo} 
evolution\footnote{Direct pomeron to parton splitting 
functions, leading to an inhomogeneous term 
in the DGLAP evolution equations \cite{levin,watt}, are not considered.
The presence of such a term would lead to a reduced gluon 
density.} 
to $Q^2>Q_0^2$ and 
convolution of the DPDFs with coefficient functions.
The fit is performed in the $\overline{MS}$ renormalisation 
scheme \cite{msbar}
with charm and beauty quarks
treated as massive, appearing via boson gluon fusion-type
processes up to order $\alpha_s^2$ \cite{hf}.
The heavy quark masses 
are set to world average values (see table~\ref{table:params}). 
The strong coupling is set via 
$\Lambda_{QCD}^{(3)} = 399 \pm 37 \ {\rm MeV}$ for 3 flavours,
which corresponds  \cite{mst} to the world average 
$\alpha^{(5)}_s (M_Z^2)$ for five flavours
(table~\ref{table:params}).
The effects of $F_L^{D(3)}$ are considered through its relation to
the quark and gluon densities at NLO. Since the relative normalisations
of the diffractive parton densities and the flux factor in 
equation~\ref{reggefac} are arbitrary, no momentum sum rule is imposed.

The DPDFs are modelled in terms of a
light flavour singlet distribution $\Sigma(z)$, consisting of $u$, $d$ and $s$ 
quarks and anti-quarks
with $u=d=s=\bar{u}=\bar{d}=\bar{s}$,
and a gluon distribution $g(z)$.  Here, $z$ is the longitudinal momentum
fraction of the parton entering the hard sub-process
with respect to the diffractive
exchange, such that $z=\beta$
for the lowest order quark-parton model process, 
whereas $0<\beta<z$ for higher order processes. 
The quark singlet and gluon distributions are parameterised 
at $Q_0^2$ using 
a similar approach to that commonly applied to 
hadronic parton densities \cite{buras,mrst,cteq}, such that the most general 
form is
\begin{equation}
z f_i (z,Q_0^2) = A_i \, z^{B_i} \, (1 - z)^{C_i} \ .
\label{param:general}
\end{equation}
The exact choices of terms included for the quark singlet and gluon densities
and the parameterisation scale $Q_0^2$
are determined through a
systematic investigation of the parameter space, as 
described in section~\ref{fitsysts}.
The DPDFs as defined in equation~\ref{param:general} are multiplied by a 
term $e^{- \frac{0.01}{1-z}}$ in order to ensure that they vanish
at $z = 1$,  
as required for the evolution 
equations to be solvable. The parameters $C_q$ and $C_g$ thus have the 
freedom to take negative as well as positive values.
Modifying the argument of the exponential term within reasonable limits
has no visible influence on the fit quality or the extracted 
DPDFs in the range of the measurement.

The $\xpom$ dependence is
parameterised using a flux factor 
motivated by Regge theory,
\begin{eqnarray}
f_{\pom/p}(\xpom, t) = A_\pom \cdot 
\frac{e^{B_\pom t}}{\xpom^{2\alpha_\pom (t)-1}} \ ,
\label{eq:fluxfac}
\end{eqnarray}
where  the pomeron trajectory is assumed to be linear,
$\alpha_\pom (t)= \alpha_\pom (0) + \alpha_\pom^\prime t$, and the parameters
$B_\pom$ and $\alphapom^\prime$ and their uncertainties are obtained from
fits to H1 FPS data \cite{H1FPS}. 
The values of these and other
parameters which are fixed in the fits 
are summarised in table~\ref{table:params}.
Following the convention of \cite{h1f2d94},
the value of the normalisation parameter $A_\pom$ is chosen such that  
$\xpom \cdot \int_{t_{\rm cut}}^{t_{\rm min}} f_{\pom/p} \ {\rm d} t
= 1$ at $\xpom = 0.003$, where
$|t_{\rm min}| \simeq m_p^2 \, \xpom^2 \, / \, (1 - \xpom)$ is the minimum
kinematically accessible value of $|t|$, $m_p$ is the proton mass and
$|t_{\rm cut}|= 1.0 \rm\ GeV^{2}$ is the limit of the measurement.

\renewcommand{\arraystretch}{1.35} 
\begin{table}[h]
\centering
\begin{tabular}{|c|r@{$\,$}l|c|}
\hline
Parameter & \multicolumn{2}{c|}{Value} & Source \\
\hline \hline
$\alpha_\pom'$    & $0.06$ & $^{\,+\,0.19}_{\,-\,0.06} \rm\ GeV^{-2}$ & \cite{H1FPS} \\
$B_\pom$          & $5.5$ & $^{\,-\,2.0}_{\,+\,0.7} \rm\ GeV^{-2}$    & \cite{H1FPS} \\
$\alpha_\reg(0)$  & $0.50$ & $ \pm \,0.10$                      & \cite{h1f2d94} \\
$\alpha_\reg'$    & $0.3$ & $^{\,+\,0.6}_{\,-\,0.3} \rm\ GeV^{-2}$    & \cite{H1FPS} \\
$B_\reg$          & $1.6$ & $^{\,-\,1.6}_{\,+\,0.4} \rm\ GeV^{-2}$    & \cite{H1FPS} \\
$m_c$             & $1.4$ & $ \pm \,0.2 \ {\rm GeV}$            & \cite{pdg} \\
$m_b$             & $4.5$ & $ \pm \,0.5 \ {\rm GeV}$            & \cite{pdg} \\
$\alpha_s^{(5)}(M_Z^2)$ & $0.118$ & $ \pm \,0.002$                    & \cite{pdg} \\
\hline
\end{tabular}
\caption{The values of the fixed parameters
and their uncertainties, as used in the QCD
fits. Since they are strongly 
anti-correlated when extracted 
from fits to the FPS data, $\alpha_\pom'$ and
$B_\pom$ are varied simultaneously to obtain the 
theoretical errors on the fit
results, as are $\alpha_\reg'$ and
$B_\reg$. The remaining parameters are varied independently.
The theoretical uncertainties on the free parameters of the fit
also contain a contribution from variation of 
the parameterisation scale $Q_0^2$, as described in section~\ref{fitsysts}.}
\label{table:params}
\end{table}
\renewcommand{\arraystretch}{1.}

To obtain a good description of the data, 
an additional sub-leading
exchange ($\reg$) is included, which has a lower trajectory intercept than the
pomeron and which
contributes significantly only at low $\beta$ and large $\xpom$.
As in \cite{h1f2d94,H1FPS} this contribution is assumed to 
factorise in the 
same way as the pomeron term, such that equation~\ref{reggefac} is modified to
\begin{equation}
f_i^D(x,Q^2,\xpom,t) = f_{\pom/p}(\xpom,t) \cdot f_i (\beta,Q^2) \ + \
n_\reg \cdot f_{\reg/p}(\xpom,t) \cdot f_i^\reg (\beta,Q^2) \ .
\label{reggefac2}
\end{equation}
The flux factor $f_{\reg/p}$ takes the
form of equation~\ref{eq:fluxfac}, 
normalised via a parameter $A_\reg$ in the same manner as for the pomeron
contribution and with fixed parameters
$\alpha_\reg(0)$, $\alpha_\reg^\prime$ and
$B_\reg$ obtained from other H1 measurements (see table~\ref{table:params}). 
The parton densities $f_i^\reg$ of the sub-leading exchange are
taken from a parameterisation derived from fits to pion 
structure function data~\cite{owens}.
Choosing a different 
parameterisation \cite{grvpion}
does not affect the fit results significantly.

The free parameters of the fit are the $A$, $B$ and $C$ parameters which
determine the quark singlet and gluon 
distributions (equation~\ref{param:general}), together with 
$\alphapom(0)$, which controls the $\xpom$ dependence and 
$n_{_\reg}$, which controls the
normalisation of the sub-leading exchange contribution.
In order to constrain these parameters,
a $\chi^2$ function as defined in \cite{h1f29697} is minimised. This
function
involves the combined statistical and uncorrelated systematic errors for
each data point and also takes account of
correlations between data points caused by systematic uncertainties by
allowing variations in each systematic error source
at the expense of 
increases in the $\chi^2$ variable.
Ten correlated systematic error parameters are considered for each of the 
SpaCal and LAr electron data sets, one for each of the error sources 
listed in tables~\ref{spacal:data} and~\ref{lar:data}.
In this procedure, the sources of correlated uncertainty
are taken to be common for the 
`1997 MB' and `1997 all' data,
whereas it is assumed that there are no correlations between the uncertainties
on the SpaCal and the LAr electron data.  
A further systematic error parameter controls the relative
normalisation of the LAr electron data set with respect to the SpaCal 
electron data, for which the uncertainty is $9.8\%$ 
(section~\ref{sec:syst}). 
The central results for the 
DPDFs and other parameters obtained from 
the fit are not altered significantly if all systematic uncertainties
leading to correlations between the data points are ignored.

The statistical and experimental systematic errors on the data points
and their correlations
are propagated \cite{pascaudzomerlal} to obtain 
experimental uncertainties on the
DPDFs and other fit parameters, which correspond to
increases in the $\chi^2$ variable by one unit. 
The theoretical error is obtained from 
variations of the assumed parameters as given in 
table~\ref{table:params},
with an additional contribution 
expressing the sensitivity to the choice of DPDF
parameterisation, obtained by varying $Q_0^2$ 
as discussed in section~\ref{fitsysts}. 
Since the pomeron flux factor is constrained simultaneously with the
parton densities, the possible 
influence of interference between the pomeron and sub-leading
exchange contributions
cannot be assessed. However, in previous similar fits in which $\alphapom(0)$
was extracted separately from the parton 
densities \cite{h1f2d94}, $\alphapom(0)$ changed by less
than $0.01$ between the cases of no interference and maximum constructive
interference.

\subsection{Choices of Fit Parameterisation and Kinematic Range}
\label{fitsysts}

In order to optimise the results of the fit, the sensitivity
to variations in the details of the parameterisation is investigated.
With the small numbers of parameters used to describe the parton densities,
the $\chi^2$ values and the results of the fits are sensitive to the choice
of the parameterisation scale $Q_0^2$ \cite{h1f29697}, so that its value must
be optimised by $\chi^2$ minimisation
for each parameterisation choice. The $Q_0^2$-optimised results are then
compared in order to make the final parameterisation choice.
To ensure that the results of the fit are not sensitive to the 
kinematic range of the  
data included in the fit, the sensitivity of this procedure to 
variations in the kinematic cuts is also tested.
All parton density parameterisation changes and kinematic 
range variations lead 
to extracted
values of $\alphapom(0)$ which are within the experimental uncertainties
(see also section~\ref{dpdf:results}). 

The only significant sensitivity to the 
boundaries of the chosen kinematic range
occurs when the minimum $Q^2$ value of the data included
in the fit, $Q^2_{\rm min}$,
is varied. 
Whereas the quark distribution remains stable 
within uncertainties for all $Q^2_{\rm min}$ choices, 
the gluon distribution for $z \ \lapprox \ 0.5$ increases systematically 
as $Q^2_{\rm min}$ varies between $3.5 \ {\rm GeV^2}$ and 
$8.5 \ {\rm GeV^2}$,
changing by about $40\%$ in total.
The $\chi^2$ per degree of freedom also improves steadily as $Q^2_{\rm min}$
varies over this range. There is no evidence for
any further variation in the gluon density
for $Q^2_{\rm min} > 8.5 \ {\rm GeV^2}$. 
The lowest $Q^2$ data are therefore omitted from the fit
and $Q^2_{\rm min} = 8.5 \ {\rm GeV^2}$ is chosen.
The $Q^2_{\rm min}$ dependence 
is reflected in 
figures~\ref{q2dep1}-\ref{q2dep5}
as a tendency for the fit result extrapolated
to $Q^2 < 8.5 \ {\rm GeV^2}$ to lie below the data. 
The dependence of the gluon density on 
$Q^2_{\rm min}$ may indicate inadequacies in the 
adopted formalism at the lowest $Q^2$ values.

The fit results are not 
sensitive to variations in the  
minimum or maximum $\beta$ values of the data included, although
the minimum $\beta$ is correlated with $Q^2_{\rm min}$
through the kinematics. 
There is similarly
no significant change in the fit results when the minimum $\mx$ value of
the data included 
is increased or when the highest $\xpom$ data are omitted.

For the quark singlet distribution, 
the data require the inclusion of all three parameters
$A_q$, $B_q$ and $C_q$ in equation~\ref{param:general}.
By comparison, the gluon density is weakly 
constrained by the data, which are found to be 
insensitive to the $B_g$ parameter. The gluon density is thus
parameterised at $Q_0^2$ using only the $A_g$ and $C_g$ parameters.
With this parameterisation, a value of $Q_0^2 = 1.75 \ {\rm GeV^2}$ 
yields the minimum $\chi^2$ value of $158$ for
$183$ degrees of freedom. This fit is referred to as 
the `H1 2006 DPDF Fit A' in the
figures. As a measure of the parameterisation
uncertainty, $Q_0^2$ is varied between $1.15 \ {\rm GeV^2}$ and 
$2.05 \ {\rm GeV^2}$, for which the $\chi^2$ variable increases by one unit. 
The correlated systematic error parameters have a mean close to 
zero and 
the largest shift in a correlated error source is  
$1.0 \, \sigma$. The fit shifts the normalisation of the LAr data 
relative to the SpaCal data by 
$0.45 \, \sigma$. 
Both the DPDFs and the $\chi^2$ per degree of
freedom of this fit
can be reproduced closely using 
the approach based on Chebyshev polynomials in \cite{h1f2d94}.

As discussed in section~\ref{sec:betaq2}
(figure~\ref{fig:logderiv}b), the $Q^2$ dependence of the
data at fixed $\beta$ and $\xpom$ determines the gluon density well at low
$\beta$. However, as $\beta$ increases the $\ln Q^2$ derivative becomes 
smaller and the
fractional error on the
gluon density becomes correspondingly larger. At the highest $\beta$
values, where the $Q^2$ evolution is driven by quarks,
the $Q^2$ dependence of $\sigma_r^{D(3)}$ becomes insensitive
to the gluon density.
The results for the gluon density at large $z$ 
are thus determined principally by the data at lower
$z$ coupled with the parameterisation choice. This lack of sensitivity is 
confirmed by repeating the fit with the parameter $C_g$, which determines
the high $z$ behaviour, set to zero. Apart from the exponential term,
the gluon density is then a simple constant
at the starting scale for evolution, which is chosen to be 
$Q_0^2 = 2.5 \ {\rm GeV^2}$ by 
$\chi^2$ minimisation. Even with this very simple
parameterisation of the gluon density, the $\chi^2$ variable increases
only slightly to $\chi^2 = 164$, with $184$ degrees of freedom.
This fit is referred to as the
`H1 2006 DPDF Fit B' in the figures.

\subsection{Diffractive Parton Distributions and Effective Pomeron 
Intercept}
\label{dpdf:results}

A good description of the data is obtained
throughout the fitted range 
$Q^2 \geq 8.5 \ {\rm GeV^2}$, $\beta \leq 0.8$ and
$\mx > 2 \ {\rm GeV}$
by both H1 2006 DPDF Fit A and Fit B. 
The results of Fit A
are compared with the measured reduced
cross section in 
figures~\ref{q2dep1}-\ref{q2dep5}.
The results for the fit parameters are given in table~\ref{table:fitresults}. 
They can also be found together with the
correlation coefficients between the parameters at \cite{fit:table}.

\renewcommand{\arraystretch}{1.35} 
\begin{table}[h]
\centering
\begin{tabular}{| c | r@{$\pm$}l | r@{$\pm$}l |}
\hline
Fit Parameter & \multicolumn{2}{c|}{Fit A} & \multicolumn{2}{c|}{Fit B} \\
\hline \hline
$\alphapom(0)$   & $1.118$ & $0.008$               & $1.111$ & $0.007$ \\
$n_{_\reg}$      & $(1.7$  & $0.4) \times 10^{-3}$ & $(1.4$  & $0.4) \times 10^{-3}$ \\
$A_q$            & $1.06$  & $0.32$                & $0.70$  & $0.11$  \\
$B_q$            & $2.30$  & $0.36$                & $1.50$  & $0.12$  \\
$C_q$            & $0.57$  & $0.15$                & $0.45$  & $0.09$ \\
$A_g$            & $0.15$  & $0.03$                & $0.37$  & $0.02$ \\
$C_g$            & $-0.95$ & $0.20$                & \multicolumn{2}{c|}{$0$
\ (fixed)} \\
\hline
\end{tabular}
\caption{The central values 
of the parameters extracted in the 
`H1 2006 DPDF Fit A' and `B', and the corresponding
experimental uncertainties.}
\label{table:fitresults}
\end{table}
\renewcommand{\arraystretch}{1.} %

\begin{figure}[h]
\centering \epsfig{file=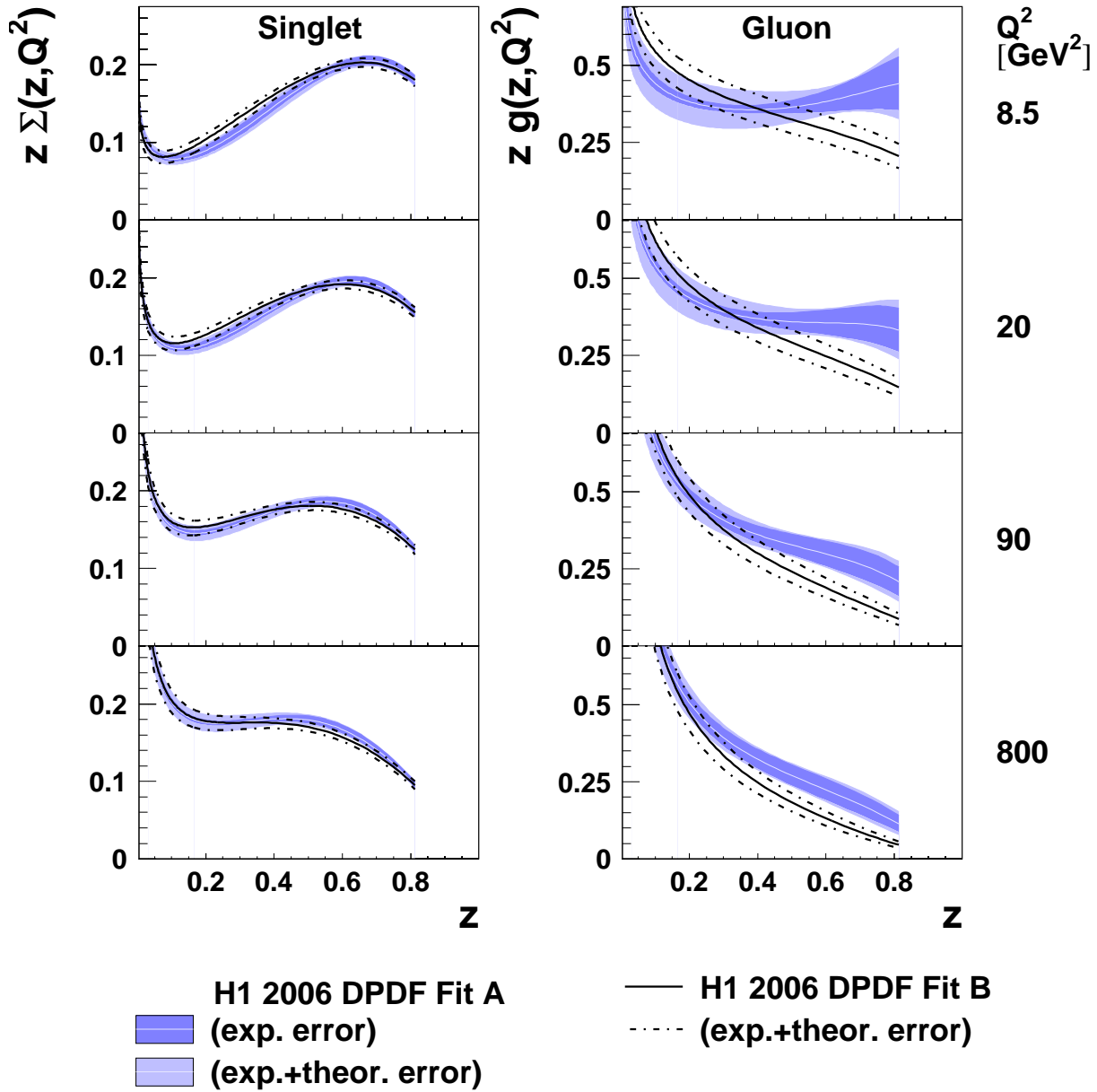,width=\linewidth}
\caption{Comparison on a linear $z$ scale between the total quark singlet and
gluon distributions
obtained from the `H1 2006 DPDF Fit A' and 
the `H1 2006 DPDF Fit B'. These two fits differ
in the parameterisation chosen for the gluon density 
at the starting scale for QCD evolution.
The DPDFs are shown
at four different values of $Q^2$ for the 
range $0.0043 < z < 0.8$, corresponding approximately to that of
the measurement. For `Fit A', the central result is 
shown as a light coloured
central line, which is surrounded by inner error bands corresponding to
the experimental uncertainties 
and outer error bands corresponding to the
experimental and theoretical uncertainties added in quadrature.
For `Fit B', only the total uncertainty is shown.}
\label{pdfparam}
\end{figure}

The diffractive quark singlet and gluon distributions from Fit A 
are shown together with their uncertainties
on a logarithmic $z$ scale in
figure~\ref{pdfplot}.
In order to illustrate the high $z$ region in more
detail, they are also shown on a linear $z$ scale in 
figure~\ref{pdfparam}, where they are compared 
with the results from Fit B. 
At low $Q^2$, 
both the quark singlet and the gluon densities remain large up to
the highest $z$ values accessed.
The quark singlet distribution is well constrained,
with an uncertainty of typically $5 -10\%$ and good agreement between
the results of Fit A and Fit B. 
The gluon distribution has a larger uncertainty of typically $15\%$ at low to
moderate $z$ and low $Q^2$, dominated by the influence of the $Q_0^2$ 
variation. For $z \ \gapprox \ 0.5$, 
where the sensitivity to the gluon density 
becomes poor, the level of agreement between Fit A and Fit B worsens.

\begin{figure}[h]
\centering \epsfig{file=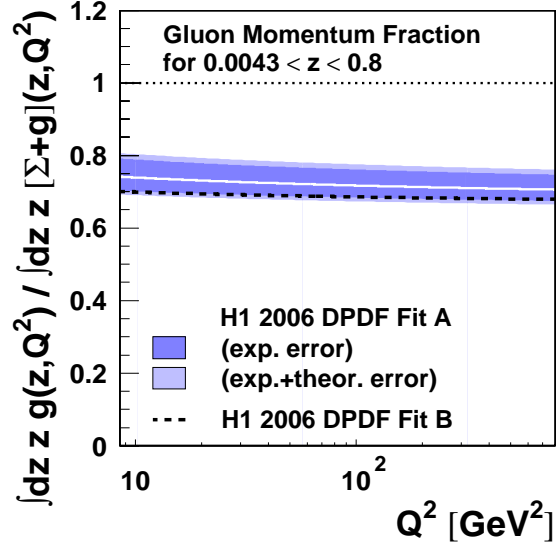,width=0.45\linewidth}
\caption{The $Q^2$ dependence of the 
fraction of the longitudinal momentum of the diffractive
exchange which is carried by gluons according to the 
`H1 2006 DPDF Fit A', 
integrated over the 
range $0.0043 < z < 0.8$, corresponding approximately to that of
the measurement.
The light coloured
central line is surrounded by an inner error band corresponding to
the experimental uncertainty
and outer error band corresponding to the
experimental and theoretical uncertainties added in quadrature.
The central result from the `H1 2006 DPDF Fit B' is also indicated.}
\label{gluplot}
\end{figure}

As shown in figure~\ref{gluplot}, the fraction of the exchanged momentum
carried by gluons integrated over the range $0.0043<z<0.8$,
corresponding approximately to that of the measurement,
is around $70 \%$ throughout the $Q^2$ range studied, 
confirming the conclusion from earlier
\linebreak
work \cite{h1f2d94,zeus:gpjets}. 
The integrated gluon fraction is somewhat
smaller for Fit B due to the lower gluon density at large $z$ values,
though the results from the two fits are consistent within the uncertainties.

Fit A yields an effective pomeron intercept of 
\begin{eqnarray}
\alphapom(0) = 1.118 \ \pm 0.008 \ \mathrm{(exp.)}  \ 
^{+ 0.029}_{- 0.010} \ \mathrm{(model)} \ , 
\label{alpom:answer}
\end{eqnarray}
where the first error is the
full experimental uncertainty
and the second expresses the model dependence. 
This model dependence uncertainty arises dominantly from
the variation of
$\alphapom^\prime$, which is strongly positively 
correlated
with $\alphapom(0)$, such that $\alphapom(0)$ increases to around $1.15$
if $\alphapom^\prime$ is set to $0.25$.
The intercept has also been shown to be strongly sensitive to 
the value of $F_L^D$ \cite{paul:budapest}, though with the NLO treatment
adopted here, $F_L^D$ is determined in the fit and no additional uncertainty
is included. The influence of $F_L^D$ on the reduced cross section according
to the fit is shown for $\xpom = 0.003$ in figure~\ref{q2dep3}a.
It is similar at other values of $\xpom$.

\begin{figure}[h]
\centering \epsfig{file=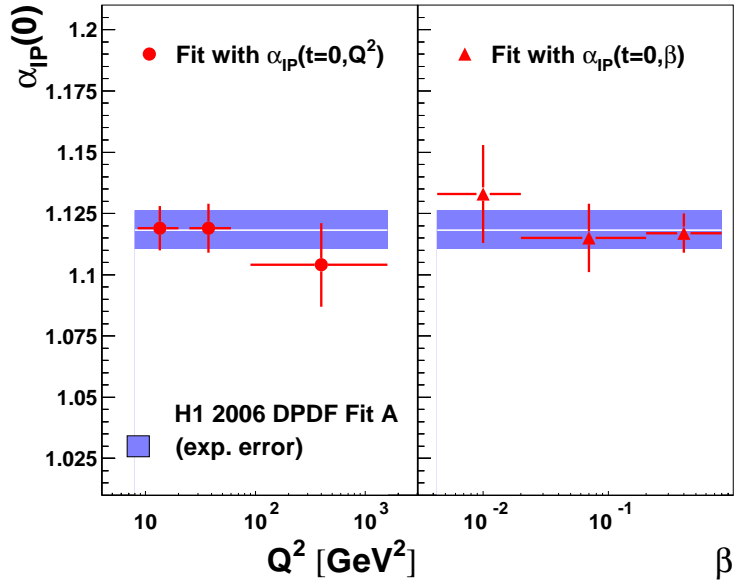,width=0.6\linewidth}
\caption{Results from modified versions of the
`H1 2006 DPDF Fit A' in which
additional free parameters are included, corresponding to the value of
$\alphapom(0)$ in a number of different ranges of $Q^2$ (left) or 
$\beta$ (right). 
The data
points correspond to the results for $\alphapom(0)$ in each interval
of $Q^2$ or $\beta$, with error bars corresponding to the full
experimental uncertainties. 
The bands show the result and experimental uncertainty from the 
standard fit in which a single parameter is used for $\alphapom(0)$.}
\label{pomq2}
\end{figure}

The extracted $\alphapom(0)$ is slightly higher than
the value $\alphapom(0) \simeq 1.08$ expected for the `soft' pomeron
\cite{DL:stot} describing long distance hadronic 
interactions. The result  
is compatible with that obtained from ZEUS $F_2^D$ data for 
$Q^2 \ \lapprox \ 20 \ {\rm GeV^2}$ \cite{ZEUS:97}. However,
in \cite{ZEUS:97}, evidence was reported for an increase of $\alphapom(0)$
for $Q^2 \ \gapprox \ 20 \ {\rm GeV^2}$.
In some models \cite{bekw,kgb}, a $\beta$ dependent $\alphapom(0)$ has
also been suggested.
Any such dependence of $\alphapom(0)$ on $Q^2$ or $\beta$ implies a breakdown 
of proton vertex factorisation. 
In order to test for such effects in the present data, the QCD
fit is repeated with additional free parameters 
corresponding to independent values of
$\alphapom(0)$ in different ranges of $Q^2$ or $\beta$.  
As can be seen from the results in figure~\ref{pomq2},
there is no evidence 
for any variation of $\alphapom(0)$ with either variable
within the kinematic range of the fit.
This remains the case if the data with $Q^2 < 8.5 \ {\rm GeV^2}$
are included.  
This lack of $Q^2$ dependence of $\alphapom(0)$
contrasts with the $Q^2$ dependent
effective pomeron intercept extracted in a Regge
approach to inclusive small $x$ proton structure function data,
as discussed further in section~\ref{sec:incdif}.

The presence of the sub-leading exchange term 
is required by the data, the $\chi^2$ increasing
by approximately 40 units if only the pomeron contribution is included. 
Its relative size is expressed through the normalisation
parameter 
$n_{_\reg} = [1.7 \pm 0.4 \ \mathrm{(exp.)} \ ^{+ 1.5}_{-0.8}
\ \mathrm{(model)}] \times 10^{-3}$ (Fit A), where the dominant 
uncertainty arises from the correlation with $\alpha_\reg(0)$.
The sub-leading exchange plays a significant role
at high $\xpom$ and low $\beta$, as shown in
figures~\ref{q2dep4}a and~\ref{q2dep5}a. It accounts
for around $30\%$ ($10\%$) of the cross section
at $\xpom = 0.03 \ (0.01)$ and is negligible at lower $\xpom$.


\section{The Diffractive Charged Current Cross Section}

The diffractive 
charged current process $e^+ p \rightarrow \bar{\nu}_e XY$ is sensitive to 
the diffractive $d$, $\overline{u}$, $s$ and $\overline{c}$
densities at large scales. Assuming factorisation, the measurement of the
charged current cross section
thus tests the assumed
flavour decomposition of the quark singlet component of the DPDFs,
which is completely unconstrained by the neutral current data.
The total charged current cross section integrated over the range
$Q^2 > 200 \ {\rm GeV^2}$, $y < 0.9$ and $\xpom < 0.05$ 
at $\sqrt{s} = 319 \ {\rm GeV}$ is measured to be
\begin{eqnarray}
  \sigma^{\rm diff}_{\rm CC} = 390 \pm 120 \ {\rm (stat.)} 
\pm 70 \ {\rm (syst.)} \ {\rm fb} \ , 
\end{eqnarray}
corresponding to 
$2.2 \pm 0.7 \ {\rm (stat.)} \pm 0.4 \ {\rm (syst.)} \%$ of the
total charged current cross section \cite{h1nc9900}
for the same $Q^2$ and $y$ ranges,
with $x < 0.05$. 

The measured charged current cross section is compared with the
prediction of the `2006 DPDF Fit A' 
to the neutral current diffractive DIS data
described in section~\ref{dpdf:results}. The prediction
is obtained by implementing the DPDFs extracted from the 
neutral current data in the RAPGAP \cite{rapgap} Monte Carlo generator.
The light quark-initiated contributions are 
calculated at lowest order and the 
$\bar{c} \rightarrow \bar{s}$ contribution is calculated using the 
${\cal O} (\alpha_s)$ matrix element.
Leading log ($Q^2$) parton showers are used to approximate higher
order QCD radiation.  
The resulting cross section prediction is 
$500 \ {\rm fb}$,
which is compatible with the measurement.
The experimental uncertainties on the DPDFs and the theoretical uncertainties 
detailed in table~\ref{table:params} lead to negligible errors on the 
predictions by comparison with the statistical error on the measured
cross section.

\begin{figure}[h] \unitlength=0.9mm
 \begin{picture}(160,170)
    \put(2,-2){\epsfig{file=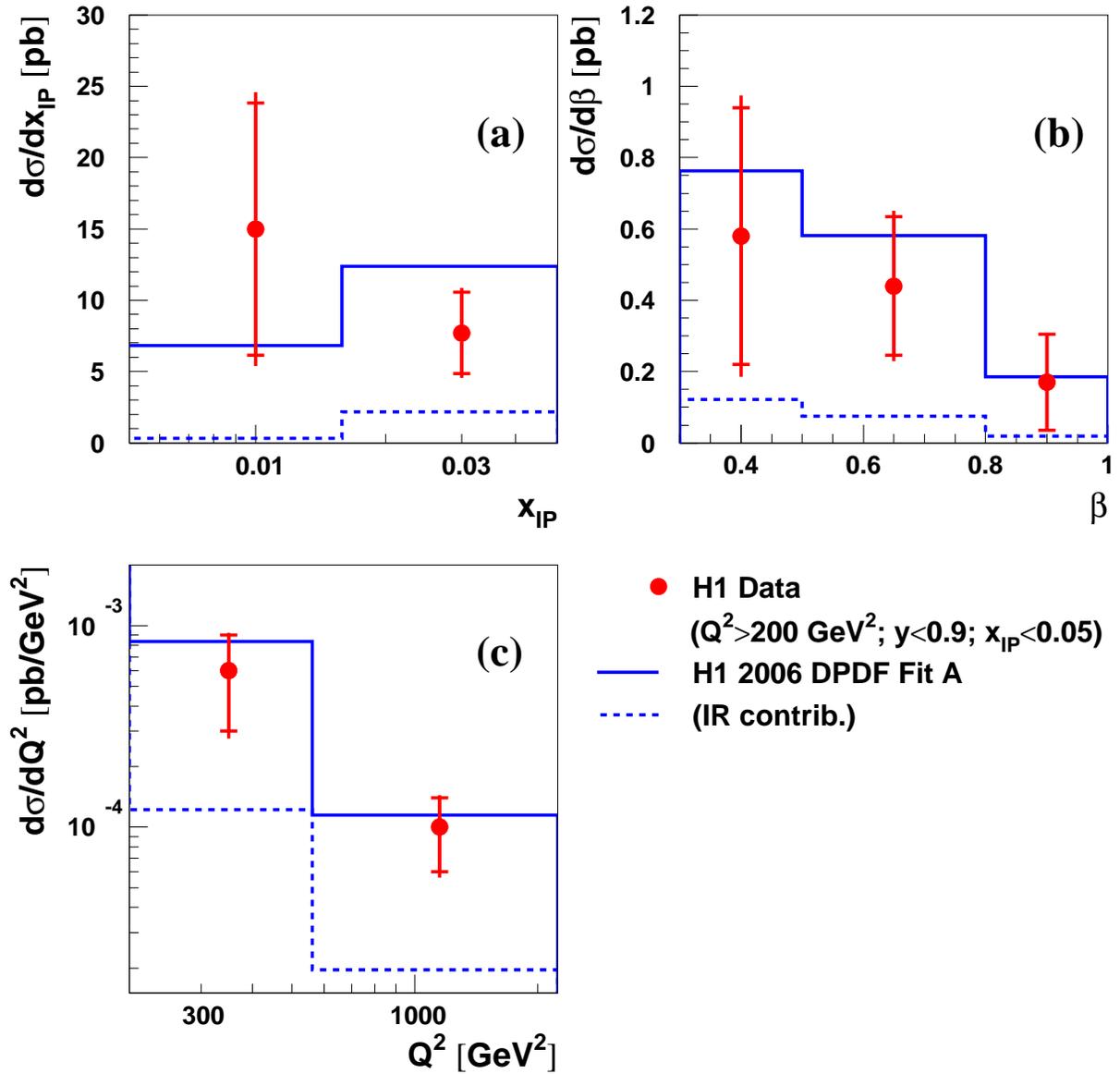,width=1.1\linewidth}}
    \put(77,153){\bf{\Large{(a)}}}
    \put(166,153){\bf{\Large{(b)}}}
    \put(77,70){\bf{\Large{(c)}}}
  \end{picture}
\caption{The cross section for the diffractive process
$e^+ p \rightarrow \bar{\nu}_e XY$ at $E_p = 920 \ {\rm GeV}$, 
shown differentially in (a) $\xpom$, (b) $\beta$
and (c) $Q^2$, for $Q^2 > 200 \ {\rm GeV^2}$, $y < 0.9$ and $\xpom < 0.05$.
The data
are compared with the predictions of the `H1 2006 DPDF Fit A'
to the neutral current data, obtained using the RAPGAP Monte Carlo 
generator. The contribution to the predictions from
the sub-leading exchange is also shown.
The data points correspond to average values 
of the differential cross sections over the regions shown by the
histograms.
The inner and outer error bars on the data points
represent the statistical and total uncertainties, respectively. 
Normalisation uncertainties are not shown.}
\label{ccplot}
\end{figure}

The charged current cross section 
measurement is shown differentially in $\xpom$,
$\beta$ and $Q^2$ in figures~\ref{ccplot}a-c, respectively. 
The numerical values are given in table~\ref{cc:tab}.
In all cases, the predictions
derived from the DPDFs of section~\ref{dpdf:results}
are in agreement with the measurements.
The charged current data are thus consistent with the singlet quark 
distribution assumed in the DPDF fit, 
where all light quark and antiquark densities are taken to be equal,
although the large statistical uncertainties preclude strong conclusions.
The contribution in the model from the
sub-leading exchange is shown as a dashed line in the figures. It
contributes at the $15\%$ level for $\xpom > 0.015$ and is negligible
at lower $\xpom$.

\section{Comparison between Diffractive and Inclusive DIS}
\label{sec:incdif}

In hadronic scattering, close connections have been drawn
between the diffractive
and the total cross sections, for example via the 
generalisation of the optical theorem to diffractive dissociation
processes \cite{triple:regge}. These connections are carried forward
into many models of low $x$ DIS \cite{sci,semicl,kgb}. Comparing the $Q^2$ and 
$x$ dynamics of the diffractive with the inclusive cross section is therefore
a powerful means of developing our understanding
of high energy QCD, comparing the properties of the
DPDFs with their inclusive counterparts and testing models.

\begin{landscape}
\begin{figure}[p] \unitlength 0.9mm
 \begin{center}
 \begin{picture}(210,160)
  \put(-10,85){\epsfig{file=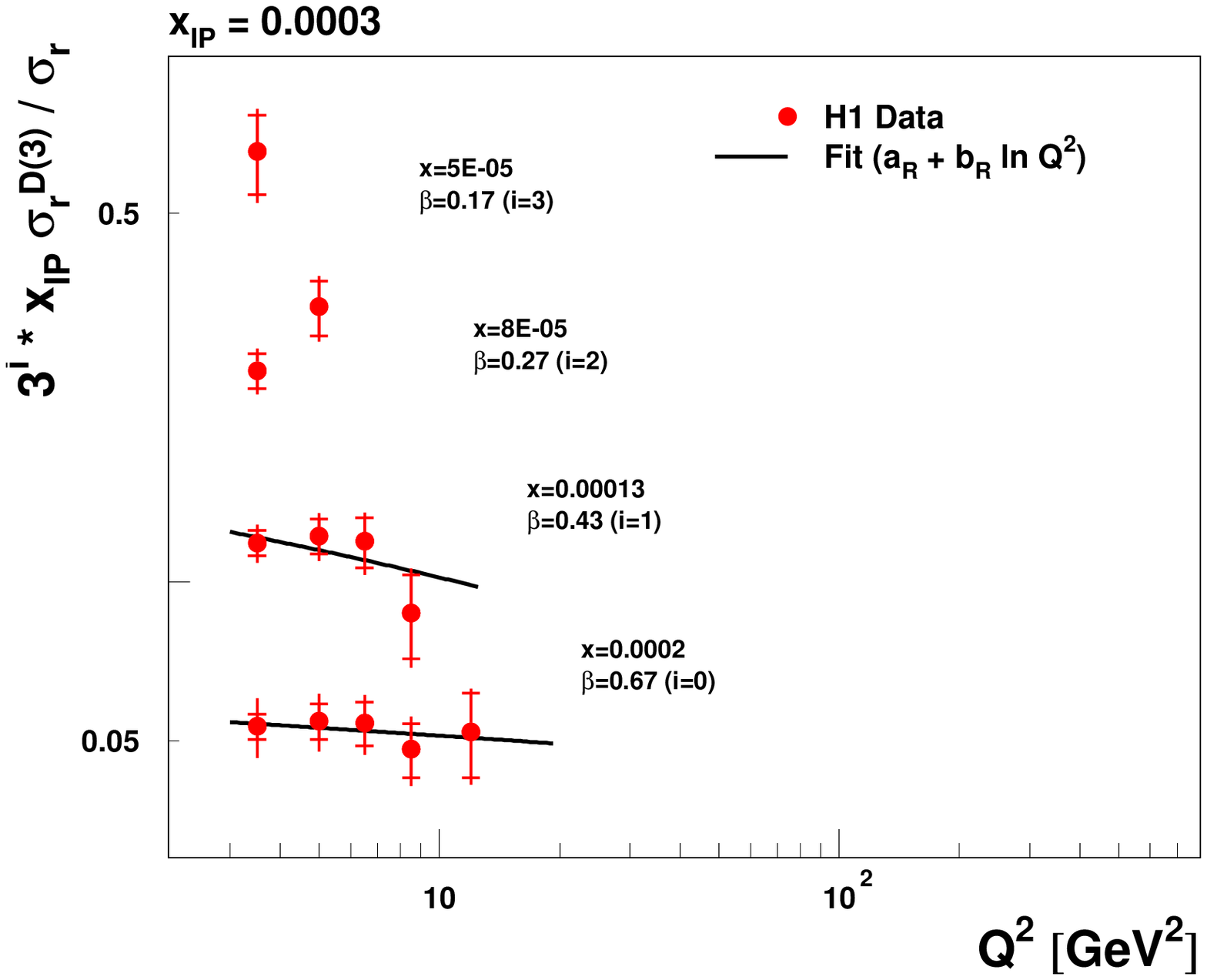,width=0.55\textwidth}}
  \put(-10,-5){\epsfig{file=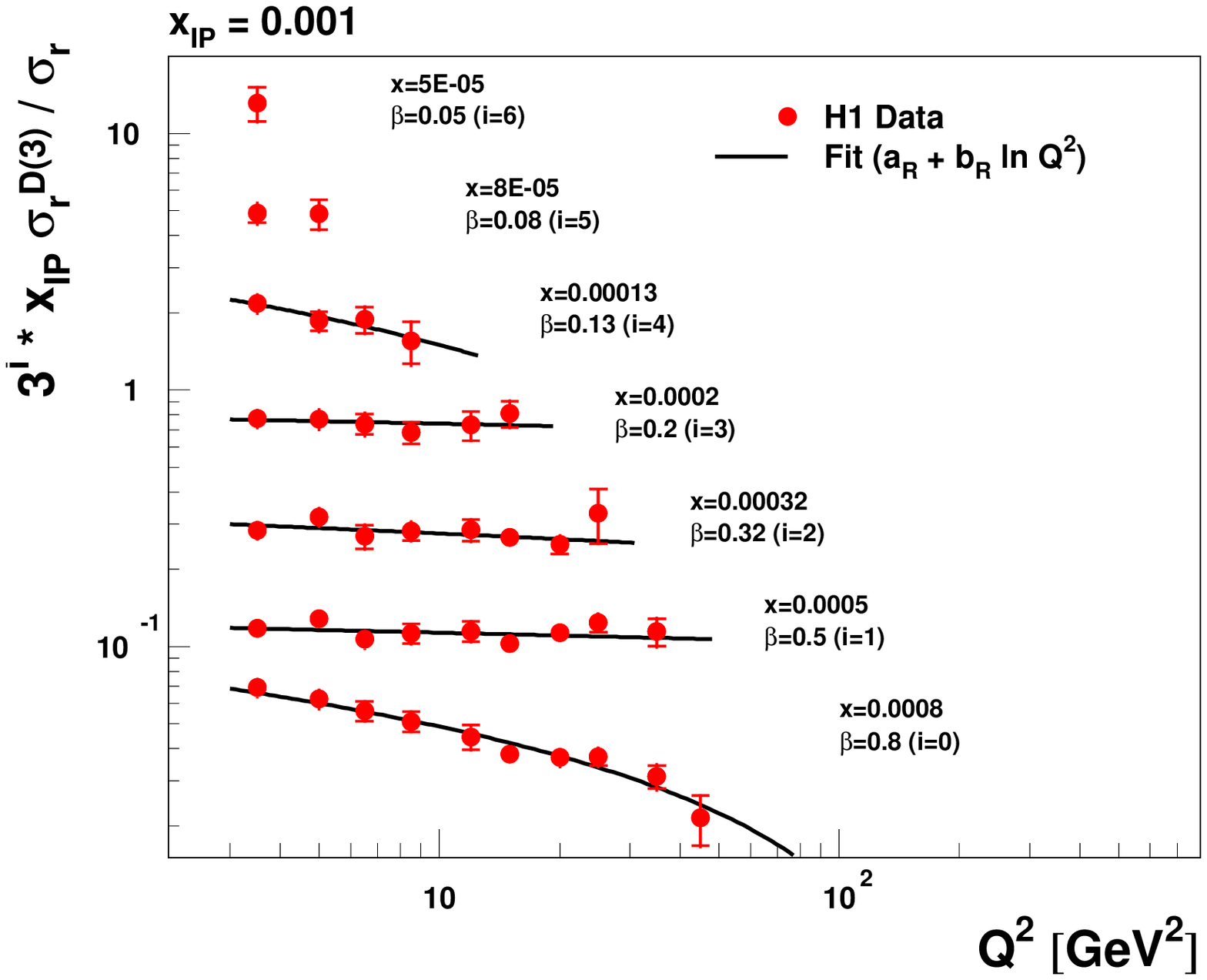,width=0.55\textwidth}}
  \put(120,10){\epsfig{file=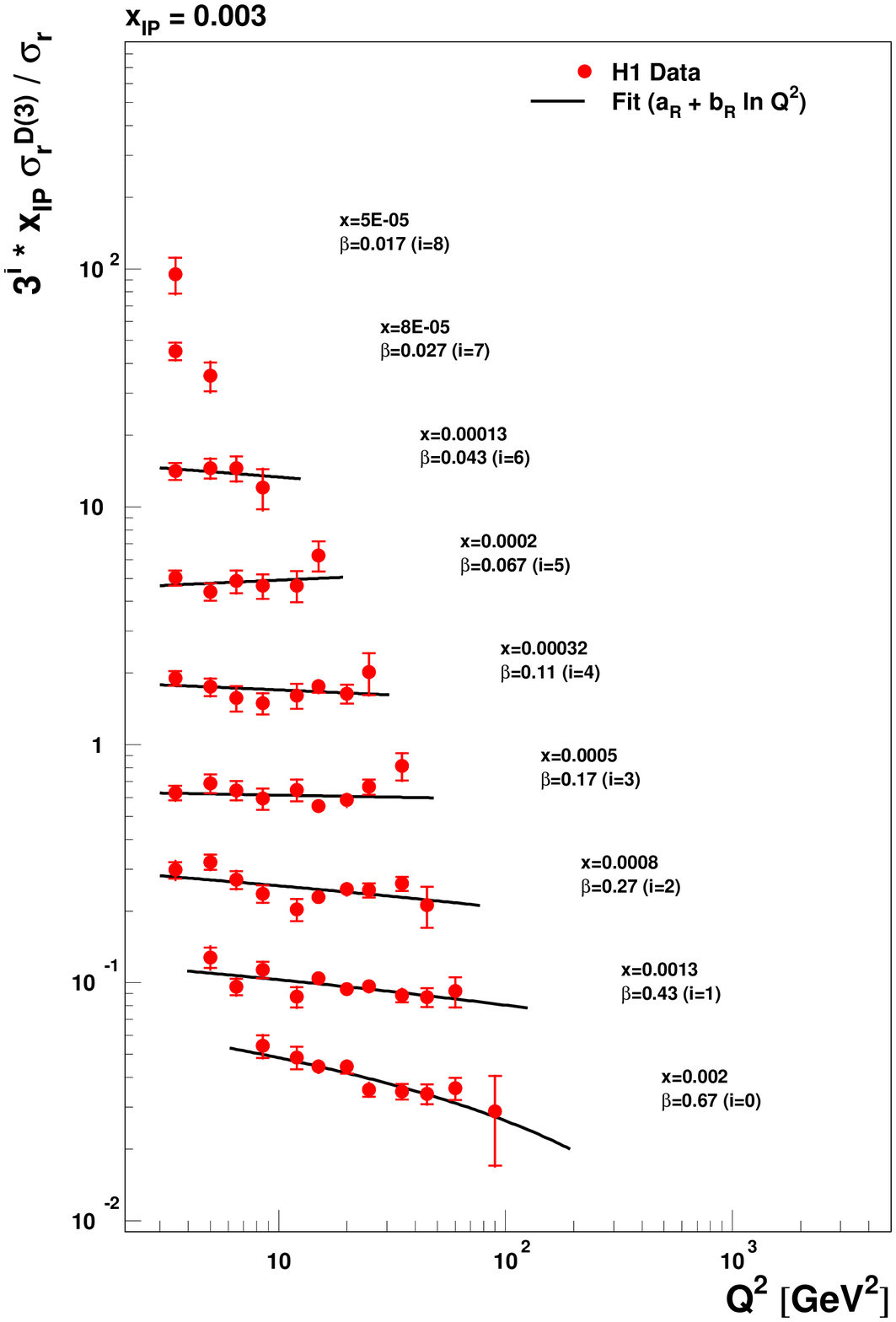,width=0.55\textwidth}}
 \end{picture}
 \end{center}
 \caption{The ratio of the diffractive to the inclusive reduced cross 
section, multiplied by $\xpom$ and shown as a function of $Q^2$ 
for fixed $x$ and fixed 
$\xpom = 0.0003$, $0.001$ and $0.003$. The data are multiplied
by a further factor of $3^i$ for visibility, with $i$ as indicated.
The inner and outer error bars 
represent the statistical and total uncertainties, respectively. 
Normalisation uncertainties are not shown. The results of fits 
of a linear dependence on $\log Q^2$ (equation~\ref{logderiv:ratio}) to the
data are also shown.}
\label{ratio1:q2}
\end{figure}
\end{landscape}

\begin{landscape}
\begin{figure}[p] \unitlength 0.9mm
 \begin{center}
 \begin{picture}(210,160)
  \put(-10,5){\epsfig{file=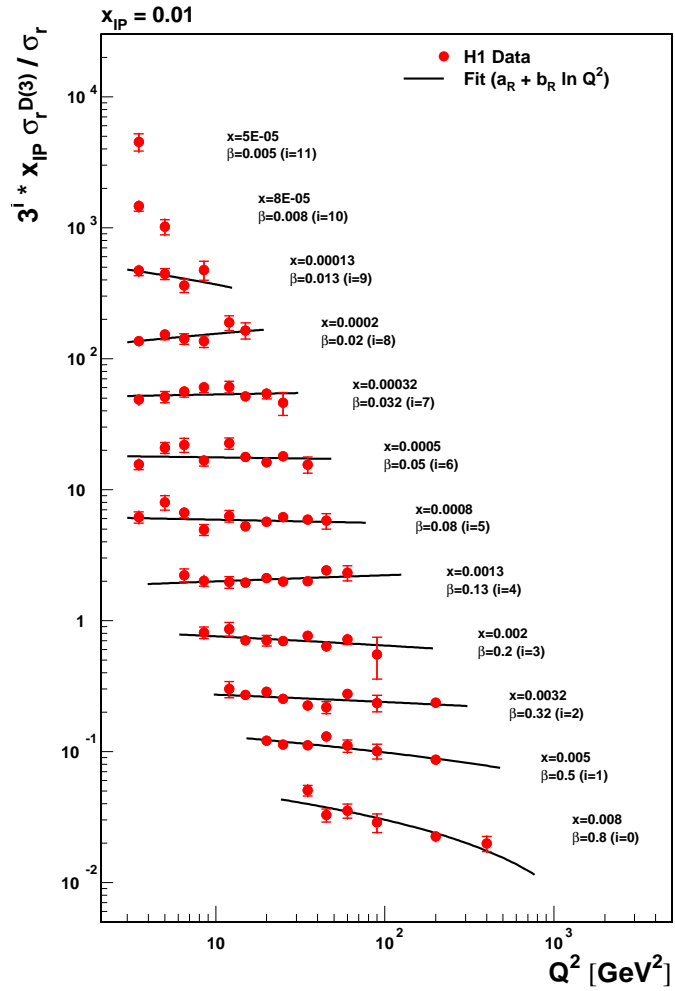,width=0.55\textwidth}}
  \put(120,5){\epsfig{file=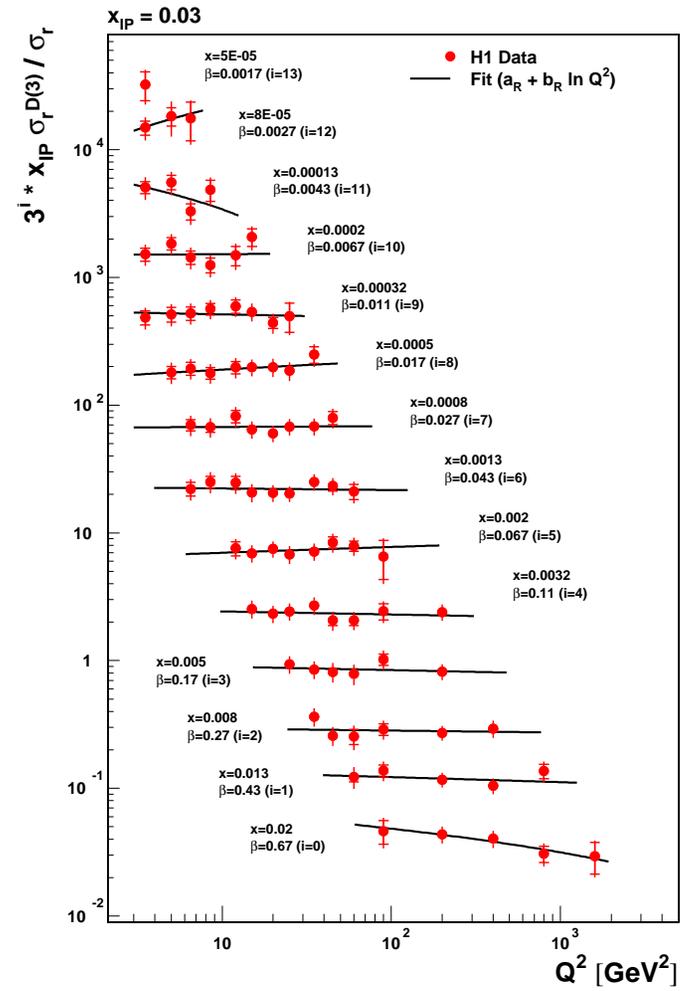,width=0.55\textwidth}}
 \end{picture}
 \end{center}
 \caption{The ratio of the diffractive to the inclusive reduced cross 
section, multiplied by $\xpom$ and shown as a function of $Q^2$ 
for fixed $x$ and fixed 
$\xpom = 0.01$ and $0.03$. 
See the caption of figure~\ref{ratio1:q2} for further details.}
\label{ratio2:q2}
\end{figure}
\end{landscape}

\begin{figure}[p]
\centering \epsfig{file=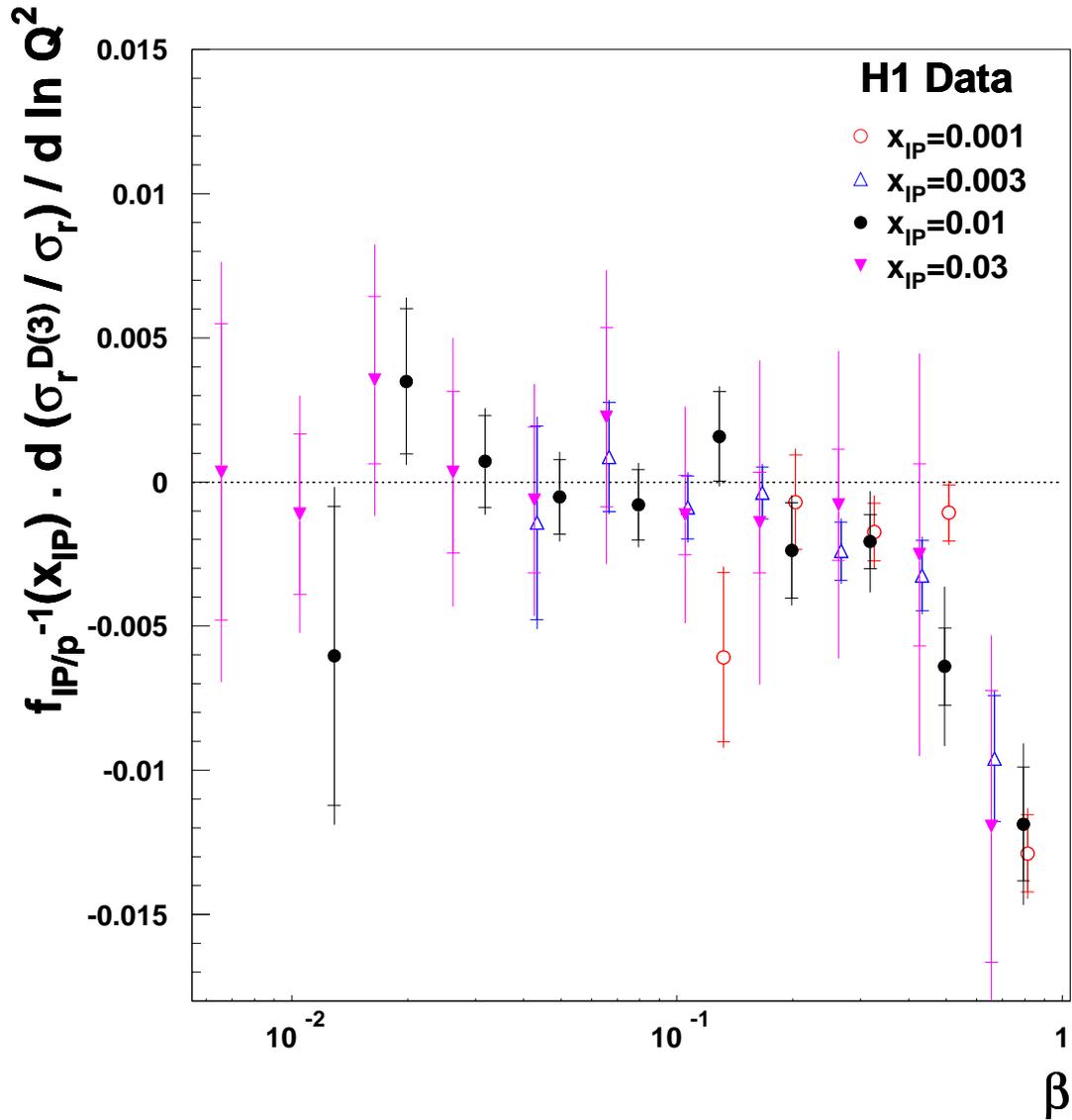,width=\linewidth}
\caption{The logarithmic $Q^2$ derivative of the ratio of the
reduced diffractive cross section to the reduced inclusive cross section
at different fixed values of $\xpom$ and
$\beta$ (see equation~\ref{logderiv:ratio}). 
The derivatives are divided by the diffractive flux factor as defined
in equation~\ref{eq:fluxfac}. The inner and outer error bars 
represent the statistical and total uncertainties, respectively.
Normalisation uncertainties are not shown.}
\label{ratderiv}
\end{figure}

The evolution of the diffractive reduced cross section with $Q^2$ is
compared with that of the inclusive DIS reduced cross 
section
$\sigma_r$ by forming the ratio
\begin{eqnarray}
\left . \frac{\sigma_r^{D(3)}(\xpom, x, Q^2)}{\sigma_r(x, Q^2)} 
\right |_{x, \xpom} \ ,
\label{ratiodef}
\end{eqnarray}
at fixed $x$ and $\xpom$, using parameterisations of the $\sigma_r$
data from\footnote{The inclusive reduced cross section is denoted
$\tilde{\sigma}_{\rm NC}$ in \cite{h1nc9900}.} \cite{h1f29697,h1nc9900}.
This ratio, which was also studied in \cite{zeuslps},  
is shown multiplied by $\xpom$ in figures~\ref{ratio1:q2} 
and~\ref{ratio2:q2}, 
as a function of $Q^2$ for all measured
$\xpom$ and $x = \beta \, \xpom$ values.
In order to compare the $Q^2$ dependences of the
diffractive and the inclusive cross sections quantitatively, 
the logarithmic derivative of their ratio,
$b_{\rm R}(x,x_\pom) =  \frac{\partial}{\partial \ln Q^2}
(\sigma_r^{D(3)} / \sigma_r)_{x, \, \xpom}$,
is extracted through fits
of a similar form to equation~\ref{eqn:logderiv}, whereby
\begin{eqnarray}
\left . \frac{  \sigma_r^{D(3)}(\xpom,x,Q^2) }{ \sigma_r(x,Q^2) } \right |_{x,\xpom} = a_{\rm R}(x,\xpom
) + b_{\rm R}(x,\xpom) \ln Q^2 \ .
\label{logderiv:ratio}
\end{eqnarray}
The fits are overlayed on the data in figures~\ref{ratio1:q2}
and~\ref{ratio2:q2}. The 
resulting values of $b_{\rm R}$ are shown in
figure~\ref{ratderiv}, where they are 
divided by the flux factor $f_{\pom / p} (\xpom)$
(equation~\ref{eq:fluxfac}), to allow comparisons between different
$\xpom$ values.
Since the dominant uncertainties arise from the diffractive data,
the statistical fluctuations in
figure~\ref{ratderiv} reproduce those of figure~\ref{fig:logderiv}a.

The ratio of the diffractive to the inclusive cross section is remarkably
flat as a function of $Q^2$ for most $x$ and $\xpom$ values, such that
the $\ln Q^2$ derivative of the ratio is consistent with zero.
At the highest $\beta \ \gapprox \ 0.3$, where $x$ approaches $\xpom$,
the ratio falls with increasing $Q^2$ and the $\ln Q^2$ derivative 
becomes negative.
This occurs in a manner which depends to good
approximation on $\beta$ only; at fixed $\beta$, there is
no significant dependence of the logarithmic derivative on $\xpom$.

\begin{figure}[h]
\centering \epsfig{file=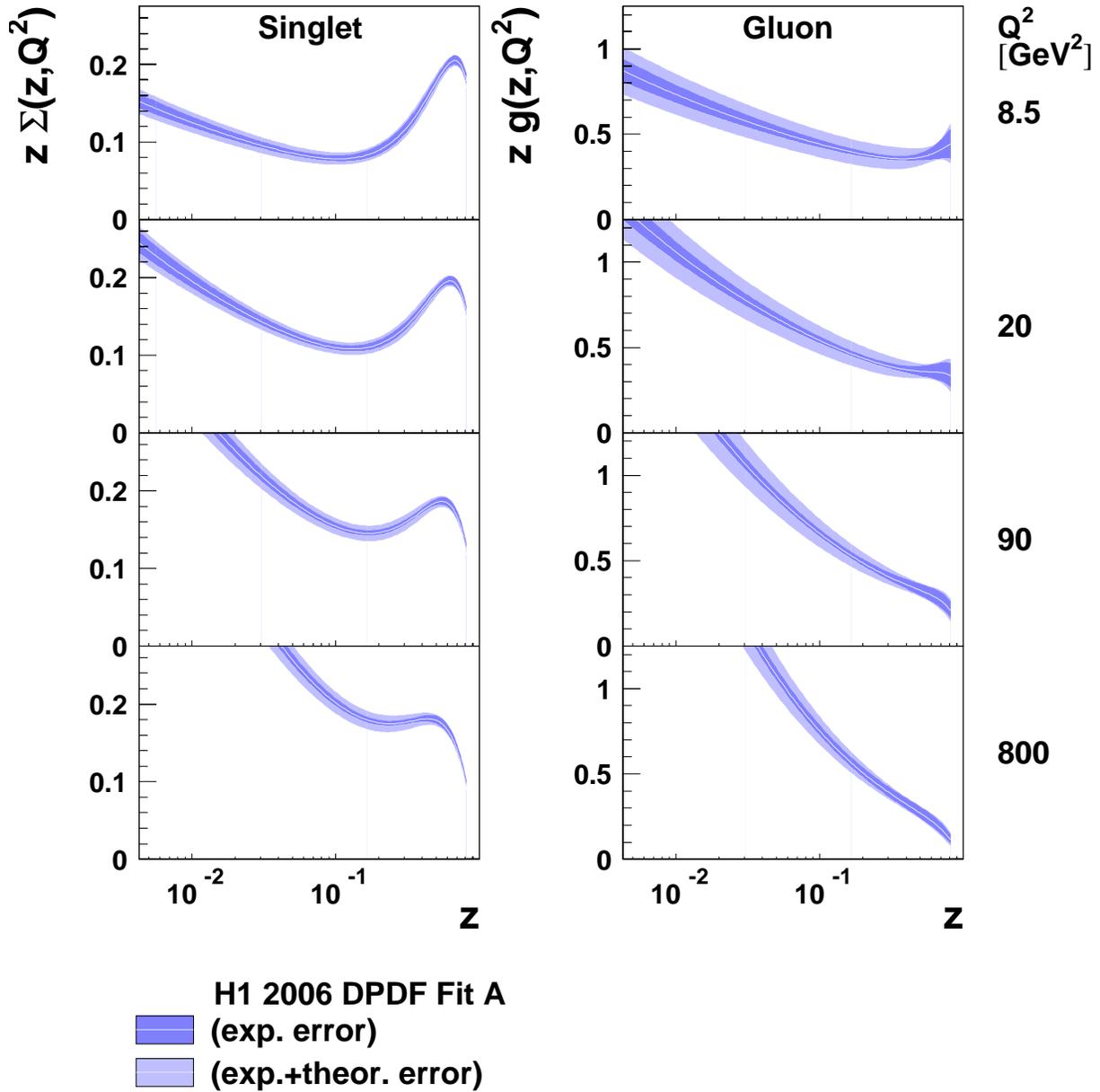,width=\linewidth}
\caption{The total quark singlet 
and gluon distributions obtained from the NLO 
QCD `H1 2006 DPDF Fit A', 
shown at four different values of $Q^2$ for the 
range $0.0043 < z < 0.8$, corresponding approximately to that of
the measurement. The light coloured
central lines are surrounded by inner error bands corresponding to
the experimental uncertainties 
and outer error bands corresponding to the
experimental and theoretical uncertainties added in quadrature.}
\label{pdfplot}
\end{figure}

The compatibility of the $\ln Q^2$ derivative of the ratio with zero over
much of the kinematic range implies that 
$1 / \sigma_r^{D(3)} \cdot \partial \sigma_r^{D(3)} / \partial \ln Q^2 \simeq
1 / \sigma_r \cdot \partial \sigma_r / \partial \ln Q^2$. 
Whereas the diffractive and inclusive 
reduced cross sections are closely related to their
respective quark densities, the
$\ln Q^2$ derivatives are approximately proportional to the relevant gluon 
densities in regions where the $Q^2$ evolution is dominated by the
$g \rightarrow q \bar{q}$ splitting 
(see section~\ref{sec:betaq2} and \cite{prytz}). 
The compatibility of $b_{\rm R}$ with zero 
for $\beta \ \lapprox \ 0.3$ thus
implies that the ratio of the quark to the gluon
density is similar in the diffractive and inclusive cases
when considered at the same $x$ values. 
Indeed, global fits to 
inclusive DIS data \cite{h1nc9900,mrst,cteq} yield
gluon fractions of approximately 70\% at low $x$, compatible with the
results of section~\ref{dpdf:results}.
At higher $\beta \ \gapprox \ 0.3$, where the DPDFs develop a more complicated
structure (see figure~\ref{pdfplot})
and the $q \rightarrow qg$ splitting becomes important in the evolution
(see section~\ref{sec:betaq2}),
$b_{\rm R}$ becomes negative.

The ratio defined in equation~\ref{ratiodef} can also be 
plotted as a function of $x$ (or $\beta$) with $\xpom$ and $Q^2$ fixed.
However, this results in a complicated dependence, which is driven
by the high $\beta$ structure of the diffractive reduced cross section
(figures~\ref{q2dep1}-\ref{q2dep5}a).

\begin{figure}[p]
\centering \epsfig{file=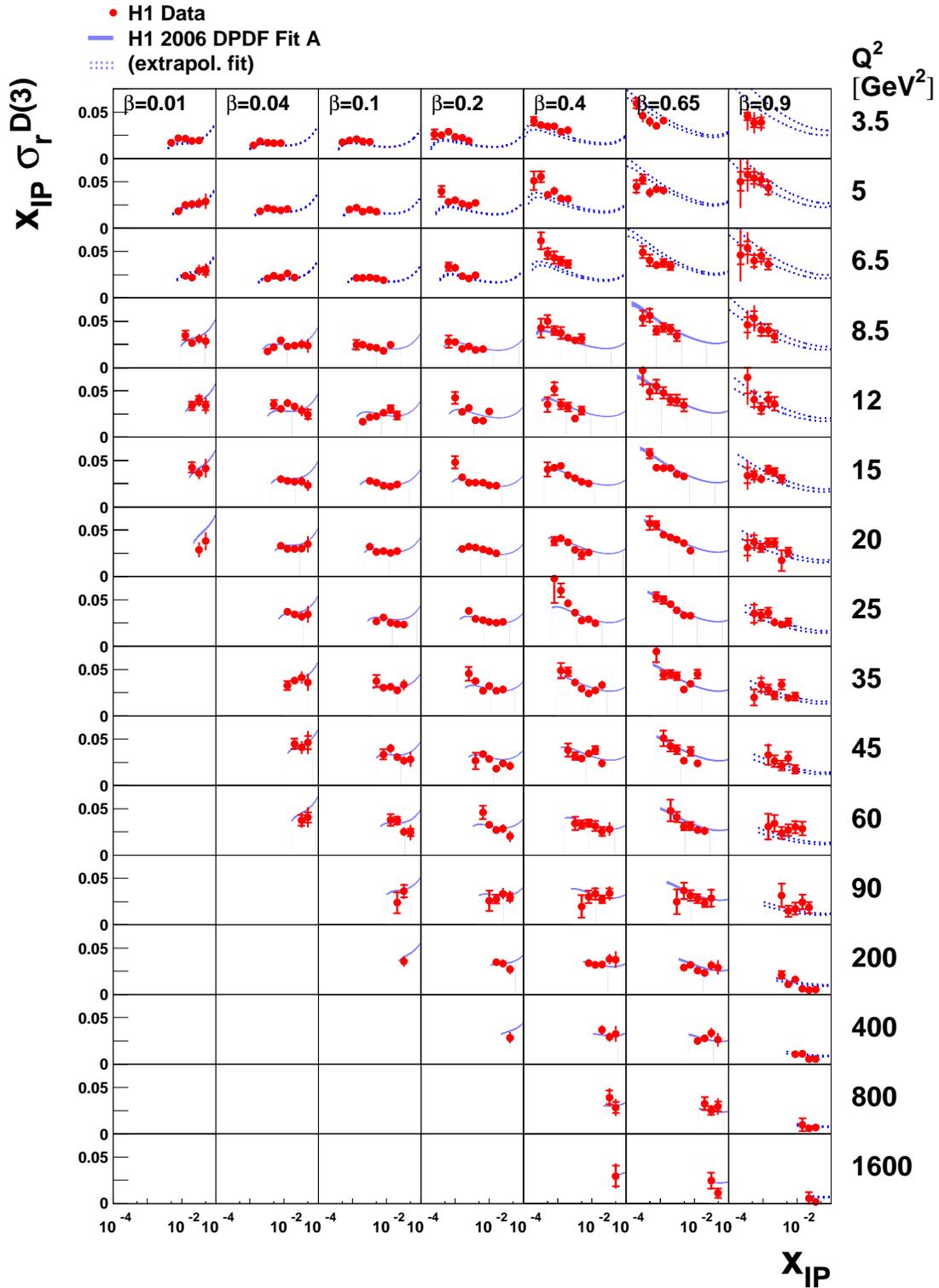,width=0.9\linewidth}
\caption{The $\xpom$ dependence of the
diffractive reduced cross section, multiplied by $\xpom$,
at fixed values of $\beta$ and $Q^2$. 
The inner and outer error bars on the data points
represent the statistical and total uncertainties, respectively. 
Normalisation uncertainties are not shown. The data
are compared with the results of the `H1 2006 DPDF Fit A'
for $E_p = 820 \ {\rm GeV}$, which is shown as a 
shaded error band (experimental uncertainties only) in 
kinematic regions which
are included in the fit and as a pair of dashed lines in regions which
are excluded from the fit.}
\label{stampa}
\end{figure}

The $x$ dependence of the ratio of the diffractive to 
the inclusive cross section
has been studied previously\footnote{The analyses 
in \cite{ZEUS:94,ZEUS:97,zeus02} 
differ from that described here in that the 
data were plotted as a function of
$W \simeq \sqrt{Q^2 / x}$. In \cite{ZEUS:94,ZEUS:97}, the results were also
shown integrated over ranges in $\mx$, whereas they
are presented here at fixed $\mx$ values.} at fixed $\mx$ 
\cite{ZEUS:94,ZEUS:97,zeus02} 
rather than fixed $\xpom$. Using the diffractive data as presented in 
figure~\ref{stampa}, the quantity
\begin{eqnarray}
\left( 1 - \beta \right) \cdot 
\frac{\xpom \cdot \sigrdarg}{\sigma_r (x, Q^2)} \ \equiv \
\mx^2 \ \frac{{\rm d} \sigma (\gamma^* p \rightarrow X Y)}{{\rm d} \mx^2}
\ \ / \ \ \sigma_{\rm tot} (\gamma^* p \rightarrow X^\prime) 
\label{eqn:trb}
\end{eqnarray}
is formed. 
Assuming proton vertex factorisation and neglecting contributions from
sub-leading exchanges, the generalised optical 
theorem \cite{triple:regge} predicts that
this ratio is independent of $Q^2$ and  
depends only weakly on 
$\beta = Q^2 / (Q^2 + \mx)$ and $x \simeq Q^2 / W^2$ for sufficiently large 
$\mx$. In models in which both the diffractive and the inclusive
cross sections are governed by a universal
pomeron \cite{DL:stot},
the remaining weak $x$ dependence of the ratio arises due to 
the deviations from unity of the pomeron trajectory.

\begin{figure}[p]
\centering \epsfig{file=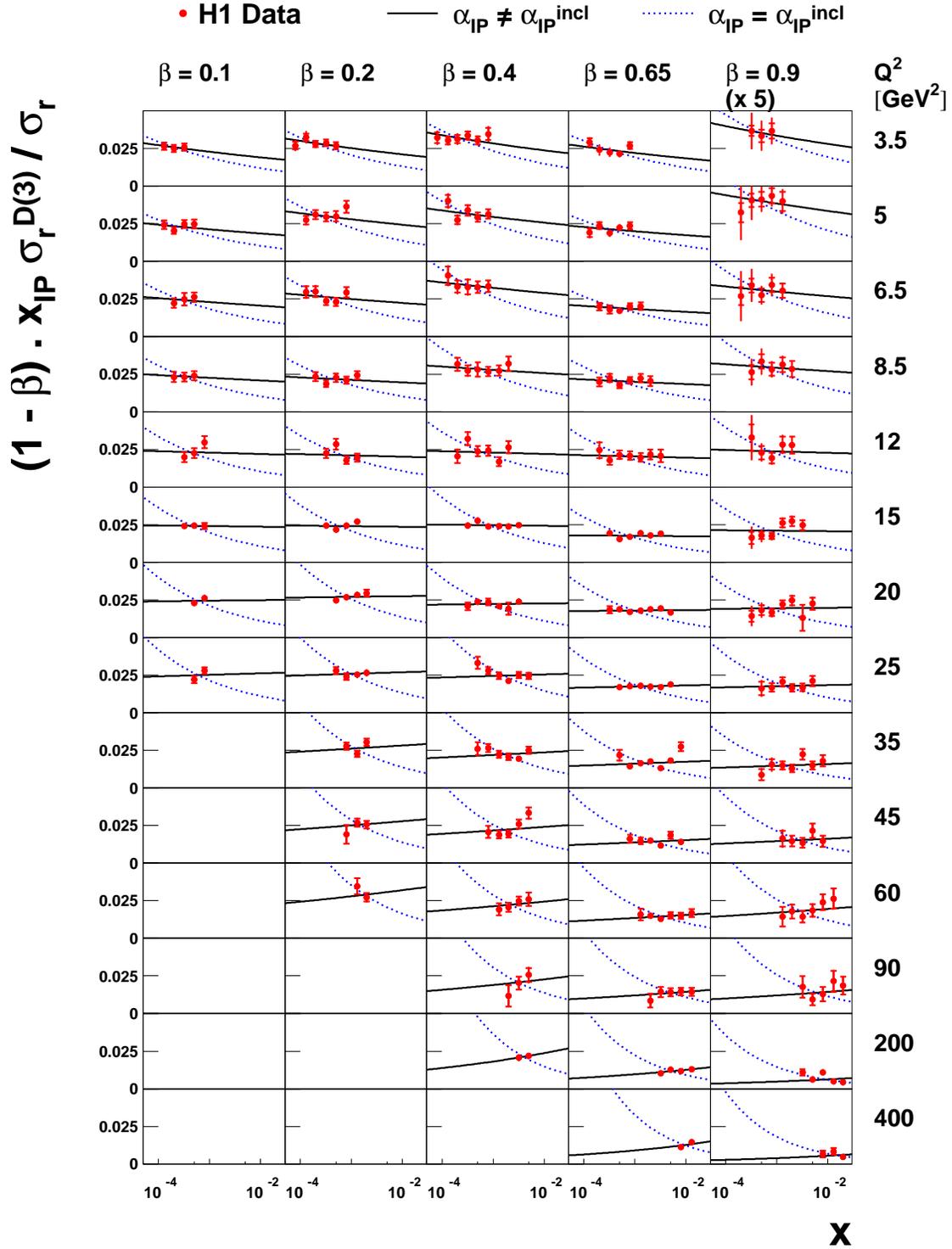,width=0.935\linewidth}
\caption{The ratio of the diffractive to the inclusive reduced cross section,
multiplied by $(1-\beta) \, \xpom$.
Data at $\beta = 0.9$ are multiplied by a 
further factor of 5 for visibility.
Data which are significantly influenced by the
sub-leading exchange or the 
longitudinal structure function according to the NLO QCD fit are excluded. 
The remaining data are compared with 
models motivated by Regge phenomenology as described in the text, for which
the diffractive and inclusive effective pomeron trajectories are the
same (`$\alphapom = \alphapom^{\rm incl}$') and for which they are different 
(`$\alphapom \neq \alphapom^{\rm incl}$').}
\label{rhod3}
\end{figure}

The ratio defined in equation~\ref{eqn:trb} is shown in figure~\ref{rhod3} as
a function of $x$ in bins of fixed $Q^2$, $\beta$ and hence $\mx$.
In order to simplify the interpretation, 
data points are excluded if, according to the QCD fit in 
section~\ref{sec:qcdfit},
the sub-leading exchange contribution is larger
than 10\% or $\sigma_r^{D(3)}$ differs by more than 10\% from 
$F_2^{D(3)}$
due to the influence of $F_L^{D(3)}$. 
Only $Q^2$ and
$\beta$ values for which there are at least two remaining 
data points are shown. 
The ratio in equation~\ref{eqn:trb}
is indeed approximately constant throughout the full kinematic 
range,
except at large $\beta$ values (the low 
$\mx$ ``non-triple-Regge'' region in which \cite{triple:regge} 
is not applicable). 
In particular, the $x$ (and hence the $W$) dependence at
fixed $Q^2$, $\beta$ and $\mx$ is strikingly flat, substantiating the 
conclusions of \cite{ZEUS:94,ZEUS:97,zeus02}. 

Expressed in terms of Regge trajectories, 
the ratio of cross sections shown in
figure~\ref{rhod3} is proportional to $x^{-\kappa}$, where 
$\kappa = 2 \langle \alphapom (t) \rangle - 
\alphapom^{\rm incl}(0) - 1$. Here
$\alphapom(t)$ is the effective pomeron trajectory for diffractive DIS and 
$\alphapom^{\rm incl} (0)$ is the effective pomeron intercept 
governing inclusive scattering. 
Analysis of inclusive DIS data has shown
that $\alphapom^{\rm incl} (0)$ is not universal, but varies with $Q^2$ 
according to
$\alphapom^{\rm incl} (0) \simeq 1 + 0.048 
\ln \left( Q^2 \, / \, [0.292 \ {\rm GeV}]^2 \right)$ \cite{f2rise}.
A prediction is overlaid in figure~\ref{rhod3},
where it is assumed that the diffractive
$\alphapom(0)$ depends on $Q^2$ in the same way
as $\alphapom^{\rm incl} (0)$ and that 
$\alphapom^\prime = 0.06 \ {\rm GeV^{-2}}$ 
(see section~\ref{fitproc}). Similar results are obtained with 
$\alphapom^\prime = 0$ or $\alphapom^\prime = 0.25$. 
The normalisation of the prediction is obtained
from separate fits to the data for each pair of $\beta$ and $Q^2$ values. 
The description of the data is poor ($\chi^2 = 876$ based on statistical
and uncorrelated systematic errors for $223$ degrees of freedom).
The ratio shown in figure~\ref{rhod3} is 
also compared with a prediction where the
inclusive data are described by the same $Q^2$ dependent 
$\alphapom^{\rm incl} (0)$ of \cite{f2rise} and the 
diffractive $x$ dependence is determined by 
the flux factor defined in equation~\ref{eq:fluxfac}, such that 
$\alphapom(0) = 1.118$ (equation~\ref{alpom:answer})
independently of $Q^2$. 
A much improved description is 
obtained ($\chi^2 = 254$), with an $x$ dependence
of the ratio 
which changes slowly with $Q^2$, being approximately flat in the region
of $Q^2 \sim 15 \ {\rm GeV^2}$. 

The ratio of cross sections shown in figure~\ref{rhod3} is incompatible
with a simple Regge approach to $\gamma^* p$ scattering,
where both the diffractive and the inclusive cross sections are driven 
by the exchange of the same
pomeron trajectory, even if that trajectory is allowed
to change with $Q^2$. There is thus no simple relationship within a
Regge model between the diffractive and inclusive cross sections.
The flatness of the ratio of cross sections
is natural if rapidity gap 
formation is a random process \cite{sci,semicl} and has 
also been
interpreted in colour dipole models \cite{kgb}.


\section{Summary}
\label{sec:summary}

The reduced semi-inclusive diffractive cross section 
$\sigrd (\xpom, \beta, Q^2)$ is measured for the process
$ep \rightarrow eXY$ under the 
conditions $\my < 1.6 \ {\rm GeV}$ and $|t| < 1 \ {\rm GeV^2}$ for various 
fixed values of
$\xpom$ in the range $0.0003 < \xpom < 0.03$. 
The data span nearly three orders of magnitude
in $Q^2$ from $3.5 \ {\rm GeV^2}$ to $1600 \ {\rm GeV^2}$ 
and cover the range $0.0017 \leq \beta \leq 0.8$. In the best measured
region,
the statistical and point-to-point systematic uncertainties are at the
level of $5\%$ each, with an additional normalisation uncertainty of $6\%$.
The kinematic dependences of the ratio of the diffractive to the inclusive 
reduced cross section are also studied.

Up to small deviations at large $\xpom$ and low
$\beta$, which are consistent with expectations from 
the presence of a sub-leading exchange, 
the $\beta$ and $Q^2$ dependences of the 
diffractive data change only in normalisation
at different $\xpom$ values.
This remarkable feature is compatible with a factorisable
proton vertex. The variation of $\sigrd$ with $\xpom$ can be expressed in 
terms of an effective pomeron trajectory with intercept 
$\alphapom(0) = 1.118 \ \pm 0.008 \ \mathrm{(exp.)}  \ 
^{+ 0.029}_{- 0.010} \ \mathrm{(model)}$ if 
$\alphapom^\prime = 0.06 \ ^{+0.19}_{-0.06} \ {\rm GeV^{-2}}$ is taken from
H1 Forward Proton Spectrometer data.
The $x$ (or $\xpom$) dependence of the 
diffractive cross
section at fixed $\beta$ and $Q^2$ is similar to
that of the inclusive cross section. The diffractive and inclusive $x$ 
dependences cannot be interpreted with a single $Q^2$ dependent effective
pomeron trajectory. 

The $\beta$ and $Q^2$ dependences 
of $\sigrd$ are interpreted in terms of 
diffractive parton distribution functions (DPDFs),
obtained through an NLO DGLAP QCD fit.
The DPDFs correspond to integrals over the measured $\my$ 
and $t$ ranges and are valid
in the region $Q^2 \geq 8.5 \ {\rm GeV^2}$, $\beta \leq 0.8$ and
$\mx > 2 \ {\rm GeV}$.
At low $\beta \ \lapprox \ 0.3$, the diffractive data exhibit a rather
fast rise with increasing $Q^2$ at fixed
$\xpom$ and $x$. This rise of the diffractive cross section  
is very similar to that of the inclusive cross
section at the same $x$ values, implying that the ratio of quarks to gluons is
similar in the diffractive and inclusive cases. 
The low $\beta$ data give good constraints on 
the diffractive quark singlet and gluon
densities at low momentum fractions $z$, with combined experimental and 
theoretical uncertainties of typically $5-10\%$ and $15\%$, respectively.
The gluon density dominates the DPDFs in this region.
At larger 
$z$, the diffractive quark density remains 
well constrained by the fit, whereas the sensitivity to the gluon
density becomes increasingly poor.

At high $Q^2$, charged current scattering 
is used to test the assumptions on the
quark flavour decomposition of the DPDFs.
Total and single differential 
diffractive $e^+ p$ charged current 
\linebreak
cross sections
are measured and are
well described by predictions based on the DPDFs 
extracted from the neutral current data, though the current level of 
experimental precision
($35 \%$ for the total cross section) is low.
The DPDFs will provide 
important input to future tests of the factorisation properties
of diffraction and to the prediction of cross sections for
diffractive processes at HERA, the LHC and elsewhere. 


\section*{Acknowledgements}
We are grateful to the HERA machine group whose outstanding efforts
have made this experiment possible. We thank the engineers and
technicians for their work in constructing and now maintaining the H1
detector, our funding agencies for financial support, the DESY
technical staff for continual assistance and the DESY directorate for
support and for the hospitality which they extend to the non DESY
members of the collaboration. 


\pagebreak


\begin{landscape} 

\begin{table} 
\begin{footnotesize} 
\begin{center}
\begin{tabular}{|c | c | c| c | c | c | c | c | c | c | c | c | c | c | c | c | c | c |} 
\hline 
\multicolumn{1}{|c|}{$\xpom$} & \multicolumn{1}{c|}{$Q^2$} & \multicolumn{1}{c|}{$\beta$} & \multicolumn{1}{c|}{$\xpom \sigma_r^{D(3)}$} & \multicolumn{1}{c|}{$\delta_{stat}$} & \multicolumn{1}{c|}{$\delta_{sys}$} & \multicolumn{1}{c|}{$\delta_{tot}$} & \multicolumn{1}{c|}{$\delta_{unc}$} & \multicolumn{1}{c|}{$\delta_{lar}$} & \multicolumn{1}{c|}{$\delta_{ele}$} & \multicolumn{1}{c|}{$\delta_{\theta}$} & \multicolumn{1}{c|}{$\delta_{noise}$} & \multicolumn{1}{c|}{$\delta_{\xpom}$}  & \multicolumn{1}{c|}{$\delta_{\beta}$}  & \multicolumn{1}{c|}{$\delta_{bg}$}  & \multicolumn{1}{c|}{$\delta_{Plug}$} & \multicolumn{1}{c|}{$\delta_{Q^2}$} & \multicolumn{1}{c|}{$\delta_{spa}$} \\ 
 & \multicolumn{1}{c|}{$[\rm{GeV^2}]$} & & & \multicolumn{1}{c|}{[\%]} & \multicolumn{1}{c|}{[\%]} & \multicolumn{1}{c|}{[\%]} & \multicolumn{1}{c|}{[\%]} & \multicolumn{1}{c|}{[\%]} & \multicolumn{1}{c|}{[\%]} & \multicolumn{1}{c|}{[\%]} & \multicolumn{1}{c|}{[\%]} & \multicolumn{1}{c|}{[\%]} & \multicolumn{1}{c|}{[\%]} & \multicolumn{1}{c|}{[\%]} & \multicolumn{1}{c|}{[\%]} & \multicolumn{1}{c|}{[\%]} & \multicolumn{1}{c|}{[\%]}  \\\hline\hline
0.0003 & 3.5 & 0.17 & 0.0224 & 17.2 & 10.8 & 20.3 & 1.6 & 0.2 & 5.3 & -1.7 & -4.7 & -0.6 & -0.8 & 0.0 & -0.0 & 3.0 & 1.7 \\ 
0.0003 & 3.5 & 0.27 & 0.0262 & 7.6 & 6.4 & 9.9 & 1.6 & 0.2 & 1.3 & 2.7 & -4.6 & -0.4 & -0.5 & -0.1 & -0.1 & 1.0 & 1.8 \\ 
0.0003 & 3.5 & 0.43 & 0.0351 & 5.5 & 6.2 & 8.3 & 1.6 & 0.2 & 0.2 & 1.6 & 4.6 & -1.6 & -0.2 & 0.0 & -0.1 & 1.2 & 0.7 \\ 
0.0003 & 3.5 & 0.67 & 0.0443 & 5.5 & 11.7 & 13.0 & 1.6 & 0.3 & -1.0 & 1.9 & 11.2 & -0.3 & 0.7 & -0.1 & 0.1 & 0.9 & 0.8 \\ 
0.0003 & 5.0 & 0.27 & 0.0392 & 11.9 & 7.7 & 14.2 & 1.6 & 0.4 & 3.8 & 1.4 & 4.7 & -0.3 & -0.8 & 0.0 & 0.1 & 1.0 & 2.4 \\ 
0.0003 & 5.0 & 0.43 & 0.0422 & 7.7 & 7.0 & 10.4 & 1.6 & 0.4 & -0.8 & 1.5 & 4.9 & -1.7 & -0.1 & 0.0 & -0.2 & 1.1 & 3.0 \\ 
0.0003 & 5.0 & 0.67 & 0.0528 & 7.9 & 9.8 & 12.6 & 1.6 & -0.2 & -0.6 & 2.6 & 9.0 & -0.2 & 0.7 & 0.0 & -0.0 & 1.0 & 1.4 \\ 
0.0003 & 6.5 & 0.43 & 0.0452 & 11.0 & 7.8 & 13.5 & 1.6 & -0.8 & -2.7 & 1.9 & 5.3 & -1.2 & 0.1 & 0.0 & -0.4 & 0.9 & 3.4 \\ 
0.0003 & 6.5 & 0.67 & 0.0580 & 9.5 & 8.9 & 13.0 & 1.6 & 0.1 & 1.2 & 0.9 & 7.6 & -0.8 & 0.9 & 0.0 & 0.3 & 0.7 & 2.6 \\ 
0.0003 & 8.5 & 0.43 & 0.0353 & 18.2 & 11.0 & 21.3 & 1.6 & -0.5 & 6.5 & 1.3 & 5.5 & -0.7 & 0.0 & 0.0 & -0.5 & 0.3 & 5.8 \\ 
0.0003 & 8.5 & 0.67 & 0.0570 & 11.7 & 9.5 & 15.0 & 1.6 & 0.2 & -0.5 & 0.6 & 6.9 & -1.6 & 0.8 & 0.0 & -0.2 & 0.4 & 4.6 \\ 
0.0003 & 12.0 & 0.67 & 0.0670 & 18.1 & 10.0 & 20.7 & 1.6 & 0.2 & 0.3 & -1.4 & 8.5 & -0.8 & 1.2 & 0.0 & 0.0 & 0.6 & 3.6 \\ 
0.001 & 3.5 & 0.05 & 0.0167 & 15.5 & 6.5 & 16.8 & 1.7 & -0.7 & 1.0 & -1.6 & -2.6 & -1.0 & -2.2 & 0.0 & 0.4 & -0.1 & 0.5 \\ 
0.001 & 3.5 & 0.08 & 0.0189 & 8.2 & 6.9 & 10.7 & 1.7 & 0.5 & 2.5 & 1.8 & -5.3 & -0.9 & -1.3 & 0.0 & -0.3 & 0.8 & -0.8 \\ 
0.001 & 3.5 & 0.13 & 0.0239 & 6.6 & 6.9 & 9.5 & 1.7 & 0.7 & 0.3 & 2.0 & -5.8 & -1.1 & -1.2 & 0.0 & 0.4 & 0.6 & -0.8 \\ 
0.001 & 3.5 & 0.2 & 0.0239 & 6.2 & 6.9 & 9.3 & 1.7 & 0.5 & 0.7 & 0.7 & -6.2 & -1.2 & -1.1 & 0.0 & 0.2 & 0.5 & -0.9 \\ 
0.001 & 3.5 & 0.32 & 0.0243 & 5.6 & 6.8 & 8.8 & 1.7 & -0.1 & -2.1 & 1.7 & -5.7 & -0.9 & -0.9 & 0.0 & -0.2 & 0.2 & 0.3 \\ 
0.001 & 3.5 & 0.5 & 0.0281 & 5.5 & 4.5 & 7.1 & 1.7 & 0.2 & -1.2 & 2.1 & 2.1 & -0.2 & -0.7 & 0.0 & -0.2 & 0.1 & 0.6 \\ 
0.001 & 3.5 & 0.8 & 0.0456 & 7.4 & 4.9 & 8.9 & 1.7 & 0.3 & -1.4 & 1.3 & 3.0 & 1.5 & 0.8 & -0.2 & 0.1 & -0.8 & 0.7 \\ 
0.001 & 5.0 & 0.08 & 0.0213 & 13.2 & 6.3 & 14.6 & 1.7 & 0.3 & 4.0 & -1.4 & -2.1 & -1.1 & -1.8 & 0.0 & 0.2 & 0.8 & 1.8 \\ 
0.001 & 5.0 & 0.13 & 0.0238 & 8.6 & 7.0 & 11.0 & 1.7 & 1.0 & 1.1 & 1.4 & -5.9 & -0.7 & -1.3 & 0.0 & 0.5 & 0.8 & 1.5 \\ 
0.001 & 5.0 & 0.2 & 0.0277 & 7.4 & 6.9 & 10.1 & 1.7 & -0.3 & -1.3 & 2.7 & -5.4 & -1.0 & -1.1 & 0.0 & -0.2 & 0.8 & -1.0 \\ 
0.001 & 5.0 & 0.32 & 0.0317 & 6.5 & 6.5 & 9.2 & 1.7 & 0.4 & -1.2 & 1.7 & -5.5 & -0.8 & -0.9 & 0.0 & 0.1 & 0.6 & 0.7 \\ 
0.001 & 5.0 & 0.5 & 0.0350 & 6.0 & 3.9 & 7.2 & 1.7 & 0.6 & -1.1 & 1.5 & 1.6 & -0.4 & -0.5 & 0.0 & 0.0 & 0.7 & 0.5 \\ 
0.001 & 5.0 & 0.8 & 0.0467 & 7.5 & 6.1 & 9.6 & 1.7 & 0.2 & -2.3 & 2.8 & 3.7 & 1.2 & 1.0 & 0.0 & 0.3 & 0.5 & 0.7 \\ 
0.001 & 6.5 & 0.13 & 0.0264 & 11.7 & 6.5 & 13.4 & 1.7 & -0.7 & 1.8 & 0.5 & -5.0 & -0.7 & -1.1 & 0.0 & -0.1 & 0.7 & 2.0 \\ 
\hline 
\end{tabular}
\end{center} 
\end{footnotesize} 
\caption{Results for $\xpom \sigma_r^{D(3)}$ at fixed
$Q^2$, $\beta$ and $\xpom$ (columns 1-4) using data with SPACAL electrons
and $E_p = 820 \ {\rm GeV}$. Columns 5-7 contain the
percentage statistical, systematic and total uncertainties.
The remaining columns contain the contributions to
the systematic
uncertainty from sources which are uncorrelated between data points
($\delta_{unc}$) and the 10 correlated sources leading to the largest
uncertainties. 
These are the LAr hadronic energy scale ($\delta_{lar}$),
the SPACAL electromagnetic energy scale ($\delta_{ele}$), the scattered
electron angle
measurement ($\delta_{\theta}$), the calorimeter noise
treatment ($\delta_{noise}$),
reweighting the simulation in $\xpom$ ($\delta_{\xpom}$) and $\beta$
($\delta_{\beta}$), the background subtraction using DJANGO ($\delta_{bg}$),
the plug energy scale ($\delta_{Plug}$), reweighting the simulation in $Q^2$
($\delta_{Q^2}$) and the SPACAL hadronic energy scale ($\delta_{spa}$).
Minus signs appear for these
systematics if the shift in a variable is
anti-correlated rather than correlated with the shift in the cross section.
The table continues on the next 7 pages.}
\label{spacal:data}
\end{table} 
\begin{table} 
\begin{footnotesize}
\begin{center} 
\begin{tabular}{|c | c | c | c | c | c | c | c | c | c | c | c | c | c | c | c | c | c |} 
\hline 
\multicolumn{1}{|c|}{$\xpom$} & \multicolumn{1}{c|}{$Q^2$} & \multicolumn{1}{c|}{$\beta$} & \multicolumn{1}{c|}{$\xpom \sigma_r^{D(3)}$} & \multicolumn{1}{c|}{$\delta_{stat}$} & \multicolumn{1}{c|}{$\delta_{sys}$} & \multicolumn{1}{c|}{$\delta_{tot}$} & \multicolumn{1}{c|}{$\delta_{unc}$} & \multicolumn{1}{c|}{$\delta_{lar}$} & \multicolumn{1}{c|}{$\delta_{ele}$} & \multicolumn{1}{c|}{$\delta_{\theta}$} & \multicolumn{1}{c|}{$\delta_{noise}$} & \multicolumn{1}{c|}{$\delta_{\xpom}$}  & \multicolumn{1}{c|}{$\delta_{\beta}$}  & \multicolumn{1}{c|}{$\delta_{bg}$}  & \multicolumn{1}{c|}{$\delta_{Plug}$} & \multicolumn{1}{c|}{$\delta_{Q^2}$} & \multicolumn{1}{c|}{$\delta_{spa}$} \\ 
 & \multicolumn{1}{c|}{$[\rm{GeV^2}]$} & & & \multicolumn{1}{c|}{[\%]} & \multicolumn{1}{c|}{[\%]} & \multicolumn{1}{c|}{[\%]} & \multicolumn{1}{c|}{[\%]} & \multicolumn{1}{c|}{[\%]} & \multicolumn{1}{c|}{[\%]} & \multicolumn{1}{c|}{[\%]} & \multicolumn{1}{c|}{[\%]} & \multicolumn{1}{c|}{[\%]} & \multicolumn{1}{c|}{[\%]} & \multicolumn{1}{c|}{[\%]} & \multicolumn{1}{c|}{[\%]} & \multicolumn{1}{c|}{[\%]} & \multicolumn{1}{c|}{[\%]}  \\\hline\hline
0.001 & 6.5 & 0.2 & 0.0294 & 9.2 & 7.2 & 11.7 & 1.7 & 0.2 & -1.7 & 1.5 & -6.1 & -1.1 & -1.3 & 0.0 & 0.1 & 0.4 & 1.1 \\ 
0.001 & 6.5 & 0.32 & 0.0295 & 10.6 & 6.3 & 12.4 & 1.7 & 0.3 & 0.8 & 1.6 & -5.4 & -0.9 & -1.0 & 0.0 & 0.3 & 0.3 & 0.9 \\ 
0.001 & 6.5 & 0.5 & 0.0321 & 7.5 & 4.6 & 8.8 & 1.7 & 0.5 & -1.1 & 1.2 & 2.9 & -0.7 & -0.5 & 0.0 & 0.2 & 0.3 & 0.7 \\ 
0.001 & 6.5 & 0.8 & 0.0461 & 8.7 & 5.3 & 10.2 & 1.7 & 0.8 & -1.7 & 1.4 & 3.8 & 0.6 & 0.9 & 0.0 & -0.2 & 0.3 & 0.9 \\ 
0.001 & 8.5 & 0.13 & 0.0232 & 18.6 & 8.6 & 20.5 & 1.7 & 0.9 & -3.0 & 2.6 & -6.5 & -0.8 & -1.6 & 0.0 & -0.2 & 0.6 & 2.4 \\ 
0.001 & 8.5 & 0.2 & 0.0298 & 9.6 & 5.2 & 10.9 & 1.7 & -0.6 & 1.2 & 1.0 & -4.0 & -0.6 & -1.2 & 0.0 & -0.2 & 0.2 & -1.3 \\ 
0.001 & 8.5 & 0.32 & 0.0341 & 8.4 & 5.6 & 10.1 & 1.7 & 0.5 & -1.7 & 2.4 & -4.0 & -0.7 & -0.9 & 0.0 & 0.3 & 0.2 & -0.6 \\ 
0.001 & 8.5 & 0.5 & 0.0372 & 8.8 & 5.1 & 10.2 & 1.7 & 0.3 & -2.8 & 1.9 & -2.4 & -0.5 & -0.5 & 0.0 & 0.1 & 0.2 & 0.8 \\ 
0.001 & 8.5 & 0.8 & 0.0457 & 9.1 & 4.8 & 10.2 & 1.7 & -0.3 & -1.3 & 2.6 & 2.4 & 0.9 & 1.0 & 0.0 & -0.0 & 0.1 & 0.6 \\ 
0.001 & 12.0 & 0.2 & 0.0349 & 13.0 & 6.0 & 14.3 & 1.7 & 1.4 & -1.7 & 1.1 & -2.9 & -0.5 & -1.3 & 0.0 & 0.6 & 0.7 & 3.8 \\ 
0.001 & 12.0 & 0.32 & 0.0385 & 9.6 & 5.6 & 11.1 & 1.7 & 0.6 & -1.7 & 1.2 & -3.8 & -0.8 & -1.0 & 0.0 & -0.2 & 0.6 & -0.3 \\ 
0.001 & 12.0 & 0.5 & 0.0426 & 9.0 & 5.4 & 10.5 & 1.7 & 0.5 & -1.0 & 2.9 & 3.3 & -0.3 & -0.4 & 0.0 & 0.2 & 0.5 & 1.2 \\ 
0.001 & 12.0 & 0.8 & 0.0445 & 11.0 & 5.6 & 12.4 & 1.7 & 0.3 & -2.0 & -0.7 & 3.7 & 1.1 & 0.9 & 0.0 & 0.2 & 0.4 & 1.2 \\ 
0.001 & 15.0 & 0.2 & 0.0395 & 11.6 & 6.3 & 13.2 & 1.7 & 0.3 & 4.0 & -0.9 & -3.3 & -0.3 & -1.9 & 0.0 & 0.2 & 0.1 & 1.9 \\ 
0.001 & 15.0 & 0.32 & 0.0384 & 5.3 & 4.5 & 7.0 & 1.7 & 0.4 & 0.7 & 0.8 & -3.4 & -0.6 & -1.0 & 0.0 & -0.1 & 0.1 & 1.3 \\ 
0.001 & 15.0 & 0.5 & 0.0410 & 4.9 & 5.3 & 7.2 & 1.7 & 0.5 & -3.1 & 1.7 & 2.9 & -0.3 & -0.5 & 0.0 & 0.0 & 0.0 & 0.5 \\ 
0.001 & 15.0 & 0.8 & 0.0409 & 6.4 & 4.7 & 7.9 & 1.7 & 0.1 & -2.1 & 1.3 & 2.6 & 0.8 & 1.0 & 0.0 & -0.1 & -0.2 & 0.7 \\ 
0.001 & 20.0 & 0.32 & 0.0379 & 8.0 & 4.9 & 9.4 & 1.7 & 0.4 & 1.1 & 0.5 & -2.8 & -0.4 & -1.1 & -0.4 & -0.1 & 0.4 & 2.7 \\ 
0.001 & 20.0 & 0.5 & 0.0488 & 5.5 & 4.4 & 7.0 & 1.7 & 0.5 & -2.3 & 0.4 & 2.3 & -0.3 & -0.3 & 0.0 & 0.1 & 0.3 & 1.0 \\ 
0.001 & 20.0 & 0.8 & 0.0431 & 6.9 & 5.8 & 9.0 & 1.7 & 0.3 & -2.9 & 1.5 & 3.3 & 0.9 & 1.0 & 0.0 & 0.1 & 0.2 & 1.1 \\ 
0.001 & 25.0 & 0.32 & 0.0505 & 23.9 & 6.9 & 24.8 & 1.7 & 0.5 & 4.6 & 1.9 & -1.5 & -0.6 & -1.4 & 0.0 & -0.6 & 0.1 & 3.3 \\ 
0.001 & 25.0 & 0.5 & 0.0562 & 8.4 & 4.6 & 9.6 & 1.7 & 0.5 & -1.8 & 1.0 & 0.9 & 0.1 & -0.4 & 0.0 & 0.2 & 0.1 & 2.8 \\ 
0.001 & 25.0 & 0.8 & 0.0460 & 7.8 & 5.9 & 9.8 & 1.7 & 0.8 & -2.3 & 1.1 & 3.5 & 0.3 & 1.1 & -0.3 & -0.2 & -0.2 & 1.2 \\ 
0.001 & 35.0 & 0.5 & 0.0534 & 12.4 & 7.1 & 14.3 & 1.7 & 0.6 & 2.2 & 1.0 & 2.4 & -0.5 & -0.5 & 0.0 & 0.7 & 0.5 & 5.3 \\ 
0.001 & 35.0 & 0.8 & 0.0416 & 10.4 & 7.0 & 12.6 & 1.7 & 0.3 & -4.0 & 1.0 & 2.9 & 0.8 & 1.0 & 0.0 & 0.0 & 0.2 & 2.1 \\ 
0.001 & 45.0 & 0.8 & 0.0297 & 22.1 & 8.8 & 23.8 & 1.7 & 0.5 & 2.1 & 1.8 & 3.9 & 1.3 & 0.8 & 0.0 & 0.0 & 0.3 & 4.5 \\ 
0.003 & 3.5 & 0.017 & 0.0134 & 17.2 & 6.8 & 18.5 & 2.1 & 0.5 & 2.9 & 0.7 & 0.4 & -0.4 & -1.3 & 0.0 & -0.3 & 0.5 & 0.8 \\ 
0.003 & 3.5 & 0.027 & 0.0194 & 8.4 & 4.6 & 9.5 & 2.1 & 0.7 & 1.4 & 1.5 & -2.2 & -0.6 & -1.2 & -0.2 & -0.1 & 0.8 & 0.6 \\ 
0.003 & 3.5 & 0.043 & 0.0172 & 8.1 & 4.0 & 9.0 & 2.1 & 0.1 & 0.6 & 2.8 & -0.9 & -0.2 & -0.5 & -0.2 & 0.1 & 0.6 & 0.9 \\ 
0.003 & 3.5 & 0.067 & 0.0172 & 7.4 & 3.7 & 8.3 & 2.1 & 0.3 & 0.7 & 1.2 & -1.1 & -0.9 & -1.0 & -0.2 & -0.3 & 0.5 & -1.0 \\ 
0.003 & 3.5 & 0.11 & 0.0181 & 6.9 & 3.6 & 7.8 & 2.1 & 1.0 & -0.2 & 1.1 & -1.7 & -0.8 & -0.9 & -0.2 & 0.2 & 0.2 & -0.3 \\ 
0.003 & 3.5 & 0.17 & 0.0166 & 7.1 & 4.3 & 8.3 & 2.1 & 0.7 & -0.8 & 1.2 & -2.2 & -1.0 & -1.0 & -0.3 & 0.2 & 0.3 & 0.5 \\ 
0.003 & 3.5 & 0.27 & 0.0218 & 8.0 & 6.1 & 10.1 & 2.1 & 0.7 & -1.0 & 0.8 & -1.5 & -1.1 & -1.2 & -0.3 & 0.1 & -3.3 & -0.8 \\ 
0.003 & 5.0 & 0.027 & 0.0173 & 13.9 & 7.5 & 15.8 & 2.1 & -0.3 & 5.3 & 1.2 & -2.7 & -0.5 & -1.3 & -0.3 & 0.2 & 0.5 & -1.4 \\ 
0.003 & 5.0 & 0.043 & 0.0207 & 9.6 & 4.9 & 10.7 & 2.1 & 1.3 & 2.8 & 0.8 & -1.8 & -0.6 & -1.1 & -0.3 & 0.4 & 0.9 & 1.1 \\ 
0.003 & 5.0 & 0.067 & 0.0175 & 8.4 & 4.2 & 9.4 & 2.1 & 0.7 & 0.9 & 1.7 & -1.9 & -0.6 & -1.1 & -0.4 & -0.2 & 0.7 & 0.9 \\ 
\hline 
\end{tabular} 
\end{center}
\end{footnotesize} 
\end{table} 
\begin{table} 
\begin{footnotesize}
\begin{center}  
\begin{tabular}{|c | c | c | c | c | c | c | c | c | c | c | c | c | c | c | c | c | c |} 
\hline 
\multicolumn{1}{|c|}{$\xpom$} & \multicolumn{1}{c|}{$Q^2$} & \multicolumn{1}{c|}{$\beta$} & \multicolumn{1}{c|}{$\xpom \sigma_r^{D(3)}$} & \multicolumn{1}{c|}{$\delta_{stat}$} & \multicolumn{1}{c|}{$\delta_{sys}$} & \multicolumn{1}{c|}{$\delta_{tot}$} & \multicolumn{1}{c|}{$\delta_{unc}$} & \multicolumn{1}{c|}{$\delta_{lar}$} & \multicolumn{1}{c|}{$\delta_{ele}$} & \multicolumn{1}{c|}{$\delta_{\theta}$} & \multicolumn{1}{c|}{$\delta_{noise}$} & \multicolumn{1}{c|}{$\delta_{\xpom}$}  & \multicolumn{1}{c|}{$\delta_{\beta}$}  & \multicolumn{1}{c|}{$\delta_{bg}$}  & \multicolumn{1}{c|}{$\delta_{Plug}$} & \multicolumn{1}{c|}{$\delta_{Q^2}$} & \multicolumn{1}{c|}{$\delta_{spa}$} \\ 
 & \multicolumn{1}{c|}{$[\rm{GeV^2}]$} & & & \multicolumn{1}{c|}{[\%]} & \multicolumn{1}{c|}{[\%]} & \multicolumn{1}{c|}{[\%]} & \multicolumn{1}{c|}{[\%]} & \multicolumn{1}{c|}{[\%]} & \multicolumn{1}{c|}{[\%]} & \multicolumn{1}{c|}{[\%]} & \multicolumn{1}{c|}{[\%]} & \multicolumn{1}{c|}{[\%]} & \multicolumn{1}{c|}{[\%]} & \multicolumn{1}{c|}{[\%]} & \multicolumn{1}{c|}{[\%]} & \multicolumn{1}{c|}{[\%]} & \multicolumn{1}{c|}{[\%]}  \\\hline\hline
0.003 & 5.0 & 0.11 & 0.0193 & 8.4 & 5.2 & 9.9 & 2.1 & -0.3 & -1.5 & 2.0 & -3.2 & -0.9 & -0.9 & -0.1 & -0.6 & 0.6 & -0.3 \\ 
0.003 & 5.0 & 0.17 & 0.0209 & 9.0 & 4.7 & 10.2 & 2.1 & -0.4 & -0.7 & 2.0 & -2.8 & -1.0 & -1.0 & -0.7 & -0.4 & 0.7 & -0.3 \\ 
0.003 & 5.0 & 0.27 & 0.0268 & 7.3 & 4.2 & 8.4 & 2.1 & 1.1 & -0.9 & 1.3 & -1.4 & -1.0 & -0.9 & -0.2 & 0.2 & 0.2 & 0.1 \\ 
0.003 & 5.0 & 0.43 & 0.0290 & 9.9 & 7.3 & 12.3 & 2.1 & 0.3 & 0.4 & 0.8 & -1.1 & -0.9 & -0.7 & 0.0 & 0.1 & -4.8 & -0.2 \\ 
0.003 & 6.5 & 0.043 & 0.0226 & 12.1 & 4.9 & 13.0 & 2.1 & 0.4 & 2.6 & 1.2 & -2.3 & -0.5 & -0.9 & -0.2 & 0.3 & 0.3 & -0.9 \\ 
0.003 & 6.5 & 0.067 & 0.0215 & 10.9 & 5.1 & 12.1 & 2.1 & 0.4 & -2.6 & 2.9 & -1.0 & -0.7 & -1.2 & -0.2 & -0.3 & 0.3 & -1.0 \\ 
0.003 & 6.5 & 0.11 & 0.0191 & 12.2 & 4.6 & 13.1 & 2.1 & 0.6 & 2.6 & 2.2 & -0.9 & -0.6 & -0.7 & 0.0 & 0.3 & 0.5 & -1.2 \\ 
0.003 & 6.5 & 0.17 & 0.0215 & 9.4 & 3.9 & 10.2 & 2.1 & 0.4 & -1.3 & 2.3 & -0.9 & -0.5 & -0.8 & -0.2 & 0.2 & 0.3 & 0.5 \\ 
0.003 & 6.5 & 0.27 & 0.0247 & 8.7 & 5.0 & 10.1 & 2.1 & 0.6 & -1.2 & 3.1 & -2.5 & -0.3 & -0.7 & -0.1 & -0.1 & 0.4 & 0.8 \\ 
0.003 & 6.5 & 0.43 & 0.0238 & 8.2 & 4.3 & 9.2 & 2.1 & 1.3 & -1.3 & 2.4 & -1.0 & 0.4 & -0.3 & 0.0 & 0.3 & 0.2 & 0.2 \\ 
0.003 & 8.5 & 0.043 & 0.0201 & 19.2 & 8.2 & 20.9 & 2.1 & 1.9 & 5.3 & 2.1 & 2.8 & -0.1 & -1.2 & -0.9 & 0.6 & 0.4 & 1.7 \\ 
0.003 & 8.5 & 0.067 & 0.0226 & 11.8 & 4.7 & 12.7 & 2.1 & -0.6 & 1.8 & -1.6 & -2.4 & -0.6 & -1.2 & -0.5 & -0.2 & 0.2 & 1.0 \\ 
0.003 & 8.5 & 0.11 & 0.0201 & 10.2 & 5.0 & 11.3 & 2.1 & -1.8 & -1.9 & 1.3 & -2.3 & -0.7 & -1.1 & -0.2 & -0.0 & 0.2 & -1.9 \\ 
0.003 & 8.5 & 0.17 & 0.0218 & 10.1 & 3.9 & 10.8 & 2.1 & 0.3 & -0.5 & 1.3 & -1.5 & -0.6 & -0.9 & -0.4 & -0.4 & 0.2 & 0.5 \\ 
0.003 & 8.5 & 0.27 & 0.0236 & 8.2 & 3.8 & 9.1 & 2.1 & -0.2 & -0.6 & 1.7 & -1.8 & -0.6 & -0.9 & -0.2 & 0.4 & 0.1 & -0.3 \\ 
0.003 & 8.5 & 0.43 & 0.0305 & 8.0 & 3.6 & 8.8 & 2.1 & 0.9 & -0.9 & 0.7 & -1.4 & -0.1 & -0.6 & -0.2 & -0.1 & 0.1 & -0.3 \\ 
0.003 & 8.5 & 0.67 & 0.0397 & 10.9 & 5.7 & 12.4 & 2.1 & 0.5 & -2.4 & 1.9 & 3.1 & 1.5 & 0.5 & -0.4 & 0.4 & 0.0 & 0.9 \\ 
0.003 & 12.0 & 0.067 & 0.0247 & 15.0 & 4.9 & 15.8 & 2.1 & 0.2 & -0.8 & 1.4 & -2.6 & -0.5 & -1.3 & -0.5 & -0.5 & 0.5 & 1.4 \\ 
0.003 & 12.0 & 0.11 & 0.0242 & 12.1 & 4.2 & 12.8 & 2.1 & 1.5 & 1.7 & 0.5 & 0.2 & -0.4 & -1.1 & -0.4 & 0.4 & 0.6 & 0.4 \\ 
0.003 & 12.0 & 0.17 & 0.0266 & 10.4 & 4.3 & 11.3 & 2.1 & -0.3 & -2.4 & 1.7 & -1.6 & -0.4 & -0.9 & -0.2 & 0.5 & 0.4 & 0.5 \\ 
0.003 & 12.0 & 0.27 & 0.0226 & 10.5 & 3.1 & 11.0 & 2.1 & 0.3 & -1.0 & -0.4 & -1.3 & -0.6 & -0.8 & 0.0 & -0.2 & 0.4 & 0.5 \\ 
0.003 & 12.0 & 0.43 & 0.0260 & 9.9 & 5.2 & 11.2 & 2.1 & 1.5 & -2.3 & 3.1 & -1.3 & -0.3 & -0.5 & 0.0 & 0.2 & 0.5 & 0.7 \\ 
0.003 & 12.0 & 0.67 & 0.0392 & 10.8 & 4.6 & 11.7 & 2.1 & 0.9 & 0.7 & 1.4 & 1.2 & 1.5 & 0.4 & 0.0 & -0.4 & 0.5 & 1.1 \\ 
0.003 & 15.0 & 0.067 & 0.0338 & 14.4 & 5.0 & 15.3 & 2.1 & 1.5 & 1.7 & 1.4 & 1.1 & -0.4 & -1.7 & 0.0 & 0.7 & -0.1 & 2.2 \\ 
0.003 & 15.0 & 0.11 & 0.0282 & 6.3 & 3.6 & 7.3 & 2.1 & 0.7 & 1.3 & 1.0 & -1.0 & -0.4 & -1.0 & -0.5 & -0.0 & 0.2 & 0.5 \\ 
0.003 & 15.0 & 0.17 & 0.0245 & 5.5 & 3.5 & 6.5 & 2.1 & 0.4 & -2.1 & 0.6 & -0.6 & -0.6 & -1.0 & 0.0 & 0.2 & 0.2 & 0.2 \\ 
0.003 & 15.0 & 0.27 & 0.0273 & 4.9 & 3.5 & 6.0 & 2.1 & 0.3 & -1.0 & 1.1 & -1.6 & -0.5 & -0.8 & -0.2 & 0.2 & -0.0 & 0.2 \\ 
0.003 & 15.0 & 0.43 & 0.0331 & 4.9 & 3.6 & 6.1 & 2.1 & 0.5 & -1.2 & 1.9 & 0.7 & -0.4 & -0.5 & -0.2 & -0.1 & 0.1 & 0.4 \\ 
0.003 & 15.0 & 0.67 & 0.0381 & 5.8 & 4.0 & 7.1 & 2.1 & 0.5 & -1.5 & 1.6 & 1.7 & 1.0 & 0.5 & -0.1 & 0.7 & 0.1 & 0.6 \\ 
0.003 & 20.0 & 0.11 & 0.0276 & 9.0 & 4.1 & 9.9 & 2.1 & 0.2 & 2.5 & 0.4 & -1.5 & -0.4 & -1.2 & -0.4 & 0.2 & 0.4 & 1.3 \\ 
0.003 & 20.0 & 0.17 & 0.0281 & 6.3 & 3.5 & 7.2 & 2.1 & 0.5 & -1.5 & 1.0 & -0.7 & -0.2 & -1.1 & -0.9 & 0.2 & 0.4 & -0.3 \\ 
0.003 & 20.0 & 0.27 & 0.0321 & 5.9 & 3.6 & 6.9 & 2.1 & -0.1 & -1.4 & 0.8 & -1.8 & -0.4 & -0.9 & -0.1 & -0.1 & 0.3 & -0.3 \\ 
0.003 & 20.0 & 0.43 & 0.0324 & 5.2 & 3.5 & 6.3 & 2.1 & 0.4 & -1.7 & 1.2 & -0.6 & -0.2 & -0.4 & -0.2 & -0.1 & 0.3 & 0.4 \\ 
0.003 & 20.0 & 0.67 & 0.0411 & 6.3 & 3.6 & 7.3 & 2.1 & 0.7 & -1.6 & 1.1 & 1.4 & 0.8 & 0.5 & -0.1 & 0.1 & 0.3 & 0.3 \\ 
0.003 & 25.0 & 0.11 & 0.0343 & 20.2 & 6.3 & 21.2 & 2.1 & -1.3 & -4.2 & 1.4 & -2.1 & -0.3 & -1.6 & 0.0 & -0.0 & 0.1 & 2.0 \\ 
0.003 & 25.0 & 0.17 & 0.0335 & 7.2 & 3.1 & 7.8 & 2.1 & -0.0 & 0.8 & 0.4 & 0.9 & -0.5 & -1.0 & 0.0 & 0.5 & 0.1 & 1.1 \\ 
\hline 
\end{tabular}
\end{center} 
\end{footnotesize} 
\end{table} 
\begin{table} 
\begin{footnotesize}
\begin{center}  
\begin{tabular}{|c | c | c | c | c | c | c | c | c | c | c | c | c | c | c | c | c | c |} 
\hline 
\multicolumn{1}{|c|}{$\xpom$} & \multicolumn{1}{c|}{$Q^2$} & \multicolumn{1}{c|}{$\beta$} & \multicolumn{1}{c|}{$\xpom \sigma_r^{D(3)}$} & \multicolumn{1}{c|}{$\delta_{stat}$} & \multicolumn{1}{c|}{$\delta_{sys}$} & \multicolumn{1}{c|}{$\delta_{tot}$} & \multicolumn{1}{c|}{$\delta_{unc}$} & \multicolumn{1}{c|}{$\delta_{lar}$} & \multicolumn{1}{c|}{$\delta_{ele}$} & \multicolumn{1}{c|}{$\delta_{\theta}$} & \multicolumn{1}{c|}{$\delta_{noise}$} & \multicolumn{1}{c|}{$\delta_{\xpom}$}  & \multicolumn{1}{c|}{$\delta_{\beta}$}  & \multicolumn{1}{c|}{$\delta_{bg}$}  & \multicolumn{1}{c|}{$\delta_{Plug}$} & \multicolumn{1}{c|}{$\delta_{Q^2}$} & \multicolumn{1}{c|}{$\delta_{spa}$} \\ 
 & \multicolumn{1}{c|}{$[\rm{GeV^2}]$} & & & \multicolumn{1}{c|}{[\%]} & \multicolumn{1}{c|}{[\%]} & \multicolumn{1}{c|}{[\%]} & \multicolumn{1}{c|}{[\%]} & \multicolumn{1}{c|}{[\%]} & \multicolumn{1}{c|}{[\%]} & \multicolumn{1}{c|}{[\%]} & \multicolumn{1}{c|}{[\%]} & \multicolumn{1}{c|}{[\%]} & \multicolumn{1}{c|}{[\%]} & \multicolumn{1}{c|}{[\%]} & \multicolumn{1}{c|}{[\%]} & \multicolumn{1}{c|}{[\%]} & \multicolumn{1}{c|}{[\%]}  \\\hline\hline
0.003 & 25.0 & 0.27 & 0.0338 & 6.8 & 3.7 & 7.7 & 2.1 & 0.3 & -1.9 & 1.1 & -1.3 & -0.3 & -0.9 & -0.2 & -0.3 & -0.0 & -0.4 \\ 
0.003 & 25.0 & 0.43 & 0.0354 & 5.8 & 3.7 & 6.8 & 2.1 & 0.8 & -2.0 & 0.8 & 1.1 & -0.3 & -0.4 & -0.2 & 0.4 & -0.0 & 0.3 \\ 
0.003 & 25.0 & 0.67 & 0.0349 & 6.7 & 3.9 & 7.7 & 2.1 & 0.6 & -1.8 & 1.2 & 1.4 & 0.9 & 0.5 & 0.0 & -0.2 & 0.0 & 0.4 \\ 
0.003 & 35.0 & 0.17 & 0.0423 & 13.2 & 4.8 & 14.1 & 2.1 & 0.6 & 1.1 & 0.8 & -0.9 & -0.5 & -1.5 & -0.7 & 0.2 & 0.7 & 3.1 \\ 
0.003 & 35.0 & 0.27 & 0.0387 & 6.9 & 3.0 & 7.5 & 2.1 & 0.6 & -1.1 & 0.2 & -0.2 & -0.4 & -1.0 & -0.4 & -0.3 & 0.4 & 0.7 \\ 
0.003 & 35.0 & 0.43 & 0.0353 & 6.7 & 3.8 & 7.6 & 2.1 & 0.6 & -2.3 & 1.0 & -0.6 & -0.2 & -0.5 & -0.4 & 0.3 & 0.3 & 0.1 \\ 
0.003 & 35.0 & 0.67 & 0.0371 & 7.7 & 3.7 & 8.5 & 2.1 & 0.5 & -1.9 & 0.9 & 0.9 & 0.9 & 0.5 & -0.2 & 0.3 & 0.3 & 0.8 \\ 
0.003 & 45.0 & 0.27 & 0.0326 & 19.7 & 3.9 & 20.1 & 2.1 & -0.3 & -0.8 & 0.8 & -1.3 & -0.4 & -1.2 & 0.0 & -0.0 & 0.3 & 1.8 \\ 
0.003 & 45.0 & 0.43 & 0.0365 & 9.1 & 3.9 & 9.9 & 2.1 & 0.7 & -1.6 & 1.3 & 0.4 & 0.2 & -0.4 & -0.6 & 0.1 & 0.2 & 0.9 \\ 
0.003 & 45.0 & 0.67 & 0.0383 & 9.5 & 4.7 & 10.6 & 2.1 & 0.6 & -3.4 & 0.9 & 1.2 & 0.7 & 0.5 & -0.3 & 0.2 & 0.0 & 0.7 \\ 
0.003 & 60.0 & 0.43 & 0.0406 & 14.5 & 5.6 & 15.5 & 2.1 & 1.2 & -3.8 & 1.0 & 1.3 & 0.1 & -0.7 & 0.0 & -0.6 & 0.5 & 1.6 \\ 
0.003 & 60.0 & 0.67 & 0.0429 & 10.6 & 4.0 & 11.3 & 2.1 & 0.4 & -2.2 & 0.6 & 1.4 & 0.6 & 0.5 & 0.0 & 0.1 & 0.1 & 0.8 \\ 
0.003 & 90.0 & 0.67 & 0.0366 & 40.9 & 10.0 & 42.1 & 2.1 & 1.0 & 3.8 & 4.8 & 4.2 & -0.9 & 0.7 & 0.0 & 0.0 & -0.4 & 1.3 \\ 
0.01 & 3.5 & 0.005 & 0.0237 & 15.3 & 8.8 & 17.6 & 3.6 & 1.1 & 1.7 & 3.4 & -0.6 & -0.9 & -2.4 & -1.2 & 0.2 & 0.6 & -3.2 \\ 
0.01 & 3.5 & 0.008 & 0.0233 & 8.7 & 6.4 & 10.8 & 3.6 & 0.9 & 1.8 & 0.8 & 0.5 & -0.1 & -2.0 & -2.7 & 0.5 & 0.6 & 1.8 \\ 
0.01 & 3.5 & 0.013 & 0.0213 & 8.7 & 6.5 & 10.8 & 3.6 & 0.7 & 3.3 & 0.9 & 0.2 & -0.5 & -1.6 & -2.3 & 0.5 & 0.7 & -2.1 \\ 
0.01 & 3.5 & 0.02 & 0.0173 & 7.2 & 5.3 & 9.0 & 3.6 & 1.2 & 1.8 & 0.7 & 1.2 & -0.2 & -1.2 & -1.7 & -0.6 & 0.3 & 1.4 \\ 
0.01 & 3.5 & 0.032 & 0.0172 & 8.4 & 6.0 & 10.3 & 3.6 & 2.5 & 0.7 & 2.4 & 1.0 & -0.6 & -0.9 & -1.3 & 0.2 & 0.1 & -0.5 \\ 
0.01 & 3.5 & 0.05 & 0.0153 & 8.7 & 6.2 & 10.7 & 3.6 & 0.8 & -2.2 & 1.3 & 0.4 & 0.1 & -0.8 & -2.9 & 0.9 & -0.2 & 0.1 \\ 
0.01 & 3.5 & 0.08 & 0.0167 & 9.9 & 7.2 & 12.3 & 3.6 & 2.4 & 0.5 & 2.6 & 1.6 & 0.2 & -1.3 & -2.4 & 0.3 & -0.6 & -1.1 \\ 
0.01 & 5.0 & 0.008 & 0.0183 & 13.1 & 8.5 & 15.6 & 3.6 & 2.4 & 3.8 & 2.3 & -2.3 & -0.1 & -1.7 & -1.6 & 0.9 & 2.0 & 0.3 \\ 
0.01 & 5.0 & 0.013 & 0.0234 & 9.4 & 5.5 & 10.9 & 3.6 & -1.0 & 2.6 & 1.4 & 0.6 & -0.1 & -1.1 & -1.6 & 0.4 & 1.3 & -0.3 \\ 
0.01 & 5.0 & 0.02 & 0.0226 & 8.6 & 5.5 & 10.2 & 3.6 & 1.7 & -1.3 & 1.9 & -0.4 & -0.1 & -1.2 & -2.1 & 0.4 & 0.3 & 0.7 \\ 
0.01 & 5.0 & 0.032 & 0.0208 & 9.9 & 5.6 & 11.4 & 3.6 & 1.4 & 1.2 & 1.2 & 1.9 & -0.1 & -0.8 & -1.9 & 1.3 & 0.5 & -1.6 \\ 
0.01 & 5.0 & 0.05 & 0.0235 & 9.5 & 6.3 & 11.4 & 3.6 & -0.8 & -2.7 & 2.2 & -0.7 & -0.3 & -0.9 & -2.4 & 0.4 & 0.2 & 0.5 \\ 
0.01 & 5.0 & 0.08 & 0.0245 & 12.5 & 7.8 & 14.7 & 3.6 & 0.8 & -1.7 & 2.1 & -1.2 & -1.3 & -1.1 & -2.2 & 0.9 & -1.1 & -0.5 \\ 
0.01 & 6.5 & 0.013 & 0.0209 & 11.7 & 5.5 & 12.9 & 3.6 & 1.7 & 2.1 & -0.6 & 0.5 & -0.3 & -1.5 & -1.8 & 0.5 & 0.3 & 0.3 \\ 
0.01 & 6.5 & 0.02 & 0.0232 & 9.5 & 5.1 & 10.8 & 3.6 & 0.9 & -1.0 & 0.8 & 1.2 & -0.1 & -1.1 & -1.5 & 0.5 & 0.3 & -0.2 \\ 
0.01 & 6.5 & 0.032 & 0.0253 & 9.1 & 5.8 & 10.8 & 3.6 & 1.1 & -2.2 & 1.3 & -2.2 & -0.2 & -1.0 & -1.5 & 0.8 & 0.3 & -1.1 \\ 
0.01 & 6.5 & 0.05 & 0.0272 & 12.5 & 6.7 & 14.2 & 3.6 & 0.3 & 0.5 & 0.5 & -1.0 & -1.0 & -1.5 & -1.7 & 0.9 & -1.2 & 0.6 \\ 
0.01 & 6.5 & 0.08 & 0.0225 & 8.6 & 6.1 & 10.6 & 3.6 & -1.7 & -1.0 & 2.5 & 1.1 & 0.2 & -0.8 & -2.7 & -0.1 & 0.3 & -0.2 \\ 
0.01 & 6.5 & 0.13 & 0.0203 & 11.2 & 6.1 & 12.8 & 3.6 & 1.1 & 1.6 & -2.0 & 1.0 & 0.3 & -1.0 & -1.8 & 0.6 & 0.1 & 1.2 \\ 
0.01 & 8.5 & 0.013 & 0.0293 & 16.6 & 7.8 & 18.3 & 3.6 & 1.0 & 4.4 & -2.7 & -1.1 & -0.4 & -1.6 & -1.0 & 1.8 & 0.4 & -1.1 \\ 
0.01 & 8.5 & 0.02 & 0.0244 & 10.2 & 5.9 & 11.8 & 3.6 & -0.9 & 0.4 & 2.8 & -1.3 & -0.4 & -1.6 & -1.7 & 0.5 & 0.1 & 0.9 \\ 
0.01 & 8.5 & 0.032 & 0.0300 & 8.7 & 5.3 & 10.2 & 3.6 & 0.7 & 1.3 & 1.3 & 1.7 & 0.2 & -1.1 & -2.3 & 0.8 & 0.0 & 0.5 \\ 
0.01 & 8.5 & 0.05 & 0.0227 & 9.0 & 5.3 & 10.5 & 3.6 & 0.8 & -1.4 & 2.4 & 1.4 & -0.2 & -0.7 & -0.9 & 0.6 & 0.2 & 0.8 \\ 
\hline 
\end{tabular}
\end{center} 
\end{footnotesize} 
\end{table} 
\begin{table} 
\begin{footnotesize}
\begin{center}  
\begin{tabular}{|c | c | c | c | c | c | c | c | c | c | c | c | c | c | c | c | c | c |} 
\hline 
\multicolumn{1}{|c|}{$\xpom$} & \multicolumn{1}{c|}{$Q^2$} & \multicolumn{1}{c|}{$\beta$} & \multicolumn{1}{c|}{$\xpom \sigma_r^{D(3)}$} & \multicolumn{1}{c|}{$\delta_{stat}$} & \multicolumn{1}{c|}{$\delta_{sys}$} & \multicolumn{1}{c|}{$\delta_{tot}$} & \multicolumn{1}{c|}{$\delta_{unc}$} & \multicolumn{1}{c|}{$\delta_{lar}$} & \multicolumn{1}{c|}{$\delta_{ele}$} & \multicolumn{1}{c|}{$\delta_{\theta}$} & \multicolumn{1}{c|}{$\delta_{noise}$} & \multicolumn{1}{c|}{$\delta_{\xpom}$}  & \multicolumn{1}{c|}{$\delta_{\beta}$}  & \multicolumn{1}{c|}{$\delta_{bg}$}  & \multicolumn{1}{c|}{$\delta_{Plug}$} & \multicolumn{1}{c|}{$\delta_{Q^2}$} & \multicolumn{1}{c|}{$\delta_{spa}$} \\ 
 & \multicolumn{1}{c|}{$[\rm{GeV^2}]$} & & & \multicolumn{1}{c|}{[\%]} & \multicolumn{1}{c|}{[\%]} & \multicolumn{1}{c|}{[\%]} & \multicolumn{1}{c|}{[\%]} & \multicolumn{1}{c|}{[\%]} & \multicolumn{1}{c|}{[\%]} & \multicolumn{1}{c|}{[\%]} & \multicolumn{1}{c|}{[\%]} & \multicolumn{1}{c|}{[\%]} & \multicolumn{1}{c|}{[\%]} & \multicolumn{1}{c|}{[\%]} & \multicolumn{1}{c|}{[\%]} & \multicolumn{1}{c|}{[\%]} & \multicolumn{1}{c|}{[\%]}  \\\hline\hline
0.01 & 8.5 & 0.08 & 0.0182 & 9.4 & 5.2 & 10.8 & 3.6 & 0.9 & 1.6 & 0.9 & -1.5 & -0.2 & -1.0 & -2.0 & 0.7 & -0.1 & 0.3 \\ 
0.01 & 8.5 & 0.13 & 0.0200 & 9.1 & 6.7 & 11.3 & 3.6 & -0.2 & -1.9 & 1.6 & -1.0 & 0.6 & -0.9 & -2.9 & 0.8 & 0.3 & -0.5 \\ 
0.01 & 8.5 & 0.2 & 0.0220 & 10.2 & 6.4 & 12.0 & 3.6 & 1.5 & -1.3 & -1.2 & -2.1 & -0.2 & -1.0 & -0.7 & 0.3 & -0.4 & -0.6 \\ 
0.01 & 12.0 & 0.02 & 0.0370 & 12.7 & 5.4 & 13.8 & 3.6 & 0.5 & 1.8 & -1.1 & -0.9 & -0.2 & -1.6 & -1.8 & 0.1 & 0.4 & 0.4 \\ 
0.01 & 12.0 & 0.032 & 0.0340 & 10.3 & 5.1 & 11.4 & 3.6 & 1.0 & -1.4 & -0.5 & 1.2 & -0.3 & -1.4 & -1.3 & 0.3 & 0.5 & -0.8 \\ 
0.01 & 12.0 & 0.05 & 0.0345 & 9.6 & 5.2 & 10.9 & 3.6 & 0.6 & -1.2 & 1.6 & -1.0 & 0.1 & -1.1 & -2.3 & 0.6 & 0.5 & 0.4 \\ 
0.01 & 12.0 & 0.08 & 0.0259 & 10.1 & 5.6 & 11.6 & 3.6 & 0.7 & 1.4 & 1.0 & -1.9 & -0.2 & -1.1 & -1.8 & 0.4 & 0.5 & -1.7 \\ 
0.01 & 12.0 & 0.13 & 0.0217 & 10.2 & 5.2 & 11.4 & 3.6 & 0.9 & -1.9 & 0.4 & 1.7 & -0.1 & -0.9 & -1.6 & 0.2 & 0.6 & 0.2 \\ 
0.01 & 12.0 & 0.2 & 0.0258 & 11.9 & 5.2 & 13.0 & 3.6 & -1.2 & -0.6 & 2.5 & -0.8 & 0.4 & -1.1 & 0.0 & -0.5 & 0.8 & 0.7 \\ 
0.01 & 12.0 & 0.32 & 0.0240 & 14.0 & 5.3 & 14.9 & 3.6 & -0.5 & -1.3 & 1.5 & -1.1 & 0.8 & -0.0 & -0.4 & -0.4 & 0.4 & -2.3 \\ 
0.01 & 15.0 & 0.02 & 0.0329 & 14.2 & 6.2 & 15.4 & 3.6 & 1.3 & -2.0 & 0.4 & -1.9 & -0.4 & -2.0 & -0.7 & -1.8 & -0.2 & 1.5 \\ 
0.01 & 15.0 & 0.032 & 0.0306 & 6.0 & 5.3 & 8.0 & 3.6 & 0.3 & 1.7 & 1.0 & -0.5 & -0.1 & -1.3 & -1.8 & 0.2 & 0.2 & 1.0 \\ 
0.01 & 15.0 & 0.05 & 0.0291 & 5.3 & 4.8 & 7.1 & 3.6 & -0.5 & -1.3 & 0.7 & -0.9 & -0.2 & -1.4 & -1.7 & 0.5 & 0.1 & -0.7 \\ 
0.01 & 15.0 & 0.08 & 0.0232 & 6.8 & 4.9 & 8.4 & 3.6 & 0.1 & 0.6 & 0.7 & -0.6 & 0.1 & -1.2 & -2.6 & 0.7 & 0.1 & 0.3 \\ 
0.01 & 15.0 & 0.13 & 0.0230 & 5.1 & 5.0 & 7.2 & 3.6 & 0.5 & -0.4 & 0.9 & -1.1 & -0.1 & -1.1 & -2.7 & 0.3 & 0.1 & 0.3 \\ 
0.01 & 15.0 & 0.2 & 0.0224 & 5.0 & 4.9 & 7.0 & 3.6 & 0.7 & 0.7 & 0.6 & -1.0 & 0.5 & -1.2 & -2.5 & 0.4 & 0.2 & 0.3 \\ 
0.01 & 15.0 & 0.32 & 0.0229 & 5.2 & 4.5 & 6.9 & 3.6 & 0.6 & -1.4 & -0.6 & -0.2 & 0.4 & -0.8 & -1.0 & 0.2 & 0.1 & 0.2 \\ 
0.01 & 20.0 & 0.032 & 0.0337 & 8.8 & 5.7 & 10.5 & 3.6 & 1.2 & -2.1 & 1.3 & -0.9 & -0.1 & -1.5 & -2.1 & 0.9 & 0.4 & 1.3 \\ 
0.01 & 20.0 & 0.05 & 0.0287 & 6.8 & 5.3 & 8.6 & 3.6 & 0.7 & -1.1 & 1.1 & -0.8 & 0.2 & -1.4 & -2.8 & 0.5 & 0.6 & -0.3 \\ 
0.01 & 20.0 & 0.08 & 0.0274 & 5.8 & 4.7 & 7.4 & 3.6 & 0.3 & -1.2 & 1.1 & 0.2 & -0.1 & -1.1 & -1.5 & -0.6 & 0.4 & -0.3 \\ 
0.01 & 20.0 & 0.13 & 0.0270 & 5.6 & 4.9 & 7.4 & 3.6 & 0.8 & -0.6 & 0.7 & -0.7 & -0.1 & -1.0 & -2.5 & 0.6 & 0.4 & -0.2 \\ 
0.01 & 20.0 & 0.2 & 0.0241 & 9.3 & 5.0 & 10.5 & 3.6 & -0.4 & -1.2 & 1.5 & -0.6 & -0.1 & -1.1 & -2.0 & 0.3 & 0.4 & -0.7 \\ 
0.01 & 20.0 & 0.32 & 0.0259 & 5.4 & 4.9 & 7.3 & 3.6 & 0.7 & 1.0 & 0.8 & 1.2 & 0.5 & -1.0 & -2.1 & 0.3 & 0.4 & 0.4 \\ 
0.01 & 20.0 & 0.5 & 0.0294 & 6.9 & 5.4 & 8.8 & 3.6 & 0.3 & -2.2 & 1.0 & -0.9 & 0.8 & -0.2 & -2.1 & -0.5 & 0.3 & 0.8 \\ 
0.01 & 25.0 & 0.032 & 0.0289 & 20.1 & 6.0 & 20.9 & 3.6 & -1.1 & -1.8 & -0.7 & 1.1 & -0.2 & -2.0 & -1.8 & 0.5 & 0.2 & -1.5 \\ 
0.01 & 25.0 & 0.05 & 0.0335 & 7.5 & 5.2 & 9.1 & 3.6 & 1.1 & 1.4 & 0.6 & -0.7 & 0.1 & -1.4 & -2.6 & 0.4 & 0.3 & 0.6 \\ 
0.01 & 25.0 & 0.08 & 0.0315 & 5.9 & 5.2 & 7.9 & 3.6 & 0.5 & -1.2 & 0.4 & -0.2 & -0.1 & -1.2 & -2.1 & 0.3 & 0.2 & 0.3 \\ 
0.01 & 25.0 & 0.13 & 0.0270 & 6.0 & 4.4 & 7.4 & 3.6 & 0.4 & -0.8 & 0.7 & 0.7 & -0.2 & -1.0 & -1.5 & 0.5 & 0.1 & -0.5 \\ 
0.01 & 25.0 & 0.2 & 0.0253 & 5.7 & 5.0 & 7.6 & 3.6 & 0.5 & 1.1 & 1.3 & -0.3 & -0.1 & -1.2 & -2.3 & 0.4 & 0.2 & 0.9 \\ 
0.01 & 25.0 & 0.32 & 0.0242 & 5.8 & 5.0 & 7.7 & 3.6 & 1.2 & -0.9 & 1.2 & 0.4 & 0.4 & -1.0 & -2.1 & 0.7 & 0.2 & -0.1 \\ 
0.01 & 25.0 & 0.5 & 0.0289 & 6.2 & 5.2 & 8.1 & 3.6 & 1.1 & -2.0 & 1.1 & 1.5 & 0.9 & -0.2 & -1.6 & 0.4 & 0.2 & 1.0 \\ 
0.01 & 35.0 & 0.05 & 0.0299 & 14.1 & 6.1 & 15.4 & 3.6 & 0.9 & 1.3 & 1.6 & -0.5 & 0.1 & -1.7 & -2.7 & 0.3 & 0.8 & 1.5 \\ 
0.01 & 35.0 & 0.08 & 0.0324 & 7.1 & 5.1 & 8.7 & 3.6 & 0.5 & -3.0 & 0.3 & 0.4 & -0.2 & -1.0 & -0.9 & 0.6 & 0.5 & 0.2 \\ 
0.01 & 35.0 & 0.13 & 0.0294 & 7.0 & 4.7 & 8.4 & 3.6 & 0.4 & -1.1 & 0.4 & 0.4 & -0.1 & -1.1 & -2.3 & -0.3 & 0.4 & 0.3 \\ 
0.01 & 35.0 & 0.2 & 0.0302 & 6.7 & 4.8 & 8.2 & 3.6 & 0.1 & -0.9 & 1.2 & 0.4 & -0.1 & -1.0 & -1.8 & 0.2 & 0.4 & -0.6 \\ 
0.01 & 35.0 & 0.32 & 0.0232 & 6.7 & 6.6 & 9.4 & 3.6 & 0.5 & -0.5 & 0.8 & 1.2 & 0.8 & -1.1 & -4.9 & 0.6 & 0.7 & 0.7 \\ 
\hline 
\end{tabular}
\end{center} 
\end{footnotesize} 
\end{table} 
\begin{table} 
\begin{footnotesize}
\begin{center}  
\begin{tabular}{|c | c | c | c | c | c | c | c | c | c | c | c | c | c | c | c | c | c |} 
\hline 
\multicolumn{1}{|c|}{$\xpom$} & \multicolumn{1}{c|}{$Q^2$} & \multicolumn{1}{c|}{$\beta$} & \multicolumn{1}{c|}{$\xpom \sigma_r^{D(3)}$} & \multicolumn{1}{c|}{$\delta_{stat}$} & \multicolumn{1}{c|}{$\delta_{sys}$} & \multicolumn{1}{c|}{$\delta_{tot}$} & \multicolumn{1}{c|}{$\delta_{unc}$} & \multicolumn{1}{c|}{$\delta_{lar}$} & \multicolumn{1}{c|}{$\delta_{ele}$} & \multicolumn{1}{c|}{$\delta_{\theta}$} & \multicolumn{1}{c|}{$\delta_{noise}$} & \multicolumn{1}{c|}{$\delta_{\xpom}$}  & \multicolumn{1}{c|}{$\delta_{\beta}$}  & \multicolumn{1}{c|}{$\delta_{bg}$}  & \multicolumn{1}{c|}{$\delta_{Plug}$} & \multicolumn{1}{c|}{$\delta_{Q^2}$} & \multicolumn{1}{c|}{$\delta_{spa}$} \\ 
 & \multicolumn{1}{c|}{$[\rm{GeV^2}]$} & & & \multicolumn{1}{c|}{[\%]} & \multicolumn{1}{c|}{[\%]} & \multicolumn{1}{c|}{[\%]} & \multicolumn{1}{c|}{[\%]} & \multicolumn{1}{c|}{[\%]} & \multicolumn{1}{c|}{[\%]} & \multicolumn{1}{c|}{[\%]} & \multicolumn{1}{c|}{[\%]} & \multicolumn{1}{c|}{[\%]} & \multicolumn{1}{c|}{[\%]} & \multicolumn{1}{c|}{[\%]} & \multicolumn{1}{c|}{[\%]} & \multicolumn{1}{c|}{[\%]} & \multicolumn{1}{c|}{[\%]}  \\\hline\hline
0.01 & 35.0 & 0.5 & 0.0305 & 6.6 & 4.8 & 8.2 & 3.6 & 0.6 & -1.5 & 0.9 & 0.8 & 0.8 & -0.3 & -1.2 & -0.2 & 0.6 & 0.3 \\ 
0.01 & 35.0 & 0.8 & 0.0362 & 9.2 & 7.0 & 11.6 & 3.6 & 1.0 & -1.1 & 1.7 & 2.4 & 2.4 & 1.2 & -1.9 & -1.3 & 0.6 & 1.6 \\ 
0.01 & 45.0 & 0.08 & 0.0329 & 13.5 & 5.3 & 14.5 & 3.6 & 1.6 & 0.8 & 1.3 & -0.7 & -0.1 & -1.4 & -1.3 & -0.6 & 0.4 & 1.9 \\ 
0.01 & 45.0 & 0.13 & 0.0377 & 7.7 & 6.0 & 9.7 & 3.6 & -0.8 & -2.5 & 0.9 & -1.3 & 0.1 & -1.0 & -2.4 & -0.4 & 0.3 & -1.5 \\ 
0.01 & 45.0 & 0.2 & 0.0264 & 8.2 & 4.7 & 9.5 & 3.6 & 0.9 & -1.5 & 0.6 & -0.5 & -0.2 & -1.0 & -1.3 & -0.9 & 0.2 & -0.7 \\ 
0.01 & 45.0 & 0.32 & 0.0237 & 10.6 & 6.1 & 12.2 & 3.6 & 1.5 & -1.6 & 0.9 & 0.4 & 0.6 & -1.0 & -3.6 & 0.1 & 0.5 & -0.4 \\ 
0.01 & 45.0 & 0.5 & 0.0373 & 8.3 & 4.9 & 9.7 & 3.6 & -0.7 & -1.1 & -0.3 & 0.9 & 0.6 & -0.4 & -1.5 & -0.2 & 0.3 & -0.8 \\ 
0.01 & 45.0 & 0.8 & 0.0245 & 11.2 & 5.9 & 12.7 & 3.6 & 1.2 & -1.4 & 1.1 & 2.3 & 1.8 & 1.0 & -0.9 & 0.4 & 0.3 & 0.4 \\ 
0.01 & 60.0 & 0.13 & 0.0379 & 13.3 & 6.5 & 14.8 & 3.6 & 1.5 & -2.6 & 0.9 & 2.9 & -0.1 & -1.2 & -1.5 & 0.9 & 0.5 & 0.8 \\ 
0.01 & 60.0 & 0.2 & 0.0318 & 8.9 & 5.4 & 10.4 & 3.6 & 1.1 & -1.3 & 1.2 & -0.5 & 0.1 & -1.2 & -2.7 & -0.2 & 0.3 & 0.7 \\ 
0.01 & 60.0 & 0.32 & 0.0319 & 8.0 & 4.9 & 9.4 & 3.6 & 0.8 & -2.1 & 0.5 & 0.6 & 0.1 & -0.8 & -1.6 & 0.5 & 0.2 & 0.1 \\ 
0.01 & 60.0 & 0.5 & 0.0335 & 10.8 & 6.9 & 12.8 & 3.6 & 0.8 & -1.5 & 0.9 & 0.5 & 1.6 & -0.2 & -5.0 & 0.2 & 0.7 & 0.8 \\ 
0.01 & 60.0 & 0.8 & 0.0277 & 12.3 & 7.0 & 14.2 & 3.6 & 0.9 & -3.5 & 0.5 & 1.5 & 1.7 & 1.0 & -2.9 & 0.3 & 0.3 & 1.5 \\ 
0.01 & 90.0 & 0.2 & 0.0259 & 35.3 & 8.9 & 36.4 & 3.6 & 0.6 & -5.4 & 3.8 & 1.2 & -0.6 & -1.6 & 0.0 & -1.2 & 0.8 & -2.3 \\ 
0.01 & 90.0 & 0.32 & 0.0292 & 14.4 & 5.5 & 15.5 & 3.6 & 1.1 & 2.3 & 1.7 & 1.5 & -0.0 & -0.7 & -0.6 & -0.5 & 0.6 & 0.8 \\ 
0.01 & 90.0 & 0.5 & 0.0326 & 12.9 & 7.3 & 14.8 & 3.6 & 1.2 & -2.7 & 1.7 & 1.8 & 0.6 & -0.2 & -2.0 & 0.8 & 0.6 & 1.6 \\ 
0.01 & 90.0 & 0.8 & 0.0240 & 16.5 & 6.2 & 17.6 & 3.6 & 0.9 & -0.3 & 0.3 & 1.7 & 2.1 & 0.8 & -2.4 & -0.9 & 0.9 & 1.5 \\ 
0.03 & 3.5 & 0.0017 & 0.0188 & 25.3 & 17.0 & 30.5 & 11.2 & -6.2 & 4.6 & -2.1 & -2.9 & 0.8 & -2.3 & -4.2 & -2.4 & 0.7 & 0.9 \\ 
0.03 & 3.5 & 0.0027 & 0.0263 & 12.7 & 16.3 & 20.7 & 11.2 & -2.1 & -1.3 & 4.4 & -2.9 & 0.7 & -2.8 & -7.1 & 4.8 & -0.7 & 1.3 \\ 
0.03 & 3.5 & 0.0043 & 0.0254 & 10.8 & 17.3 & 20.3 & 11.2 & -1.3 & 7.5 & 2.5 & -3.4 & 0.9 & -2.2 & -7.0 & 2.7 & -0.4 & 1.2 \\ 
0.03 & 3.5 & 0.0067 & 0.0214 & 11.6 & 14.5 & 18.6 & 11.2 & -2.8 & -2.2 & -1.1 & -3.3 & 1.2 & -2.2 & -6.3 & 2.2 & 0.8 & 1.1 \\ 
0.03 & 3.5 & 0.011 & 0.0190 & 12.5 & 14.6 & 19.2 & 11.2 & -2.3 & -1.9 & 3.0 & -2.8 & 1.1 & -1.1 & -6.0 & -2.9 & -1.2 & -1.2 \\ 
0.03 & 5.0 & 0.0027 & 0.0366 & 16.2 & 26.1 & 30.7 & 11.2 & -4.6 & -17.1 & -2.4 & -3.4 & 1.2 & -3.2 & -8.2 & 4.3 & -0.2 & -8.6 \\ 
0.03 & 5.0 & 0.0043 & 0.0325 & 12.9 & 15.9 & 20.5 & 11.2 & -0.7 & 8.5 & -2.4 & -2.6 & 0.6 & -1.2 & -3.9 & 1.7 & 0.4 & -1.2 \\ 
0.03 & 5.0 & 0.0067 & 0.0302 & 11.0 & 15.2 & 18.7 & 11.2 & 1.9 & -2.7 & 0.5 & -2.4 & 1.2 & -2.7 & -7.7 & 3.2 & 0.5 & 0.4 \\ 
0.03 & 5.0 & 0.011 & 0.0233 & 13.3 & 14.4 & 19.6 & 11.2 & -1.8 & 4.3 & -0.5 & -1.5 & 1.0 & -1.3 & -6.0 & 2.5 & -1.0 & -2.7 \\ 
0.03 & 5.0 & 0.017 & 0.0225 & 11.0 & 15.6 & 19.1 & 11.2 & 0.2 & -4.6 & 1.8 & -2.3 & 2.1 & -1.1 & -6.6 & 3.8 & 0.7 & 2.6 \\ 
0.03 & 6.5 & 0.0027 & 0.0366 & 33.5 & 17.5 & 37.8 & 11.2 & 0.9 & -8.9 & -1.4 & 2.7 & -0.1 & -3.4 & -3.4 & 0.1 & -0.4 & -3.5 \\ 
0.03 & 6.5 & 0.0043 & 0.0211 & 14.6 & 16.0 & 21.7 & 11.2 & 3.1 & 2.1 & 3.6 & 2.5 & 1.5 & -1.6 & -6.2 & 5.5 & 0.1 & 3.3 \\ 
0.03 & 6.5 & 0.0067 & 0.0261 & 12.0 & 15.3 & 19.4 & 11.2 & -1.7 & -2.0 & 1.5 & -3.4 & 1.3 & -2.6 & -7.5 & 1.6 & 0.3 & -1.9 \\ 
0.03 & 6.5 & 0.011 & 0.0263 & 12.0 & 14.4 & 18.7 & 11.2 & -2.3 & 0.4 & 2.4 & -2.1 & 1.0 & -0.7 & -3.9 & 5.2 & 0.2 & 2.1 \\ 
0.03 & 6.5 & 0.017 & 0.0266 & 11.6 & 13.8 & 18.0 & 11.2 & -0.8 & 0.7 & -2.0 & -1.3 & 1.3 & -0.8 & -5.7 & -1.6 & -0.1 & 0.2 \\ 
0.03 & 6.5 & 0.027 & 0.0262 & 10.3 & 15.0 & 18.2 & 11.2 & -1.6 & 0.5 & 1.2 & -2.1 & 1.5 & -1.7 & -7.4 & 2.6 & 0.5 & -0.3 \\ 
0.03 & 6.5 & 0.043 & 0.0225 & 12.1 & 14.7 & 19.1 & 11.2 & 1.3 & 2.3 & 3.8 & 2.6 & 1.4 & -0.8 & -6.9 & 0.5 & 0.0 & 1.0 \\ 
0.03 & 8.5 & 0.0043 & 0.0331 & 18.8 & 16.9 & 25.2 & 11.2 & -5.5 & -2.0 & 2.4 & -3.4 & 0.7 & -1.8 & -4.2 & 1.8 & 0.4 & -6.0 \\ 
0.03 & 8.5 & 0.0067 & 0.0250 & 13.1 & 14.7 & 19.7 & 11.2 & 0.8 & 2.8 & -0.1 & -2.6 & 1.4 & -0.9 & -4.8 & 1.8 & 0.4 & 2.8 \\ 
\hline 
\end{tabular}
\end{center} 
\end{footnotesize} 
\end{table} 
\begin{table} 
\begin{footnotesize}
\begin{center}  
\begin{tabular}{|c | c | c | c | c | c | c | c | c | c | c | c | c | c | c | c | c | c |} 
\hline 
\multicolumn{1}{|c|}{$\xpom$} & \multicolumn{1}{c|}{$Q^2$} & \multicolumn{1}{c|}{$\beta$} & \multicolumn{1}{c|}{$\xpom \sigma_r^{D(3)}$} & \multicolumn{1}{c|}{$\delta_{stat}$} & \multicolumn{1}{c|}{$\delta_{sys}$} & \multicolumn{1}{c|}{$\delta_{tot}$} & \multicolumn{1}{c|}{$\delta_{unc}$} & \multicolumn{1}{c|}{$\delta_{lar}$} & \multicolumn{1}{c|}{$\delta_{ele}$} & \multicolumn{1}{c|}{$\delta_{\theta}$} & \multicolumn{1}{c|}{$\delta_{noise}$} & \multicolumn{1}{c|}{$\delta_{\xpom}$}  & \multicolumn{1}{c|}{$\delta_{\beta}$}  & \multicolumn{1}{c|}{$\delta_{bg}$}  & \multicolumn{1}{c|}{$\delta_{Plug}$} & \multicolumn{1}{c|}{$\delta_{Q^2}$} & \multicolumn{1}{c|}{$\delta_{spa}$} \\ 
 & \multicolumn{1}{c|}{$[\rm{GeV^2}]$} & & & \multicolumn{1}{c|}{[\%]} & \multicolumn{1}{c|}{[\%]} & \multicolumn{1}{c|}{[\%]} & \multicolumn{1}{c|}{[\%]} & \multicolumn{1}{c|}{[\%]} & \multicolumn{1}{c|}{[\%]} & \multicolumn{1}{c|}{[\%]} & \multicolumn{1}{c|}{[\%]} & \multicolumn{1}{c|}{[\%]} & \multicolumn{1}{c|}{[\%]} & \multicolumn{1}{c|}{[\%]} & \multicolumn{1}{c|}{[\%]} & \multicolumn{1}{c|}{[\%]} & \multicolumn{1}{c|}{[\%]}  \\\hline\hline
0.03 & 8.5 & 0.011 & 0.0313 & 10.4 & 13.8 & 17.3 & 11.2 & -2.1 & 1.2 & -1.5 & -3.0 & 0.8 & -1.3 & -4.1 & 3.0 & 0.2 & -1.0 \\ 
0.03 & 8.5 & 0.017 & 0.0270 & 10.4 & 14.3 & 17.7 & 11.2 & -1.2 & -1.5 & 4.0 & -2.0 & 1.5 & -1.1 & -6.3 & 2.1 & 0.1 & -1.0 \\ 
0.03 & 8.5 & 0.027 & 0.0276 & 9.4 & 14.6 & 17.3 & 11.2 & -1.4 & 1.3 & -0.1 & -2.9 & 1.5 & -1.5 & -7.5 & 1.8 & 0.2 & -0.8 \\ 
0.03 & 8.5 & 0.043 & 0.0277 & 10.5 & 15.6 & 18.8 & 11.2 & 0.7 & -3.8 & 3.8 & 0.6 & 0.8 & -0.4 & -4.1 & 3.9 & 0.1 & 1.8 \\ 
0.03 & 12.0 & 0.0067 & 0.0325 & 17.1 & 16.0 & 23.5 & 11.2 & 2.5 & 1.8 & 0.5 & -5.4 & 0.9 & -1.7 & -5.2 & -0.9 & 0.7 & 0.4 \\ 
0.03 & 12.0 & 0.011 & 0.0368 & 12.0 & 12.8 & 17.5 & 11.2 & -0.7 & -1.9 & 2.3 & -2.3 & 0.7 & -1.6 & -3.9 & 1.6 & 0.6 & 1.3 \\ 
0.03 & 12.0 & 0.017 & 0.0336 & 11.0 & 14.5 & 18.2 & 11.2 & -2.1 & 0.9 & 2.0 & -3.0 & 1.4 & -1.7 & -6.6 & 3.2 & 0.6 & -0.9 \\ 
0.03 & 12.0 & 0.027 & 0.0375 & 11.1 & 14.3 & 18.1 & 11.2 & -1.5 & -3.5 & 2.5 & -1.9 & 1.3 & -1.7 & -6.5 & 2.5 & 0.5 & -0.4 \\ 
0.03 & 12.0 & 0.043 & 0.0303 & 11.9 & 14.1 & 18.5 & 11.2 & -1.3 & -0.9 & 1.4 & -2.3 & 1.5 & -0.7 & -5.3 & 2.4 & 0.8 & 1.0 \\ 
0.03 & 12.0 & 0.067 & 0.0251 & 12.7 & 13.2 & 18.3 & 11.2 & -1.4 & -1.3 & 1.9 & -2.0 & 1.6 & -0.7 & -4.9 & 1.6 & 0.5 & 0.3 \\ 
0.03 & 15.0 & 0.0067 & 0.0462 & 15.6 & 14.6 & 21.4 & 11.2 & -2.1 & 6.5 & 0.9 & -3.6 & 0.6 & -2.3 & -3.6 & 1.3 & -0.2 & -1.7 \\ 
0.03 & 15.0 & 0.011 & 0.0354 & 6.6 & 14.7 & 16.1 & 11.2 & -0.9 & -1.7 & -1.3 & -1.7 & 1.4 & -2.0 & -7.5 & 2.1 & 0.4 & -0.8 \\ 
0.03 & 15.0 & 0.017 & 0.0360 & 5.4 & 14.1 & 15.1 & 11.2 & 0.4 & 1.2 & 1.5 & -1.6 & 1.4 & -1.9 & -7.2 & 3.0 & 0.3 & 0.6 \\ 
0.03 & 15.0 & 0.027 & 0.0317 & 5.4 & 14.2 & 15.2 & 11.2 & -1.2 & -1.4 & 1.5 & -1.2 & 1.2 & -1.6 & -6.7 & 1.8 & 0.3 & 0.3 \\ 
0.03 & 15.0 & 0.043 & 0.0271 & 6.7 & 15.1 & 16.5 & 11.2 & -0.7 & 2.3 & 1.2 & -2.2 & 1.5 & -1.6 & -8.7 & 2.6 & 0.1 & 0.5 \\ 
0.03 & 15.0 & 0.067 & 0.0244 & 6.4 & 13.9 & 15.3 & 11.2 & -0.7 & 1.1 & 0.4 & -1.2 & 1.6 & -1.2 & -7.2 & 1.6 & 0.4 & -1.0 \\ 
0.03 & 15.0 & 0.11 & 0.0239 & 6.1 & 14.9 & 16.1 & 11.2 & -1.1 & -0.7 & 0.7 & -2.8 & 1.6 & -1.3 & -7.8 & 1.8 & 0.2 & -0.2 \\ 
0.03 & 20.0 & 0.011 & 0.0307 & 9.7 & 15.1 & 17.9 & 11.2 & -0.3 & -0.6 & 0.9 & -2.0 & 1.3 & -2.4 & -7.7 & 2.7 & 0.6 & 1.1 \\ 
0.03 & 20.0 & 0.017 & 0.0391 & 7.3 & 14.4 & 16.1 & 11.2 & 0.4 & -1.1 & -1.2 & -2.5 & 1.1 & -2.0 & -6.7 & 2.6 & 0.6 & 1.3 \\ 
0.03 & 20.0 & 0.027 & 0.0321 & 5.9 & 13.4 & 14.6 & 11.2 & -0.6 & -0.8 & 1.3 & -2.3 & 1.1 & -1.6 & -6.2 & 1.0 & 0.6 & 0.7 \\ 
0.03 & 20.0 & 0.043 & 0.0292 & 6.2 & 13.5 & 14.9 & 11.2 & 0.7 & -2.2 & 1.3 & -0.1 & 1.1 & -1.3 & -6.4 & 2.1 & 0.6 & -0.8 \\ 
0.03 & 20.0 & 0.067 & 0.0285 & 6.0 & 13.3 & 14.6 & 11.2 & -1.1 & -0.8 & 0.6 & -1.9 & 1.1 & -1.2 & -5.8 & 2.5 & 0.6 & -0.7 \\ 
0.03 & 20.0 & 0.11 & 0.0235 & 7.0 & 13.7 & 15.4 & 11.2 & -0.5 & -0.4 & -0.8 & -0.9 & 1.6 & -1.1 & -7.0 & 2.1 & 0.6 & -0.5 \\ 
0.03 & 25.0 & 0.011 & 0.0348 & 25.7 & 14.1 & 29.3 & 11.2 & -1.3 & 3.1 & 1.5 & -2.9 & 0.3 & -2.1 & -2.9 & 4.8 & 0.2 & 3.3 \\ 
0.03 & 25.0 & 0.017 & 0.0385 & 8.3 & 14.3 & 16.5 & 11.2 & 0.6 & 0.8 & 1.3 & -0.8 & 1.4 & -2.0 & -7.6 & 2.0 & 0.6 & 1.2 \\ 
0.03 & 25.0 & 0.027 & 0.0384 & 6.4 & 13.5 & 14.9 & 11.2 & -0.1 & -1.5 & 1.7 & -2.4 & 1.3 & -1.5 & -6.1 & 1.8 & 0.5 & 0.1 \\ 
0.03 & 25.0 & 0.043 & 0.0305 & 6.5 & 12.5 & 14.1 & 11.2 & -0.6 & -0.6 & 1.1 & -2.2 & 0.9 & -0.8 & -4.0 & 2.4 & 0.3 & 0.6 \\ 
0.03 & 25.0 & 0.067 & 0.0274 & 6.7 & 14.8 & 16.2 & 11.2 & -1.0 & -1.1 & 0.9 & -1.7 & 1.4 & -1.5 & -7.9 & 2.4 & 0.7 & -1.1 \\ 
0.03 & 25.0 & 0.11 & 0.0258 & 6.2 & 14.1 & 15.4 & 11.2 & -1.4 & -0.8 & 0.9 & -1.1 & 1.7 & -1.2 & -7.6 & 2.2 & 0.6 & -0.5 \\ 
0.03 & 25.0 & 0.17 & 0.0265 & 6.4 & 14.6 & 16.0 & 11.2 & 0.5 & 1.0 & 1.0 & -2.2 & 1.7 & -1.4 & -8.2 & 1.6 & 0.5 & -1.1 \\ 
0.03 & 35.0 & 0.017 & 0.0533 & 14.7 & 13.7 & 20.1 & 11.2 & -2.7 & 3.8 & 1.2 & -2.9 & 0.4 & -1.7 & -3.2 & -2.7 & 0.7 & 1.9 \\ 
0.03 & 35.0 & 0.027 & 0.0415 & 7.7 & 13.0 & 15.1 & 11.2 & 1.3 & 1.6 & 0.5 & -0.7 & 0.8 & -1.3 & -4.7 & 2.9 & 0.8 & 0.4 \\ 
0.03 & 35.0 & 0.043 & 0.0411 & 6.9 & 13.1 & 14.8 & 11.2 & -1.1 & -1.7 & 0.3 & -1.5 & 0.7 & -1.0 & -3.8 & 2.0 & 0.6 & -0.6 \\ 
0.03 & 35.0 & 0.067 & 0.0312 & 7.1 & 13.9 & 15.6 & 11.2 & 0.9 & -1.2 & 1.1 & -1.2 & 1.3 & -1.3 & -6.7 & 2.9 & 0.9 & 1.2 \\ 
0.03 & 35.0 & 0.11 & 0.0311 & 6.7 & 14.3 & 15.8 & 11.2 & -0.9 & -0.6 & 0.6 & -2.2 & 1.6 & -1.2 & -7.8 & 1.7 & 1.0 & -0.9 \\ 
0.03 & 35.0 & 0.17 & 0.0258 & 7.0 & 14.3 & 15.9 & 11.2 & -0.6 & 0.9 & 1.4 & -1.3 & 1.8 & -1.0 & -7.3 & 2.6 & 0.9 & -0.3 \\ 
\hline 
\end{tabular}
\end{center} 
\end{footnotesize} 
\end{table} 
\begin{table} 
\begin{footnotesize}
\begin{center}  
\begin{tabular}{|c | c | c | c | c | c | c | c | c | c | c | c | c | c | c | c | c | c |} 
\hline 
\multicolumn{1}{|c|}{$\xpom$} & \multicolumn{1}{c|}{$Q^2$} & \multicolumn{1}{c|}{$\beta$} & \multicolumn{1}{c|}{$\xpom \sigma_r^{D(3)}$} & \multicolumn{1}{c|}{$\delta_{stat}$} & \multicolumn{1}{c|}{$\delta_{sys}$} & \multicolumn{1}{c|}{$\delta_{tot}$} & \multicolumn{1}{c|}{$\delta_{unc}$} & \multicolumn{1}{c|}{$\delta_{lar}$} & \multicolumn{1}{c|}{$\delta_{ele}$} & \multicolumn{1}{c|}{$\delta_{\theta}$} & \multicolumn{1}{c|}{$\delta_{noise}$} & \multicolumn{1}{c|}{$\delta_{\xpom}$}  & \multicolumn{1}{c|}{$\delta_{\beta}$}  & \multicolumn{1}{c|}{$\delta_{bg}$}  & \multicolumn{1}{c|}{$\delta_{Plug}$} & \multicolumn{1}{c|}{$\delta_{Q^2}$} & \multicolumn{1}{c|}{$\delta_{spa}$} \\ 
 & \multicolumn{1}{c|}{$[\rm{GeV^2}]$} & & & \multicolumn{1}{c|}{[\%]} & \multicolumn{1}{c|}{[\%]} & \multicolumn{1}{c|}{[\%]} & \multicolumn{1}{c|}{[\%]} & \multicolumn{1}{c|}{[\%]} & \multicolumn{1}{c|}{[\%]} & \multicolumn{1}{c|}{[\%]} & \multicolumn{1}{c|}{[\%]} & \multicolumn{1}{c|}{[\%]} & \multicolumn{1}{c|}{[\%]} & \multicolumn{1}{c|}{[\%]} & \multicolumn{1}{c|}{[\%]} & \multicolumn{1}{c|}{[\%]} & \multicolumn{1}{c|}{[\%]}  \\\hline\hline
0.03 & 35.0 & 0.27 & 0.0289 & 7.3 & 14.6 & 16.4 & 11.2 & 0.8 & -1.8 & 0.8 & 0.2 & 1.9 & -1.1 & -8.4 & 2.6 & 0.9 & 0.3 \\ 
0.03 & 45.0 & 0.027 & 0.0504 & 11.7 & 13.3 & 17.7 & 11.2 & -1.0 & -1.8 & 1.1 & -2.4 & 0.6 & -1.3 & -3.6 & 2.0 & 0.7 & 0.9 \\ 
0.03 & 45.0 & 0.043 & 0.0402 & 8.6 & 12.6 & 15.3 & 11.2 & 1.4 & -1.1 & 0.2 & -1.6 & 0.8 & -1.1 & -4.2 & 1.3 & 0.6 & -0.8 \\ 
0.03 & 45.0 & 0.067 & 0.0390 & 10.1 & 13.7 & 17.0 & 11.2 & -1.7 & -0.3 & -1.5 & -0.7 & 1.4 & -1.2 & -6.6 & 2.4 & 0.7 & -1.0 \\ 
0.03 & 45.0 & 0.11 & 0.0250 & 8.7 & 14.5 & 16.9 & 11.2 & -0.9 & -1.1 & 1.2 & -1.1 & 1.6 & -1.2 & -8.2 & 2.0 & 0.9 & -0.6 \\ 
0.03 & 45.0 & 0.17 & 0.0260 & 8.2 & 15.3 & 17.4 & 11.2 & 2.1 & 0.5 & 1.4 & -2.3 & 1.9 & -1.3 & -9.2 & 1.5 & 1.0 & -0.1 \\ 
0.03 & 45.0 & 0.27 & 0.0215 & 8.2 & 14.7 & 16.8 & 11.2 & 0.7 & -2.3 & 0.8 & -1.3 & 1.7 & -1.4 & -7.8 & 2.3 & 1.0 & 1.0 \\ 
0.03 & 60.0 & 0.043 & 0.0382 & 13.7 & 13.6 & 19.3 & 11.2 & 1.4 & 1.8 & 2.2 & 1.1 & 1.1 & -1.5 & -5.7 & 2.6 & 1.0 & 2.3 \\ 
0.03 & 60.0 & 0.067 & 0.0387 & 9.4 & 13.4 & 16.4 & 11.2 & 0.8 & -3.1 & 1.1 & 1.3 & 1.2 & -1.2 & -5.6 & 1.7 & 0.7 & 0.5 \\ 
0.03 & 60.0 & 0.11 & 0.0265 & 9.2 & 13.5 & 16.3 & 11.2 & 0.8 & -0.3 & 0.9 & -1.0 & 1.4 & -1.0 & -6.8 & 1.9 & 0.7 & -0.8 \\ 
0.03 & 60.0 & 0.17 & 0.0264 & 8.5 & 16.2 & 18.3 & 11.2 & 1.2 & -1.2 & 1.7 & -1.8 & 2.0 & -1.4 & -10.5 & 3.2 & 1.2 & 0.7 \\ 
0.03 & 60.0 & 0.27 & 0.0222 & 13.6 & 17.0 & 21.8 & 11.2 & -0.5 & -1.0 & 1.0 & 1.8 & 2.3 & -1.3 & -11.6 & 2.3 & 1.3 & 0.3 \\ 
0.03 & 60.0 & 0.43 & 0.0277 & 8.9 & 17.2 & 19.4 & 11.2 & 1.3 & -1.6 & -1.2 & 1.3 & 2.8 & -0.9 & -11.7 & 3.0 & 1.4 & 1.2 \\ 
0.03 & 90.0 & 0.067 & 0.0341 & 33.7 & 19.9 & 39.1 & 11.2 & -0.5 & -4.3 & 1.2 & -5.7 & 1.3 & -1.7 & -5.6 & -3.1 & 1.5 & -0.8 \\ 
0.03 & 90.0 & 0.11 & 0.0337 & 14.4 & 12.7 & 19.2 & 11.2 & 1.6 & -1.3 & -1.1 & -0.5 & 1.2 & -0.7 & -3.9 & 1.8 & 1.0 & 0.3 \\ 
0.03 & 90.0 & 0.17 & 0.0368 & 10.4 & 14.1 & 17.5 & 11.2 & 2.0 & -2.3 & 1.5 & -1.9 & 1.4 & -1.1 & -6.7 & 2.1 & 1.2 & 0.7 \\ 
0.03 & 90.0 & 0.27 & 0.0269 & 10.3 & 13.5 & 17.0 & 11.2 & -0.5 & -1.3 & 1.6 & 0.7 & 1.3 & -1.1 & -6.5 & 1.5 & 1.2 & -0.5 \\ 
0.03 & 90.0 & 0.43 & 0.0328 & 10.6 & 14.0 & 17.5 & 11.2 & -0.3 & -1.6 & -1.5 & -1.3 & 1.4 & -0.6 & -6.2 & 2.2 & 1.1 & -1.3 \\ 
0.03 & 90.0 & 0.67 & 0.0289 & 20.9 & 16.8 & 26.8 & 11.2 & 2.1 & 1.7 & 1.4 & 2.3 & 3.3 & 0.8 & -9.3 & 3.5 & 1.8 & 3.0 \\ 
\hline 
\end{tabular}
\end{center} 
\end{footnotesize} 
\end{table} 
\begin{table} 
\begin{footnotesize}
\begin{center}  
\begin{tabular}{|c | c | c | c | c | c | c | c | c | c | c | c | c | c | c | c | c | c |} 
\hline 
\multicolumn{1}{|c|}{$\xpom$} & \multicolumn{1}{c|}{$Q^2$} & \multicolumn{1}{c|}{$\beta$} & \multicolumn{1}{c|}{$\xpom \sigma_r^{D(3)}$} & \multicolumn{1}{c|}{$\delta_{stat}$} & \multicolumn{1}{c|}{$\delta_{sys}$} & \multicolumn{1}{c|}{$\delta_{tot}$} & \multicolumn{1}{c|}{$\delta_{unc}$} & \multicolumn{1}{c|}{$\delta_{lar}$} & \multicolumn{1}{c|}{$\delta_{ele}$} & \multicolumn{1}{c|}{$\delta_{\theta}$} & \multicolumn{1}{c|}{$\delta_{noise}$} & \multicolumn{1}{c|}{$\delta_{\xpom}$}  & \multicolumn{1}{c|}{$\delta_{\beta}$}  & \multicolumn{1}{c|}{$\delta_{bg}$}  & \multicolumn{1}{c|}{$\delta_{Plug}$} & \multicolumn{1}{c|}{$\delta_{\beta(2)}$} & \multicolumn{1}{c|}{$\delta_{tra}$} \\ 
 & \multicolumn{1}{c|}{$[\rm{GeV^2}]$} & & & \multicolumn{1}{c|}{[\%]} & \multicolumn{1}{c|}{[\%]} & \multicolumn{1}{c|}{[\%]} & \multicolumn{1}{c|}{[\%]} & \multicolumn{1}{c|}{[\%]} & \multicolumn{1}{c|}{[\%]} & \multicolumn{1}{c|}{[\%]} & \multicolumn{1}{c|}{[\%]} & \multicolumn{1}{c|}{[\%]} & \multicolumn{1}{c|}{[\%]} & \multicolumn{1}{c|}{[\%]} & \multicolumn{1}{c|}{[\%]} & \multicolumn{1}{c|}{[\%]} & \multicolumn{1}{c|}{[\%]}  \\\hline\hline
0.01 & 200.0 & 0.32 & 0.0321 & 5.8 & 6.6 & 8.8 & 3.6 & -0.9 & -0.1 & -0.7 & -0.5 & 1.0 & -0.9 & 5.1 & -0.7 & 1.0 & -0.4 \\ 
0.01 & 200.0 & 0.5 & 0.0315 & 5.3 & 7.0 & 8.8 & 3.6 & -1.3 & 0.9 & -0.8 & -1.6 & 0.9 & -0.8 & 5.1 & -0.6 & 0.6 & -1.2 \\ 
0.01 & 200.0 & 0.8 & 0.0211 & 7.9 & 5.9 & 9.9 & 3.6 & -1.6 & 1.8 & -0.3 & -3.0 & 1.0 & -1.0 & 1.0 & -1.0 & 0.4 & -1.2 \\ 
0.01 & 400.0 & 0.8 & 0.0196 & 13.3 & 7.9 & 15.4 & 3.6 & -1.5 & 0.6 & -0.7 & -3.0 & 0.4 & -1.1 & 5.5 & -1.0 & 0.3 & -0.8 \\ 
0.03 & 200.0 & 0.11 & 0.0361 & 5.7 & 13.3 & 14.5 & 11.1 & 0.3 & 0.8 & -0.8 & 0.9 & 1.8 & -0.4 & 6.0 & -2.4 & 1.9 & 0.4 \\ 
0.03 & 200.0 & 0.17 & 0.0331 & 4.8 & 13.1 & 14.0 & 11.1 & 0.2 & 0.4 & -0.7 & 0.9 & 1.8 & -0.6 & 5.8 & -2.3 & 1.7 & 0.6 \\ 
0.03 & 200.0 & 0.27 & 0.0283 & 5.3 & 12.3 & 13.4 & 11.1 & 0.1 & 0.5 & -0.8 & 0.8 & 1.7 & -0.9 & 3.8 & -2.1 & 1.4 & 0.2 \\ 
0.03 & 200.0 & 0.43 & 0.0309 & 5.5 & 12.2 & 13.4 & 11.1 & -0.5 & 1.2 & -0.7 & 0.1 & 1.5 & -1.0 & 3.7 & -2.1 & 0.9 & -0.1 \\ 
0.03 & 200.0 & 0.67 & 0.0297 & 7.2 & 13.4 & 15.2 & 11.1 & -1.0 & 2.8 & -0.9 & -1.2 & 0.8 & -0.6 & 5.8 & -2.6 & 0.2 & 0.5 \\ 
0.03 & 400.0 & 0.27 & 0.0322 & 7.6 & 13.4 & 15.4 & 11.1 & -0.4 & 0.6 & -0.2 & 0.1 & 1.2 & -0.8 & 6.9 & -1.8 & 1.3 & -0.3 \\ 
0.03 & 400.0 & 0.43 & 0.0293 & 6.9 & 12.2 & 14.0 & 11.1 & -0.7 & -0.4 & -0.1 & -0.3 & 0.9 & -0.8 & 4.1 & -2.2 & 0.7 & -0.2 \\ 
0.03 & 400.0 & 0.67 & 0.0289 & 8.0 & 13.5 & 15.6 & 11.1 & -0.9 & 1.2 & -0.7 & -0.9 & 1.1 & -0.9 & 6.6 & -2.1 & 0.5 & -0.4 \\ 
0.03 & 800.0 & 0.43 & 0.0391 & 13.1 & 13.7 & 19.0 & 11.1 & -0.9 & 2.4 & -1.2 & -0.3 & 0.1 & -0.5 & 7.1 & -2.1 & 0.2 & -0.4 \\ 
0.03 & 800.0 & 0.67 & 0.0228 & 14.6 & 14.6 & 20.6 & 11.1 & -0.1 & -0.9 & 0.3 & -0.5 & 0.2 & -0.5 & 9.1 & -2.1 & 0.1 & -0.2 \\ 
0.03 & 1600.0 & 0.67 & 0.0214 & 27.9 & 15.5 & 31.9 & 11.1 & -1.6 & -5.7 & -3.2 & -1.7 & -0.4 & -0.1 & 7.5 & -2.6 & -0.2 & -0.3 \\ 
\hline 
\end{tabular} 
\end{center}
\end{footnotesize} 
\caption{Results for $\xpom \sigma_r^D$ at fixed
$Q^2$, $\beta$ and $\xpom$ (columns 1-4) using data with LAr electrons
and $E_p = 920 \ {\rm GeV}$. Columns 5-7 contain the
percentage statistical, systematic and total uncertainties.
The remaining columns contain the contributions to the systematic
uncertainty from sources which are uncorrelated between data points
($\delta_{unc}$) and the 10 
correlated sources
leading to the largest uncertainties. 
These are the LAr hadronic energy scale ($\delta_{lar}$),
the LAr electromagnetic energy scale ($\delta_{ele}$), the scattered
electron angle
measurement ($\delta_{\theta}$), the calorimeter noise
treatment ($\delta_{noise}$),
reweighting the simulation in $\xpom$ ($\delta_{\xpom}$) and $\beta$
($\delta_{\beta}$), the background subtraction using
the non-diffractive RAPGAP simulation ($\delta_{bg}$),
the plug energy scale ($\delta_{Plug}$), 
the $1-\beta$ reweighting of the simulation 
($\delta_{\beta(2)}$) and the contribution to the hadronic energy
from charged particle tracks ($\delta_{tra}$).
Minus signs appear for these
systematics if the shift in a variable is
anti-correlated rather than correlated with the shift in the cross section.}
\label{lar:data}
\end{table} 
\end{landscape}

\begin{footnotesize} 

\begin{table} 
\centering
\begin{tabular}{|c | c | c | c | c | c |} 
\hline 
$\xpom$ range & $\langle \xpom \rangle$ & ${\rm d} \sigma / {\rm d} \xpom$ $[{\rm pb}]$ & $\delta_{stat}$ & $\delta_{sys}$ & $\delta_{tot}$ \\ 
 & & $Q^2>200 \ {\rm GeV^2}, y<0.9, \xpom<0.05$ & (\%) & (\%) & (\%)\\
\hline \hline
0.005 - 0.016 & 0.01 & 15 & 59 & 25 & 64 \\ 
0.016 - 0.05  & 0.03 & 7.7 & 37 & 18 & 41 \\ 
 \hline 
\end{tabular} 

\vspace*{1.5cm}

\begin{tabular}{|c | c | c | c | c | c |} 
\hline 
$Q^2$ range & $\langle Q^2 \rangle$ & ${\rm d} \sigma / {\rm d} Q^2$ [${\rm pb/GeV^{2}}$] & $\delta_{stat}$ & $\delta_{sys}$ & $\delta_{tot}$ \\
$[{\rm GeV^{2}}]$ & $[{\rm GeV^{2}}]$ & $Q^2>200 \ {\rm GeV^2}, y<0.9, \xpom<0.05$  & (\%) & (\%) & (\%)\\
\hline  \hline
200 - 560 &  350  & 6.0$\;\cdot\; 10^{-4}$ & 50 & 21 & 54 \\ 
560 - 2240 & 1150 & 1.0$\;\cdot\; 10^{-4}$ & 40 & 18 & 44 \\ 
 \hline 
\end{tabular} 

\vspace*{1.5cm}

\begin{tabular}{|c | c | c | c | c | c |} 
\hline 
$\beta$ range & $\langle \beta \rangle$ & ${\rm d} \sigma / {\rm d} \beta$ [${\rm pb} $] & $\delta_{stat}$ & $\delta_{sys}$ & $\delta_{tot}$ \\
&  & $Q^2>200 \ {\rm GeV^2}, y<0.9, \xpom<0.05$  & (\%) & (\%) & (\%)\\
\hline  \hline
0.3 - 0.5 & 0.40 & 0.58 & 62 & 27 & 68 \\ 
0.5 - 0.8 & 0.65 & 0.44 & 44 & 21 & 48 \\ 
0.8 - 1.0 & 0.90 & 0.17 & 79 & 17 & 81 \\ 
 \hline 
\end{tabular} 
\caption{Measurements of the cross section for the
diffractive charged current process $e^+ p \rightarrow \bar{\nu}_e XY$
for $Q^2 > 200 \ {\rm GeV^2}$, $y < 0.9$ and $\xpom < 0.05$
with $E_p = 920 \ {\rm GeV}$, differential in $\xpom$, $Q^2$
and $\beta$. The differential
cross sections correspond to average values over the
ranges shown. The 
percentage statistical, systematic and total uncertainties 
are also given.}
\label{cc:tab}
\end{table} 

\end{footnotesize}

\end{document}